\documentclass{emulateapj} 

\usepackage{apjfonts} 
\usepackage{natbib} 
\usepackage{color}
\usepackage{multirow}
\usepackage{booktabs}
\usepackage{amssymb}
\usepackage{amsmath}
\def\cm3{cm$^{-3}$}
\def\kms{km~s$^{-1}$}
\def\vlsr{v$_{\rm{LSR}}$}
\def\thet1{${\theta}^1$~Ori~C}

\newcommand{\mum}{\ifmmode{\rm \mu m}\else{$\mu$m}\fi}

\shorttitle{C$^+$, CH$^+$, CH in Orion BN/KL}
\shortauthors{P. Morris et al. }

\begin{document}

\title{Herschel/HIFI Spectral Mapping of C$^+$, CH$^+$, and CH in Orion BN/KL:  \\
The Prevailing Role of Ultraviolet Irradiation in CH$^+$ Formation}

\author{Patrick W. Morris\altaffilmark{1},
Harshal Gupta\altaffilmark{1,2},
Zsofia Nagy\altaffilmark{3,4}, 
John C. Pearson\altaffilmark{5},
Volker~Ossenkopf-Okada\altaffilmark{3},
Edith~Falgarone\altaffilmark{6},
Dariusz~C.~Lis\altaffilmark{7,8},
Maryvonne~Gerin\altaffilmark{6},
Gary~Melnick\altaffilmark{9},
David~A.~Neufeld\altaffilmark{10},
Edwin~A.~Bergin\altaffilmark{11}
}

\email{pmorris@ipac.caltech.edu}

\altaffiltext{1}{California Institute of Technology, NASA Herschel Science Center, IPAC M/C 100-22, Pasadena, CA 91125.}
\altaffiltext{2}{Division of Astronomical Sciences, National Science Foundation, 4201 Wilson Boulevard, Suite 1045, Arlington, VA 22230}
\altaffiltext{3}{I. Physikalisches Institut, Universit\"at zu K\"oln, Z\"ulpicher Str. 77, D-50937 K\"oln, Germany}
\altaffiltext{4}{Department of Physics and Astronomy, University of Toledo, 2801 West Bancroft Street, Toledo, OH 43606}
\altaffiltext{5}{California Institute of Technology, Jet Propulsion Laboratory, 4800 Oak Grove Drive, Pasadena, CA 91109.}
\altaffiltext{6}{LERMA, Observatoire de Paris, PSL Research University, CNRS, Sorbonne Universit\'es, UPMC Univ. Paris 06, \'Ecole normale sup\'erieure, F-75005, Paris, France}
\altaffiltext{7}{LERMA, Observatoire de Paris, PSL Research University, CNRS, Sorbonne Universit\'es, UPMC Univ. Paris 06, F-75014, Paris, France}
\altaffiltext{8}{California Institute of Technology, Cahill Center for Astronomy and Astrophysics M/C 301-17, Pasadena, CA 91125.}
\altaffiltext{9}{Harvard-Smithsonian Center for Astrophysics, 60 Garden Street, Mail Stop 66, Cambridge MA 02138, USA}
\altaffiltext{10}{Department of Physics and Astronomy, Johns Hopkins University, 3400 North Charles Street, Baltimore, MD 21218, USA}
\altaffiltext{11}{Department of Astronomy, University of Michigan, 500 Church Street, Ann Arbor, MI 48109, USA}

\begin{abstract}
 
The CH$^+$ ion is a key species in the initial steps of interstellar carbon chemistry. Its formation in diverse environments where it is observed is not well understood, however, because the main production pathway is so endothermic (4280 K) that it is unlikely to proceed at the typical temperatures of molecular clouds.  We investigation CH$^+$ formation with the first velocity-resolved spectral mapping of the CH$^+$ $J=1-0, 2-1$ rotational transitions, three sets of CH $\Lambda$-doubled triplet lines, $^{12}$C$^+$ and $^{13}$C$^+$, and CH$_3$OH 835~GHz E-symmetry Q branch transitions, obtained with Herschel/HIFI over $\approx$12 arcmin$^2$ centered on the Orion BN/KL source.  We present the spatial morphologies and kinematics, cloud boundary conditions, excitation temperatures, column densities, and $^{12}$C$^+$ optical depths.  Emission from C$^+$, CH$^+$, and CH is indicated to arise in the diluted gas, outside of the explosive, dense BN/KL outflow.  Our models show that UV-irradiation provides favorable conditions for steady-state production of CH$^+$ in this environment.  Surprisingly, no spatial or kinematic correspondences of these species are found with H$_2$ S(1) emission tracing shocked gas in the outflow.  We propose that C$^+$ is being consumed by rapid production of CO to explain the lack of C$^+$ and CH$^+$ in the outflow, and that fluorescence provides the reservoir of H$_2$ excited to higher ro-vibrational and rotational levels.  Hence, in star-forming environments containing sources of shocks and strong UV radiation, a description of CH$^+$ formation and excitation conditions is incomplete without including the important --- possibly dominant --- role of UV irradiation. 
  
\end{abstract}

\keywords{ISM: abundances --- clouds --- astrochemistry --- molecules --- molecular processes --- lines and bands --- sub-millimeter}

\section{Introduction}

The diatomic hydrides CH and CH$^+$ were, along with CN, the first molecules to be identified from their optical absorption spectra in diffuse molecular clouds,  some 75 years ago (McKellar 1940; Douglas \& Herzberg 1941; Adams 1941).  Advances in detector technologies and observing techniques now make the optical lines readily observable toward bright background stars (e.g. Crane, Lambert, \& Sheffer 1995; Zsarg\'o \& Federman 2003; Le Petit et al.  2004; Shaw et al.  2006; Nehm\'e et al.  2008; Boiss\'e et al.  2009), yet fundamental carbon chemistry involving C$^+$, CH and CH$^+$ in interstellar gas is not well enough understood to explain the observed abundances of CH$^+$ in particular.  The standard gas-phase model with a chemistry in which heating is provided by a background far-ultraviolet (FUV) field from hot stars generally works well for matching the abundances of CH observed in photon-dominated regions (PDRs), but falls short by orders of magnitude for CH$^+$ (Van Dishoek \& Black 1986; Gredel et al.  1993; Crane et al.  1995; Gredel 1997; Pan et al.  2004, 2005; Sheffer et al.  2008; Godard et al.  2009).  

Direct formation pathways involving C$^+$, H, and H$_2$ fail to explain the high abundances of CH$^+$.   First, the radiative association between C$^+$ and H is too slow at the typical temperatures of molecular clouds: the rate coefficient of the process $\rm{C^+(^2{\it P}_{3/2,1/2}) + \rm{H}(^2{\it S}_{1/2}) \rightarrow \rm{CH}^+(^1{\Sigma}^+) + \it{h\nu}}$ was calculated by Kramers \& Ter Haar (1946) and Bates (1951) to be around $2 \times 10^{-18} \rm{cm}^3 \; \rm{s}^{-1}$ at 100 K, a few orders of magnitude too low to yield the observed abundances.  More detailed calculations involving coupling between spin-orbit and rotational states were carried out by Graff et al.  (1983) and Barinovs \& van Hemert (2006) over temperatures up to 1000 K, but the rate increases significantly only at temperatures below 1~K, where the frequency of resonances in the radiative association rate coefficients increases substantially.  Second, the reaction of C$^+$ with molecular hydrogen, C$^+ \; +$ H$_2 \; \rightarrow \; $ CH$^+$ + H,  is highly endothermic ($-{\Delta}E/k = 4280$ K; Hierl et al. 1997), requiring high gas temperatures, which likely result from non-equilibrium processes, or a reservoir of H$_2$ vibrationally excited by FUV fluorescence (Sternberg \& Dalgarno 1995). Additionally, the reaction must proceed fast enough to counteract the high destruction rate of CH$^+$ by collisions with H$_2$, H, and electrons.   

\begin{figure*}
  \begin{center}
  \includegraphics[width=17cm]{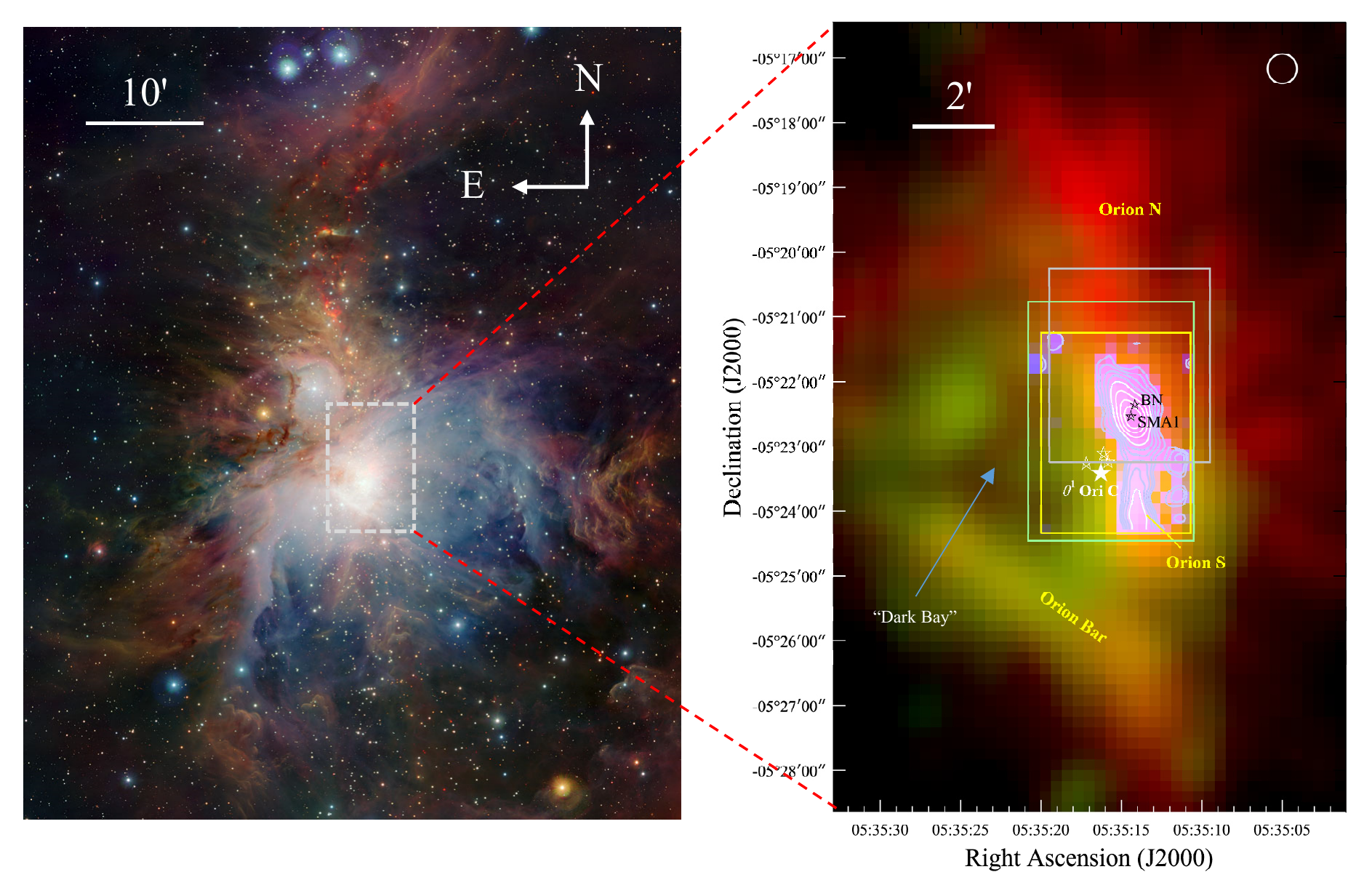}
  \end{center}
  \caption{On the left, a wide field composite image of M42 taken with the VIRCAM instrument on the European Southern Observatory's Visible and Infrared Survey for Astronomy (VISTA) telescope at 2.15 (red), 1.25 (green), and 0.88 (blue) $\mu$m (credit: ESO/J. Emerson/VISTA).  On the right, a composite image of OMC~1 using {\em{Herschel}}/SPIRE 350 $\mu$m (red), HIFI CH$^+$ $J=1-0$ (green), and HIFI 1900 GHz (158 $\mu$m) continuum (light blue), each on a normalized intensity scale.  Overlaid boxes indicate HIFI spectral mapping regions that are the subject of this study: C$^+$ (yellow), CH$^+$ $J = 1 - 0$ (green), and CH $^2{\Pi}_{3/2} - ^2{\Pi}_{1/2}$ 1657/1661 GHz and CH$^+$ $J = 2 - 1$ (gray).  Continuum contours in light blue and white indicate the upper 50\% and 30\% of emitted power, respectively, within the indicated region of C$^+$ mapping.  The HIFI beam size 25$''$.3 HPBW at 835 GHz corresponding to the CH$^+$ $J=1-0$ observations is represented in the top corner of the {\em{Herschel}} composite image.  The SPIRE beam at 350 $\mu$m (850 GHz) is very similar, 23$''$.9 FWHM.  The 7$'.5 \times 10'$ CH$^+$ $J=1-0$ map is from the Open Time program OT1\_jgoicoec\_4 (PI  J. Goicoechea), and the SPIRE scan map is from GT2\_pandre\_5 (PI: P. Andr{\'e}). 
  \label{fig:f1}}
\end{figure*}

The observed mean column density of CH$^+$ log$_{\rm{10}}$[$N$(CH$^+$)/$N$(H)] = $-8.11$$\pm$0.33 in the ISM (from a compilation of references by Godard \& Cenricharo 2013) has been proposed to trace  production through non-equilibrium processes, as may arise in relatively low velocity ($\sim$10 {\kms}) magnetohydrodynamic (MHD) waves and C-type shocks,  turbulent mixing, or dissipation of turbulence.  The importance of CH$_n^+$ chemistry in MHD shocks for the abundances of CH, OH, and HCO$^+$ in warm diffuse clouds was emphasized by Draine \& Katz (1986a,b) and Flower \& Pineau des For\^ets (1998), although both neutral and MHD shocks in the ISM have been disfavored in some observations (Gredel et al. 1993; Crawford 1995). Draine, Roberge, \& Dalgarno (1983) had earlier pointed out that ion-neutral drift in MHD waves preferentially helps drive endothermic ion-neutral reactions in molecular clouds, not just by raising the temperature but also by increasing the collision energies to values $\gg kT$.  Zsarg{\'o} \& Federman (2003) have also addressed the effects of non-thermal CH$^+$ chemistry on C-bearing molecules in diffuse clouds with low molecular abundances.    In the turbulent dissipation region (TDR) model of Godard et al.  (2009), the abundances of CH$^+$, SH$^+$, CN, H$_2$O, and many other molecules rise by more than 2 orders of magnitude when the activation barriers are overcome, as gas is heated via ion-neutral friction in the magnetized structure of a turbulent dissipative burst and during the subsequent relaxation period.   This model has been applied by Falgarone et al.  (2010) to observations of the CH$^+$ ground state rotational ($J = 1 - 0$) transition in emission and absorption towards the bright massive star-forming region DR~21(OH), providing evidence that non-equilibrium chemistry successfully explains the high CH$^+$ abundances without over-production of other molecules.    

\begin{deluxetable*}{l c c c c c r r r}
  \tablewidth{0pc}
  \tablecaption{HIFI Observations Summary.\label{obs_t}}

  \tablehead{\colhead{Obs ID} & 
  \colhead{Obs Mode\tablenotemark{a}} & 
  \colhead{Map Area} & 
  \colhead{Sampling\tablenotemark{b}} & 
  \colhead{Obs Date} & 
  \colhead{Program\tablenotemark{c}} & 
  \colhead{Obs Time} & 
  \colhead{LO freq.\tablenotemark{d,e}} & 
  \colhead{$\eta_{\rm{mb}}$\tablenotemark{f}} \\
  \colhead{ } & 
  \colhead{ } & 
  \colhead{ } & 
  \colhead{ } & 
  \colhead{ } & 
  \colhead{ } & 
  \colhead{(sec)} & 
  \colhead{(GHz)} &
  \colhead{ }
  }
  
  \startdata
 
1342190786 &	2 &	$1'.2\times1'.1$  &  H  &  	2010 Feb 18 &  	C &  	6637	 &  1896.9  & 0.60 \\
1342190787 &	1 &	$2'.2\times2'.3$  &  H  &  	2010 Feb 18 &  	C &  	17287 &   1896.9 & 0.60 \\ 
1342192562 &   5 &  \dotfill & \dotfill  & 2010 Mar 22 & H & 13336 & 1578-1698 & 0.57  \\
1342203225 &	1 &	$2'.1\times2'.4$  &  H  &  	2010 Aug 19 &   C &  	1382	 &  841.2  & 0.63   \\
1342203226 &	1 &	$2'.1\times2'.4$  &  H  &  	2010 Aug 19 &  	C &  	1051	 &  841.2  & 0.63 \\
1342203240 &	3\tablenotemark{h} &	$1'.1\times1'.2$  &  N  &  	2010 Aug 20 &  	C &  	8210	 &  1897.4 & 0.60 \\
1342203241 &	2 &	$1'.1\times1'.2$  &  N  &  	2010 Aug 20 &  	C &  	2680	 &  1897.4  & 0.60 \\
1342203242\tablenotemark{g} &	1 &	$1'.1\times1'.2$  &  N  &  	2010 Aug 20  &  C &  	8935 &  1897.4 & 0.60  \\
1342203731 &   1 & $3'.4\times4'.1$ & H & 2010 Aug 30 & H & 1114 & 543.0 & 0.61 \\
1342203926 &   6 & \dotfill & \dotfill & 2010 Sep 01 & C & 2152 & 1474.1 & 0.57 \\
1342203949 &	4 &	$2'.5\times2'.5$  &  H  &  	2010 Sep 02  &  H  &  27069 &  	1666.2  & 0.57 \\
1342205541 & 4 & $1'.8\times1'.9$ & N & 2010 Oct 02 & H & 1838.3 & 28003 & 0.60 \\
1342227557 &  1 & $1'.7\times1'.9$ & N & 2011 Aug 27 & C & 551 & 1158.2 & 0.63 \\
1342228593 &	1 &	$2'.3\times2'.7$  &  N  &  	2011 Sep 14 &   C &  	1543	 &  841.2  & 0.63 \\
1342229765\tablenotemark{g} &	2 &	$1'.9\times2'.0$  &  N  &    2011 Sep 16  &  	C  &  4000	 &  1897.4 & 0.60 \\
1342239586 &	1 &	$2'.5\times3'.2$  &  N  &  	2012 Feb 23  &  C &  	1788	 &  841.2  & 0.63 \\
1342239623 &	2 &	$2'.0\times3'.0$  &  N  &  	2012 Feb 24 &  	C &  	5488	 &  1897.4 & 0.60 \\
1342251025 &	1 &	$2'.5\times3'.2$  &  N  &  	2012 Sep 13 &  	C &  	1788	 &  841.2 & 0.63 \\
1342251084 &	2 &	$2'.0\times3'.0$  &  N  &  	2012 Sep 17 &  	C &  	5488	 &  1897.4  & 0.60 \\
1342266602 &	2 &	$2'.7\times1'.9$  &  N  &  	2013 Feb 28 &  	C &  	5488	 &  1897.4 & 0.60 \\
1342266892 &	1 &	$2'.8\times2'.3$  &  N  &  	2013 Mar 09 &  	C &  	1788	 &  841.2  & 0.63 \\

  \enddata

  \tablenotetext{a}{Observing modes: (1) OTF map; (2) OTF map with Load Chop; (3) OTF map with Frequency Switch; (4) Fast DBS Raster map; (5) Spectral Scan with Fast DBS; (6) Point DBS.}
  \tablenotetext{b}{Sampling: N = Nyquist; H = Half-Beam.}
  \tablenotetext{c}{Program: C = HIFI Calibration; H = HEXOS.}
  \tablenotetext{d}{LO frequency is the setting requested in HSpot.}
  \tablenotetext{e}{Noise requirements over a line-free 1 GHz wide WBS sub-band for both H and V polarizations combined:  300-500 mK at 1897 GHz; 100 mK at 841 GHz; 300 mK at 1666 GHz; 60 mK at 1841 GHz.} 
  \tablenotetext{f}{$\eta_{\rm{mb}}$ are the HIFI main beam efficiencies, taken from calibrations with HIPE 14.0}
  \tablenotetext{g}{Maps are oriented with a rotation angle of 90$^\circ$.}
  \tablenotetext{h}{Processed but excluded from this study; see text.}
\end{deluxetable*}

In denser gas around star forming regions with strong far-ultraviolet (FUV) radiation fields and in PDRs, the internal energy of vibrationally excited H$_2$ ($\nu > 0$) can overcome the high activation barrier to form CH$^+$ (Sternberg \& Dalgarno 1995; Hierl et al. 1997; Ag{\'u}ndez et al.  2010).   Indeed the CH$^+$ intensities up to $J = 6-5$ observed in the Orion Bar have been reproduced by Nagy et al.  (2013) with PDR models involving collisions of C$^+$ with vibrationally excited H$_2$.  Their results indicate that the CH$^+$ abundance peaks in the warm, high-pressure ($T\approx 500\text{-}1000$ K; $p/k = 10^8$ cm$^{3}$ K) surface regions of the PDR.  Godard \& Cernicharo (2013) have modeled the complete energy structure of CH$^+$ during formation, treating pumping of the vibrational and bound and unbound electronic states over all levels and including non-reactive collisions with H, H$_2$, He, and electrons.  Their results show that chemical pumping or formation in excited states drives the distribution over its rotational levels, while both photo-dissociation and radiative pumping of bound states have only a minor influence on the distribution.  One of the surprising consequences of their results is that the intensities of the rotational lines of CH$^+$ observed towards the Orion Bar and the NGC~7027 PDRs are reproduced with H densities one to two orders of magnitude lower than those inferred from previous excitation models. 

It is important to stress that while the reaction of C$^+$ with H$_2$ is the main production route to CH$^+$, it can operate through the different mechanisms mentioned above in different environments, i.e., with H$_2$ in different excited states (e.g., Hierl et al. 1997; Ag{\'u}ndez et al. 2010; Zanchet et al. 2013).  In diffuse gas of the ISM with H densities of a few tens to a few hundreds of cm$^{-3}$, the route is dominated by the intermittent dissipation regions of turbulence.  In star-forming regions where warm and dense molecular material is exposed to FUV radiation or shocked outflows, reactions may proceed much faster as C$^+$ and vibrationally-excited H$_2$ are both abundant.  In either scenario, CH$^+$ is produced with an excess energy, and should be rapidly destroyed by reacting with H$_2$ and electrons. Hence the distribution of the population in the rotational levels may not be at equilibrium, depending on the environment and dominant formation pathway, i.e., in TDRs, or through UV- or shock-driven chemistries in denser molecular clouds.  Until now, there have been very few opportunities to study the physical characteristics of CH$^+$ in dense clouds of star forming regions, in relation to shock-heated {\em{versus}} UV-irradiated gas.

Observations of the two lowest rotational transitions of CH$^+$ have been made possible with the Heterodyne Instrument for the Far Infrared (HIFI; de Graauw et al.  2010; Roelfsema et al.  2012), which flew on board the {\em{Herschel}} Space Observatory (Pilbratt et al.  2010).  The majority of these observations were taken in single pointings or as spatially composite measurements.  As such, radial velocity measurements are the only means to differentiate between ISM components in these observations.  The nearby Orion Molecular Cloud 1 (OMC~1) surrounding the BN/KL massive SFR offers a varied and complex environment, in which to study the characteristics of the CH$^+$ molecule.  The embedded BN/KL SFR is the second most luminous site of star formation in the OMC and exhibits evidence for an energetic outburst, possibly the result of a stellar merger some 500 $-$ 1000 years ago (Bally \& Zinnecker 2005; Zapata et al.  2011; Bally et al.  2011).  The feature is observed as loosely collimated bipolar outflows producing H$_2$O masers (Genzel et al.  1981; Genzel \& Stutzki 1989; Greenhill et al.  1998), shock-excited H$_2$ (Beckwith et al.  1978), and high-$J$ CO and various other molecules with broadened line wings (Zuckerman et al.  1976; Kwan \& Scoville 1976; Erickson et al.  1982; Chernin \& Wright 1996).   The striking butterfly-shaped H$_2$ outflow is extended over $\sim$30 arcsec (0.4 pc at a distance of 420 pc; Menten et al. 2007; Hirota et al. 2007) in a NW-SE direction, tracing the faster (30 - 100 km s$^{-1}$) of the two outflows detected from H$_2$O maser emission by Genzel et al.  (1981).  We should naturally expect that H$_2$ excited to sufficient vibrational or rotational levels in gas heated by the powerful molecular outflow (or more accurately an explosive, momentum-driven outburst; e.g., Bally \& Zinnecker 2005)  will react with C$^+$, if also present, to form CH$^+$.  In Orion BN/KL we must account for the strong extended thermal continuum emission from heated dust.

The principle aim of this paper is to investigate the spatial, kinematic, and physical conditions in the C$^+$ and CH$^+$ gas around Orion BN/KL, and their relations to the vibrationally-excited H$_2$ emission,  allowing us to further describe the condtions which lead to the endothermic formation of CH$^+$.  The intermediate (IF) frequency coverage of our CH$^+$ $J= 1 - 0$ observations includes several CH$_3$OH ground state $E$-symmetry transitions, allowing us to trace the highest density regions where dust grain evaporation is presumed to be taking place (Caselli et al.  1993).    We also present spectral maps of the CH $^2{\Pi}_{3/2} - ^2{\Pi}_{1/2}$ triplet lines at 537 GHz, and fixed-position observations of the $^2{\Pi}_{3/2} - ^2{\Pi}_{3/2}$ (1471/1477 GHz) and  $^2{\Pi}_{5/2} - ^2{\Pi}_{3/2}$ (1657/1661 GHz) triplets towards the dense Hot Core.   The formation of CH is believed to occur where both C$^+$ and excited H$_2$ are abundant and can react by radiative association in a UV-driven chemistry (Black \& Dalgarno 1973).  When collision temperatures are below 400~K, radiative association between C$^+$ and H$_2$ proceeds slowly to produce CH$_2^+$, which reacts with H$_2$ to produce CH$_3^+$, subsequently forming CH$_2$, CH, and C by dissociative electronic recombination of CH$_3^+$ and CH$_2^+$, and CH$^+$ by photodissociation of CH$_3^+$ (e.g., Godard et al. 2009; Ag\'undez et al. 2010; Indriolo et al. 2010). In the dense gas of SFRs, CH abundances relative to CH$^+$ range from factors 1 to 3 (Godard et al. 2012).  In more diffuse, translucent gas where high abundances of CH$^+$ are observed, non-thermal CH$^+$ chemistry may be important for production of CH and CO, as shown by Zsarg{\'o} \& Federman (2003). Their results suggest that non-thermal chemistry is necessary to account for the observed abundance of CH in these clouds, where on average some 30-40\% (but as much as 90\%) of the observed CH along some sightlines, has been attributed to originating from non-equilibrium synthesis of CH$^+$.  Through the linear relation between CH and H$_2$ column densities $N$(CH) = 3.5 $\times$ 10$^{-8} N$(H$_2$) derived by Sheffer et al.  (2008) to $A_V$ of around 5 mag.   In steady-state chemical models, on the other hand CH traces H$_2$ at hydrogen densities greater than about 100 cm$^{-3}$  (Liszt \& Lucas 2002; Levrier et al. 2012).  Thus in warm and dense environments such as Orion BN/KL, the relationship between H$_2$ and CH could be more complex, as the abundances of CH may be at least partially a result of radiative association between C$^+$ and H$_2$, which are more abundant in this environment compared to the ISM.

\section{Observations and Data Reduction}

\label{obs_s}

The observations presented here were obtained with the HIFI instrument, principally using the on-the-fly (OTF) spectral mapping mode, centered approximately on the BN source, as shown schematically in Figure~\ref{fig:f1}.  Most maps employ both the High Resolution Spectrometer (HRS) and the Wide Band Spectrometer (WBS).  The nominal WBS resolution of 1.1~MHz, with a channel spacing of 0.5 MHz, provides a maximum velocity resolution of $\sim$0.2 km~s$^{-1}$ at 1900 GHz and 0.7 km~s$^{-1}$ at 835 GHz, resolving all lines of interest to this study.  At 835 GHz in band 3a, CH$_3$OH $K=5{\rightarrow}4$ Q branch torsional ground state ($v_t = 0$) $J_{u,l}$ 5 through 18 transitions are contained in the lower sideband within the 4.0 GHz IF range. Since HRS and WBS noise is correlated (using the same mixer assemblies), co-addition of those data does not improve data quality, thus we present only the WBS observations here.
The C$^+$ $^2P_{3/2} - ^2P_{1/2}$ 1900.537 GHz and CH$^+$ J = $1-0$ 835.137 GHz maps were carried out in calibration time by the HIFI Instrument Control Center for performance validations of the HIFI Astronomical Observing Templates (AOTs).  As summarized in Table~\ref{obs_t}, the observations have been acquired using each of the three calibration schemes available with the OTF mode, and Nyquist or half-beam spacing of the map points.  While most maps were taken in a North-South line scanning direction, two maps were taken in an East-West direction, for evaluation of non-uniform beam sampling effects and repeatability.  The $^{12}$C$^+$ observations include the three $^{13}$C$^+$ $^2P_{3/2} - ^2P_{1/2}$ fine structure lines within the 2.4 GHz wide intermediate frequency (IF) band pass of band 7b.  

We also include the analysis of spectral maps taken in the {\it{Herschel}} Observations of EXtraOrdinary Sources (HEXOS) Key Program (PI E. Bergin) which target water transitions and cover the CH$^+$ $J = 2-1$ at 1669.281 GHz and the CH $^2{\Pi}_{3/2} - ^2{\Pi}_{1/2}$ triplet lines at 537 GHz.  The CH$^+$ $J= 2-1$ observation was carried out as a raster map using Fast Dual Beam Switching (DBS), which applies alternating nods of the telescope and chopping of the internal M3 pick-up mirror by 3$'$ on the sky on either side of the mapped region for calibrations of the baseline.  A chop frequency of 3.5 Hz and phase length of 0.14 sec was timed for optimum stabilization of the baseline at this LO frequency.  No contamination from CH$^+$ $J = 2-1$ emission or absorption has been detected in the alternated sky reference positions, by inspection of data from both nod positions processed through standing wave corrections in HIPE 14.0.

\begin{deluxetable*}{l l l l r r l}
  \tablewidth{0pc}
  \tablecaption{Physical parameters of the transitions observed with HIFI in this study.\label{restfreq}}

\tablehead{\colhead{Species} & 
  \colhead{Transition} & 
  \colhead{Rest Freq.} & 
  \colhead{Ref.\tablenotemark{a}} &
  \colhead{E$_{\rm{up}}$} & 
  \colhead{log$_{\rm{10}}$(A$_{ij}$)} &
  \colhead{BN/KL Line Crowding\tablenotemark{b}} \\ 
  \colhead{} & 
  \colhead{} & 
  \colhead{GHz} & 
  \colhead{} &
  \colhead{K} &
  \colhead{s$^{-1}$} & 
  \colhead{}} 
  
  \startdata

$^{12}$C$^+$  & $^2P_{3/2} - ^2P_{1/2}$ & 1900.5369 & C, 1 & 91.2 & $-$5.64 & \\
& & & & & & \\
$^{13}$C$^+$  & $^2P_{3/2} - ^2P_{1/2}$ $F = 1 - 1$ & 1900.1360 & C, 1, 2 & 91.2 & $-$6.11 & H$_2$S 7$_{1,6}$-7$_{0,7}$ 1900.14, 7$_{2,6}$-7$_{1,7}$ 1900.18 GHz \\
                        & $^2P_{3/2} - ^2P_{1/2}$ $F = 2 - 1$ & 1900.4661 & C, 1, 2 & 91.2 & $-$5.63 &  $^{12}$C$^+$ 1900.5369 GHz \\
                        & $^2P_{3/2} - ^2P_{1/2}$ $F = 1 - 0$ & 1900.9500 & C, 1, 2 & 91.2 & $-$5.81 &  H$_{2}^{18}$O 3$_{2,2}$-3$_{1,3}$ 1894.32 GHz \\
& & & & & & \\
$^{12}$CH$^+$ &  J = $1 - 0$ & 835.1369 & C, 3, 4& 40.1 & $-$2.20 &   \\
                        & J = $2 - 1$ & 1669.2810 & C, 4 & 120.2 & $-$1.21 &  \\
& & & & & & \\
$^{12}$CH &  $^2\Pi_{3/2} - ^2\Pi_{1/2}$ $F = 2^-$ - 1$^+$ & 536.76266 & J, 5, 6 & 25.8 & $-$3.20 & \\
    &  $^2\Pi_{3/2} - ^2\Pi_{1/2}$ $F = 1^-$ - 1$^+$ & 536.78354 & J, 5, 6 & 25.8 & $-$3.67 & \\
    &  $^2\Pi_{3/2} - ^2\Pi_{1/2}$ $F = 1^-$ - 0$^+$ &  536.79729 & J, 5, 6 & 25.8 & $-$3.38 & SO$_2 10_{4,6} - 9_{3,7}$ 549.30 GHz\\
   &  $^2\Pi_{3/2} - ^2\Pi_{3/2}$ $F = 1^+$ - 2$^-$ & 1470.68940 & C, 5, 6 & 96.3 & $-$3.05326 & \\
   &  $^2\Pi_{3/2} - ^2\Pi_{3/2}$ $F = 1^+$ - 1$^-$ & 1470.69169 & C, 5, 6 & 96.3 & $-$2.35435 & \\
   &  $^2\Pi_{3/2} - ^2\Pi_{3/2}$ $F = 2^+$ - 2$^-$ & 1470.73960 & C, 5, 6 & 96.3 & $-$2.32069 & \\
   &  $^2\Pi_{3/2} - ^2\Pi_{3/2}$ $F = 2^+$ - 1$^-$ & 1470.74189 & C, 5, 6 & 96.3 & $-$3.27559 & \\
   &  $^2\Pi_{3/2} - ^2\Pi_{3/2}$ $F = 1^+$ - 1$^-$ & 1477.29211 & C, 5, 6 & 96.7 & $-$2.35264 & \\
   &  $^2\Pi_{3/2} - ^2\Pi_{3/2}$ $F = 1^+$ - 2$^-$ & 1477.31293 & C, 5, 6 & 96.7 & $-$3.05513 & \\
   &  $^2\Pi_{3/2} - ^2\Pi_{3/2}$ $F = 2^+$ - 1$^-$ & 1477.36552 & C, 5, 6 & 96.7 & $-$3.27726 & \\
   &  $^2\Pi_{3/2} - ^2\Pi_{3/2}$ $F = 2^+$ - 2$^-$ & 1477.38634 & C, 5, 6 & 96.7 & $-$2.32266 & \\
    &  $^2\Pi_{5/2} - ^2\Pi_{3/2}$ $F = 3^+$ - 2$^-$ &1656.96136 & C, 5, 6 & 105.2 & $-$1.42954 & \\
    &  $^2\Pi_{5/2} - ^2\Pi_{3/2}$ $F = 2^+$ - 2$^-$ &1656.97065 & C, 5, 6 & 105.2 & $-$2.42971 & \\
    &  $^2\Pi_{5/2} - ^2\Pi_{3/2}$ $F = 2^+$ - 1$^-$ &1656.96136 & C, 5, 6 & 105.2 & $-$1.47521 & \\
    &  $^2\Pi_{5/2} - ^2\Pi_{3/2}$ $F = 3^+$ - 2$^-$ &1661.10745 & C, 5, 6 & 105.5 & $-$1.42587 & H$_2$O 2$_{2,1} - 2_{1,2}$ 1660.0 GHz \\
    &  $^2\Pi_{5/2} - ^2\Pi_{3/2}$ $F = 2^+$ - 1$^-$ &1661.11822 & C, 5, 6 & 105.5 & $-$1.47163 & H$_2$O 2$_{2,1} - 2_{1,2}$ 1660.0 GHz \\
    &  $^2\Pi_{5/2} - ^2\Pi_{3/2}$ $F = 2^+$ - 2$^-$ &1651.13903 & C, 5, 6 & 105.5 & $-$2.42603 & H$_2$O 2$_{2,1} - 2_{1,2}$ 1660.0 GHz \\
& & & & & & \\
CH$_3$OH\tablenotemark{c} & $5_{5,1} \rightarrow 5_{4,1}$ & 835.00376 & C, 7, 8 & 170.9 & $-$2.99 \\
 & $6_{5,2} \rightarrow 6_{4,2}$ & 834.95935 & C, 7, 8 & 184.8 & $-$2.80 & \\
 & $7_{5,3} \rightarrow 7_{4,3}$ & 834.90431 & C, 7, 8 & 201.1 & $-$2.71 & \\
 & $8_{5,4} \rightarrow 8_{4,4}$ & 834.83715 & C, 7, 8 & 219.6 & $-$2.66 & \\
 & $9_{5,5} \rightarrow 9_{4,5}$ & 834.75617 & C, 7, 8 & 240.5 & $-$2.62 & \\
 & $10_{5,6} \rightarrow 10_{4,6}$ &	834.65944 & C, 7, 8 & 263.7 & $-$2.60 & \\
 & $11_{5,7} \rightarrow 11_{4,7}$ & 834.54485 & C, 7, 8 & 289.2 & $-$2.59 & \\
 & $12_{5,8} \rightarrow 12_{4,8}$ & 834.41005 & C, 7, 8 & 317.1 & $-$2.58 & \\
 & $13_{5,9} \rightarrow 13_{4,9}$ & 834.25251 & C, 7, 8 & 347.2 & $-$2.57 & \\
 & $14_{5,10} \rightarrow 14_{4,10}$ & 834.06946 & C, 7, 8 & 379.7 & $-$ 2.56 & SO$_2 15_6 - 14_5$ 848.52 GHz \\
 & $15_{5,11} \rightarrow 15_{4,11}$ &  833.85794 & C, 7, 8 & 414.5 & $-$2.56 & \\
 & $16_{5,12} \rightarrow 16_{4,12}$ &  833.61481 & C, 7, 8 & 451.6 & $-$2.56 & HDO $2_{1,2} - 1_{1,1}$ 848.96 GHz \\   
 
  \enddata

  \tablenotetext{a}{References: (C) Cologne Database for Molecular Spectroscopy (M\"uller et al.  2005); (J) JPL Molecular Spectroscopy Database (Pickett et al.  1998); (1) Cooksy et al. (1986); (2) Ossenkopf et al. (2013); (3) Pearson \& Drouin (2006); (4) M\"uller (2010); (5) Brown \& Evenson (1983); (6) Davidson et al. (2001); (7) Xu \& Lovas (1997); (8) Comito et al. (2005).}
  \tablenotetext{b}{Notable line blending or crowding refers to the inner 15$''$ of the BN/KL complex centered on IRc~2.   Species separated by more than 12 GHz from the rest frequencies of column 3 are from the image side band.}
  \tablenotetext{c}{CH$_3$OH transitions are of E-symmetry $K 5 \leftrightarrow 4 \; v_t$ = 0.}
\end{deluxetable*}

Every OTF observation includes measurements at a blank sky reference position, offset from the center of the map by typically 20$'$.  The reference measurement was taken to correct for the standing waves internal to the optical paths between the instrument mixer assemblies and the telescope secondary mirror, and electrical standing waves originating in the mixer amplifier chains at LO frequencies of 1897 GHz (C$^+$ mapping) and 1666 GHz (CH $^2\Pi_{5/2} - ^2\Pi_{3/2}$ mapping).
Spectra at the OFF positions were inspected for contamination, particularly from extended  C$^+$ emission; however, no contaminations could be noticed within our adopted line detection criterion (described below) in the response-calibrated OFF data, or in the resulting spectral map in the appearance of absorption at constant strength and velocity that would occur from a contaminated sky reference subtraction.  

All data have been processed in the HIFI pipeline with the Herschel Interactive Processing Environment (HIPE), initially to Level 1 products using either version 13.0 or 14.0 which both include updated telescope pointing reconstruction\footnote{\url{http://herschel.esac.esa.int/twiki/bin/view/Public/\\ImprovedPointingGyro}}$^,$\footnote{\url{http://herschel.esac.esa.int/twiki/pub/Public/\\HifiCalibrationWeb/HIFI\_PointingAccuracyNote.pdf}}, and then using software and calibrations which correspond to the HIPE v14.0.1 User Release version.  The WBS spectrum datasets at each map position were treated interactively to inspect for LO spurs or impurities, which might affect the calibration or measurements, and to remove other baseline artifacts prior to gridding the scans into spectral data cubes.  Artifacts include standing waves and baseline offsets left as residuals of the reference sky subtraction in the pipeline.  These residuals are generally associated with the mixer currents having drifted out of phase with the calibration measurement timing loops, and are most evident in data taken in HIFI bands 6 and 7, which have shortest Allan times (Roelfsema et al. 2012).  Data in these bands often exhibit non-sinusoidal standing waves with periods around 300 MHz that are electrical in origin (from the mixer amplifier chains).  HIPE v13.0 (with updates in v14.0) includes corrections to the electrical standing waves (ESWs) and accompanying baseline drift, successfully recovering source continuum levels, based on an algorithm that fits the ESWs from a mission-wide catalog of standing wave patterns (Kester et al. 2014).  The algorithm is susceptible to line-rich spectra, especially broad or complex profiles and absorption features, and we have verified that the ESW solutions applied to our observations were well-treated with respect to selection of line-free baselines.  All data were also inspected for optical (LO to mixer path) standing waves, typically dominated by a 93 MHz ripple, and treated where necessary with a sine wave approximation using the HIPE task {\it{fitHifiFringe}}.  Remaining residuals after fitting each scan in HIPE are mitigated in maps with more than one cycle of scans.  

After baseline corrections, data from observations of the same transition (i.e., with overlapping sky frequency coverage), including the two polarizations, have been interactively merged into a single timeline product, and then into a single spectral cube for each of the upper and lower sideband frequency scales.  The {\it{doGridding}} routine in HIPE uses a Gaussian (for OTF maps) or box (for raster maps) filter function for spatially weighting signal readouts during scanning, according to the astrometric assignments to each spectrum, and frequency-dependent beam shapes which are approximated as 2-dimensional Gaussian profiles for this purpose. The gridding and convolution technique is very similar to that used in the GILDASS/CLASS\footnote{See \url{www.iram.fr/IRAMFR/GILDAS/}.} at most ground-based sub-millimeter telescopes\footnote{A detailed description is at {\url{herschel.esac.esa.int/hcss-doc-14.0/load/hifi\_um/html/hdrg\_dogridding\_detail.html}}}.  The final spectral cubes yield excellent signal-to-noise ratios (SNRs), well above 100 for the CH$^+$ $J = 1-0$ line and exceptionally high in the strong $^{12}$C$^+$ line, at the beam resolutions of 11$''$2 (1900 GHz), 12$''$.5 (1670 GHz), 14$''$.3 (1470 GHz), 25$''$.2 (835 GHz), and 38$''$.5 (537 GHz).  It is important to note that source features with angular dimensions a fraction of the beam size can be detected when SNRs are high, so that positional information with 1$''$ to 4$''$ relative uncertainty can be measured from strong sources of compact emission.   SNRs are estimated to be around 10-20 for the much weaker $^{13}$C$^+$ fine structure transitions.  For fitting the weaker CH$^+$ $J = 2-1$ and CH line emission we follow the standard detection criterion for the integrated line intensity, namely 3$\sigma$ [K {\kms}] = $3 \times \sigma_{\rm{RMS}} \sqrt{2 \Delta v_{\rm{bin}} \Delta v_{\rm{FWHM}}} $, where $\sigma_{\rm{RMS}}$ is the RMS noise of the line-free baseline,  $\Delta v_{\rm{bin}}$ is the channel binning width in km~s$^{-1}$, and $\Delta v_{\rm{FWHM}}$ is the full width at half maximum intensity of the spectral line fitted as a Gaussian profile.   

\begin{figure}
\centering
\includegraphics[width=8.5cm]{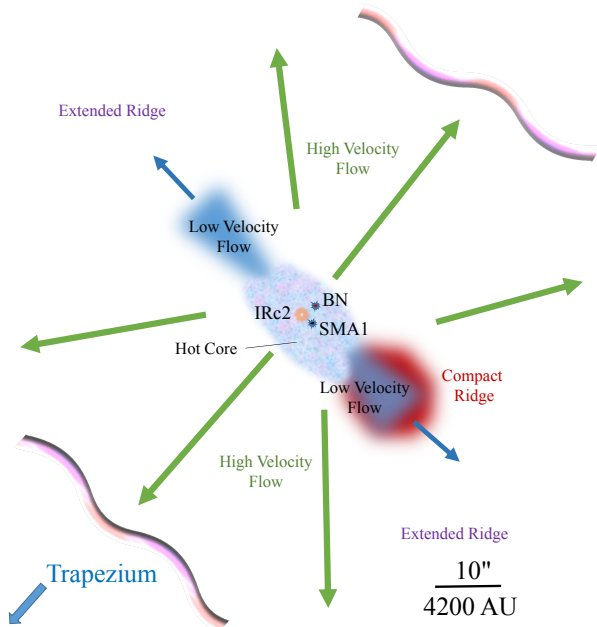}
  \caption{Schematic representation of principal kinematic features in the vicinity of the Orion Hot Core, based on Fig.~7 in Crockett et al. (2014a), representing approximate coverage of mapping in the study drawn on the plane of the sky.  Each distinct velocity component is labeled in a different color, and is summarized in Sec.~\ref{components}.  The angular scale in AU assumes a distance of 420 pc.
  \label{fig:f2}}
\end{figure}

\section{Observational Results}
\label{results}


\subsection{Overview of Orion BN/KL Kinematic Features}\label{components}

As part of the description of our observational results for the spectrally mapped region around Orion BN/KL, it is helpful to place this region in morphological context with its wider OMC~1 surroundings.   Figure~\ref{fig:f1} (right) indicates where our ``deep'' maps have been constructed from multiple observations especially for C$^+$ and CH$^+$ $J=1-0$ (see Table~\ref{obs_t}), by the boxes overlaid on the composite SPIRE 850~GHz, CH$^+$ $J=1-0$, and 1900 GHz continuum observations, over a 7$'$.5 $\times$ 10$'$ area on a normalized intensity scale.  Each of these components were observed at similar angular resolutions, $\approx 12'' - 25''$ (half power beam width; HPBW). The archival OMC~1 CH$^+$ map from the {\em{Herschel}} program OT1\_jgoicoec\_4 was observed at lower sensitivity, using a lower total integration time in a single observation (5.5 sec per map point) and uses half-beam rather than Nyquist sampling, but the SNR $\approx 50$ mK per beam at maximum WBS resolution (1.1 MHz) is clearly sufficient  to show the widespread spatial distribution of line emission.  Overall the CH$^+$ intensitiy distribution is very similar to that of C$^+$ mapped over the same area by Goicoechea et al. (2015a,b).

In comparison to the SPIRE 850 GHz map, which is dominated by CO and thermal dust emission, the relative distribution of C$^+$ is characteristic of a stratification of neutral and photo-ionized gas around the ionizing Trapezium stars.  The 1900 GHz continuum is strongest towards the dense Hot Core, and secondarily in Orion South.  The neutral/ionized stratification has been observed via comparison of integrated emission from H recombination, CO $J=2-1$, and C$^+$ (similarly mapped in OMC~1 with HIFI but at lower sensitivities) by Goicoechea et al. (2015b).   

The immediate region around the dense Hot Core contains several physical features, which give rise to spatially and kinematically distinct molecular line emission.   The main features described in Crockett et al. (2014a) are summarized here, and shown schematically in Figure~\ref{fig:f2}.  The main velocity components are identified with different colors, showing the locations of the BN object, IRc~2, and directional relation to the massive and luminous OB stars of the Trapezium.   Also shown is the sub-millimeter source SMA~1, where vibrationally excited transitions of CH$_3$OH, HC$_3$N, and SO$_2$ have been detected (Beuther et al.  2004), and interpreted to indicate the presence of an embedded protostar.  Close to SMA~1 and IRc~2 is the radio ``source I'' (not labeled, $\approx$0$''.5$ to the south), which is an embedded protostar with no detectable sub-millimeter or IR emission (Menten \& Ried 1995; Plambeck et al. 2009), and thought to be driving the so-called ``low velocity flow'' (LVF) of molecular gas, detected by Genzel \& Stutzki (1989), in a NE-SW direction; see also Crockett et al. (2014a) and references therein.  A distinct and more extended ``high velocity flow'' (HVF) in the NW-SE direction may be driven by another embedded protostar associated with SMA~1, and gives rise to the well-known butterfly-shaped bipolar outflow traced by vibrationally-excited H$_2$ (e.g., Bally et al. 2011 and references therein) and other species tracing shocked gas. The H$_2$ outflow is important to the discussion of the dominant CH$^+$ formation mechanism over our mapped region (Sec.~\ref{discussion}).  Both flows constitute part of a molecular ``plateau'' characterized by broad lines ($\Delta v \gtrsim$ 20 km s$^{-1}$) and \vlsr $\approx 7-11$ km s$^{-1}$.  To the south-west of the Hot Core region is the so-called ``Compact Ridge'', a group of quiescent gas clumps containing complex oxygen-bearing organics, with emission lines typically at velocities \vlsr  $\approx 7-9$ {\kms} and widths $\Delta v = 3-6$ {\kms} (Blake et al.  1987; Friedel \& Snyder 2008; Beuther et al.  2005; Crockett et al.  2014).   

\begin{figure*}
\centering
  \includegraphics[width=14cm]{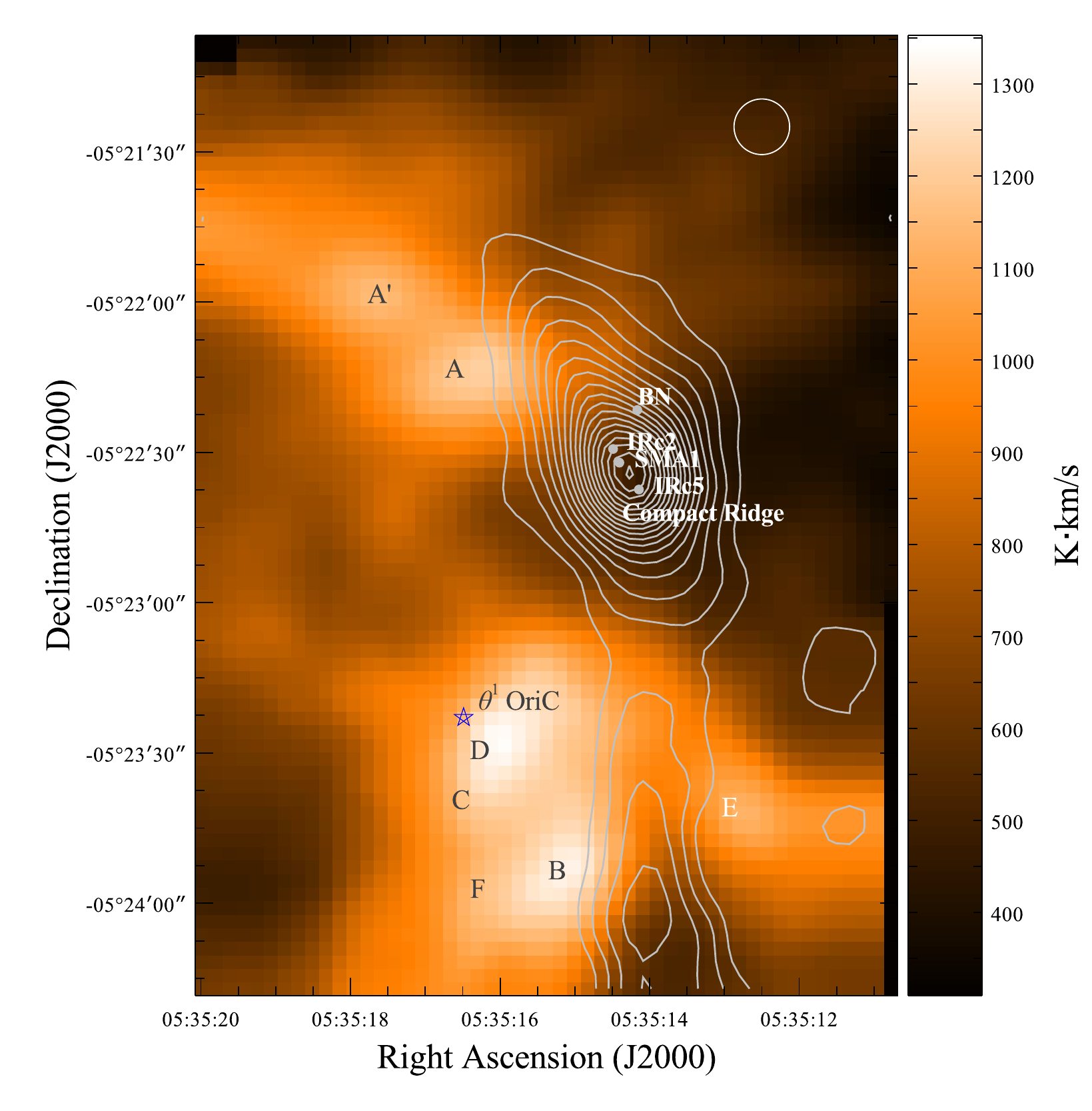}
  \caption{C$^+$ emission integrated on a main beam temperature scale over $-$10 to $+$15 {\kms}, and the continuum at 1900 GHz shown in white contours on a linear scale between 0.5 and 35 K on intervals of 1.0 K.  The 1$\sigma_{\rm{RMS}}$ noise level is 0.1 K {\kms} per beam.  The HIFI beam size at 1900.5 GHz is 11$''$.2 (HPBW), represented in the upper right.  Locations of peak emission are labeled for reference to spectral extractions presented in Fig.~\ref{fig:f4}.      
    \label{fig:f3}}
\end{figure*}

The Hot Core itself has been so named because of the hot ($T_{\rm{kin}} \geq$ 150 K) and dense ($\gtrsim$ 10$^7$ cm$^{-3}$)  material hosting massive protostars, although whether one or more protostars is the source of heating {\em{versus}} an external source is a matter of debate in the literature (cf. de Vicente et al. 2002; Zapata et al. 2011; Goddi et al. 2011).   Line widths are typically $\Delta v \approx 7-11$ {\kms} at \vlsr $\approx 4-6$ {\kms}.

\subsection{Properties of the C$^+$ Gas}\label{cplusproperties}

Emission from $^{12}\rm{C}^+$ is strong, with up to 6 velocity peaks which can be resolved at $>$5-$\sigma$ detection confidence across the region, tracing a complex photo-dissociated cloud structure.   The distribution of integrated line intensities is shown in Figure~\ref{fig:f3}, covering the range of LSR velocities $-$10 to $+$15 {\kms} (avoiding minor contributions from $^{13}$C$^+$ $F = 2-1$ emission).   The integrated emission shows maximum intensity $I(^{12}$C$^+) \approx$ 1160 K~{\kms} at 0$'$.5 SW of the {\thet1} (labeled as Peak B in Fig.~\ref{fig:f3}), in part of a loosely projected ring of gas $\sim$1$'$.4 south of IRc~2.  While there are numerous nearby young stellar objects and OB-type stars, the emission structure does not correlate with any particular stellar (e.g., the Trapezium) or protostellar source, nor is there any particular source near the projected center of the southern emission ring.  Secondary peaks of C$^+$ emission occur just to the NE of IRc~2, labeled A and A$'$, where the latter coincides more closely with the CH$^+$ peak nearest to the Hot Core, as will be shown in Section~\ref{CHplus}.  By contrast, the integrated emission towards the Hot Core is nearly a factor $\sim$4 lower,  possibly from line-of-sight absorption of the strong continuum which we shall address further below in this section. 

\begin{figure*}
 \centering
  \includegraphics[width=14cm]{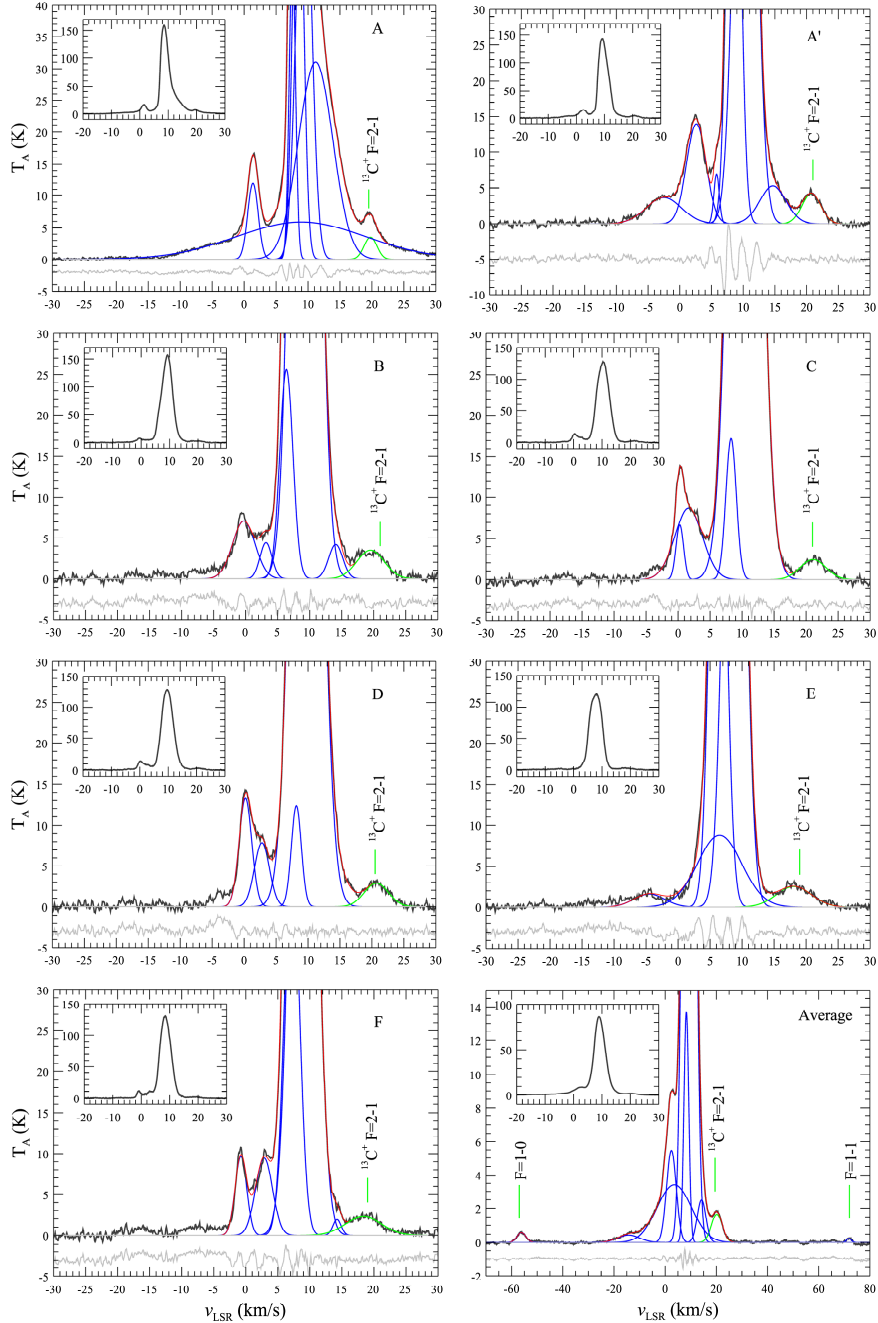}
  \caption{HIFI spectra of C$^+$ $^2\it{P}_{3/2,1/2}$ as an average and at positions marked in Fig.~\ref{fig:f3}, extracted using a circular aperture of width 11$''$.2 corresponding to the HPBW at 1900 GHz.   The labeled Average spectrum (bottom right) excludes the inner 15$''$ centered on BN/KL, which is shown in Fig.~\ref{fig:f5}. Observed spectra (black) are fitted with Gaussian profiles (blue) using ${\chi}^2$ minimization of the residual differences between the total model (red) and the observations.  Only the number of profiles is constrained; intensities and widths are free parameters during the minimization. The residuals (gray) near the bottom of each panel are offset for clarity.  Integrated fluxes of the two strongest components are given in Table~\ref{cplusline}. 
\label{fig:f4}}
\end{figure*}

\begin{figure}
 \begin{center}
  \includegraphics[width=\columnwidth]{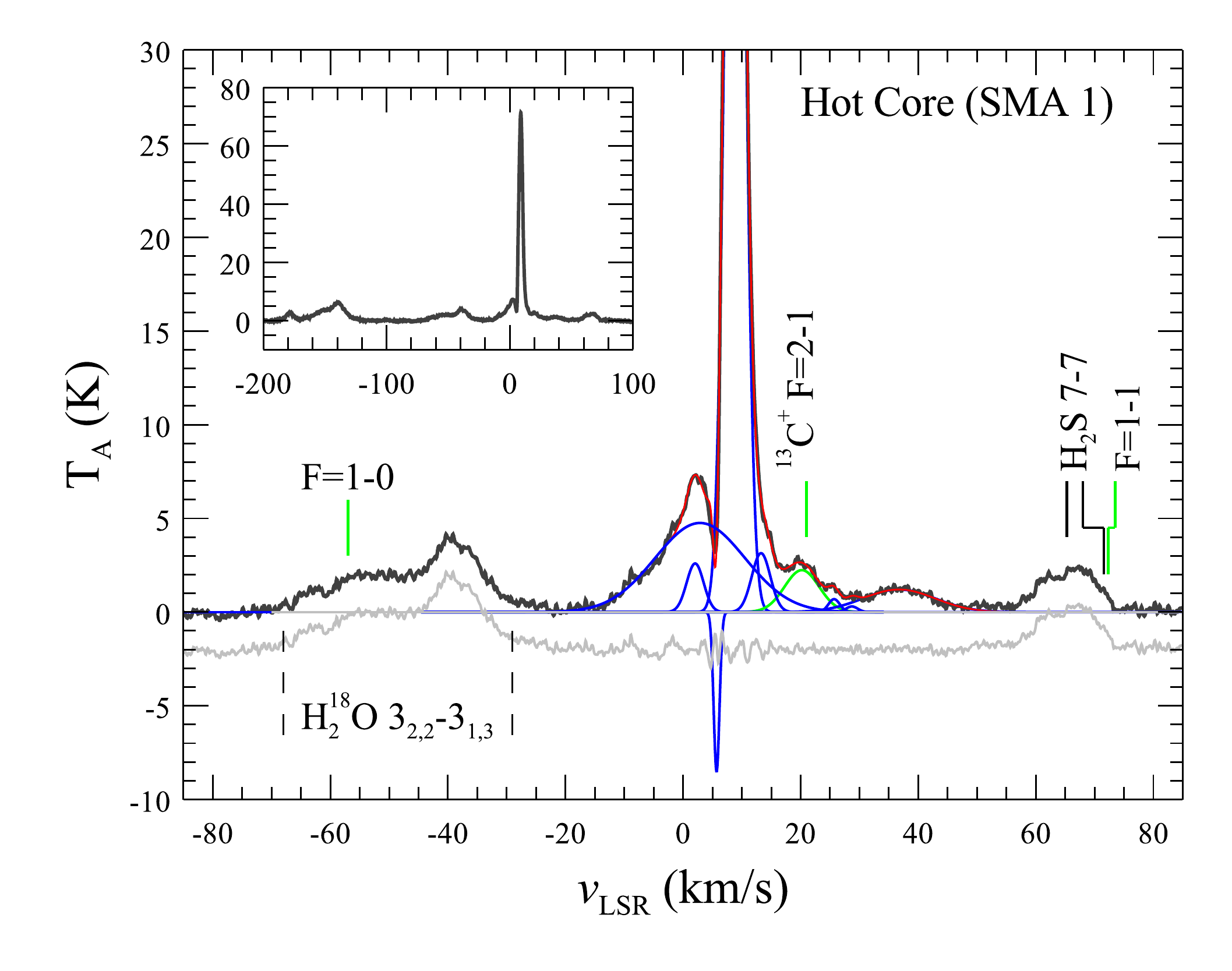}
  \end{center}
  \caption{HIFI spectrum of C$^+$ in the Hot Core region using a circular extraction aperture of width 15$''$ centered on the SMA~1 position indicated in Fig.~\ref{fig:f3}.    Observed spectra (black) are fitted with Gaussian profiles (blue) using ${\chi}^2$ minimization of the residual differences between the total model (red) and the observations.  The residuals (gray) near the bottom of each panel are offset for clarity.  Integrated fluxes of the two strongest components are given in Table~\ref{cplusline}. 
\label{fig:f5}}
\end{figure}

Figure~{\ref{fig:f4}} shows spectra extracted at the labeled locations of strongest emission, and an overall average that excludes a 25$''$ diameter region centered on IRc~2,  illustrating the range of profiles.  Each spectrum is plotted on the antenna temperature $T_{\rm{A}}$ scale, and has been extracted in synthetic circular apertures of width 11$''$.2 corresponding to the HPBW of HIFI at 1900 GHz.  In these and other figures showing local spectral extractions, we have employed a method of fitting Gaussian profiles which minimizes the residual spectrum in order to efficiently identify and characterize LSR velocities and line widths.  In reality, gas turbulence and moderate changes in optical depth (which are explored below) will cause departures from a Gaussian profile shape.   The residuals shown in Figures~\ref{fig:f4} and \ref{fig:f5} show the levels where the approximations fail, typically at the 2-4\% level of peak intensity near line center.  The residuals are non-Gaussian (thus cannot be minimized further with an arbitrary increase in the number of Gaussian fit components), and widths of the residual features are $\approx$1-2 {\kms}.  

\begin{figure*}
 \begin{center}
 \includegraphics[width=14cm]{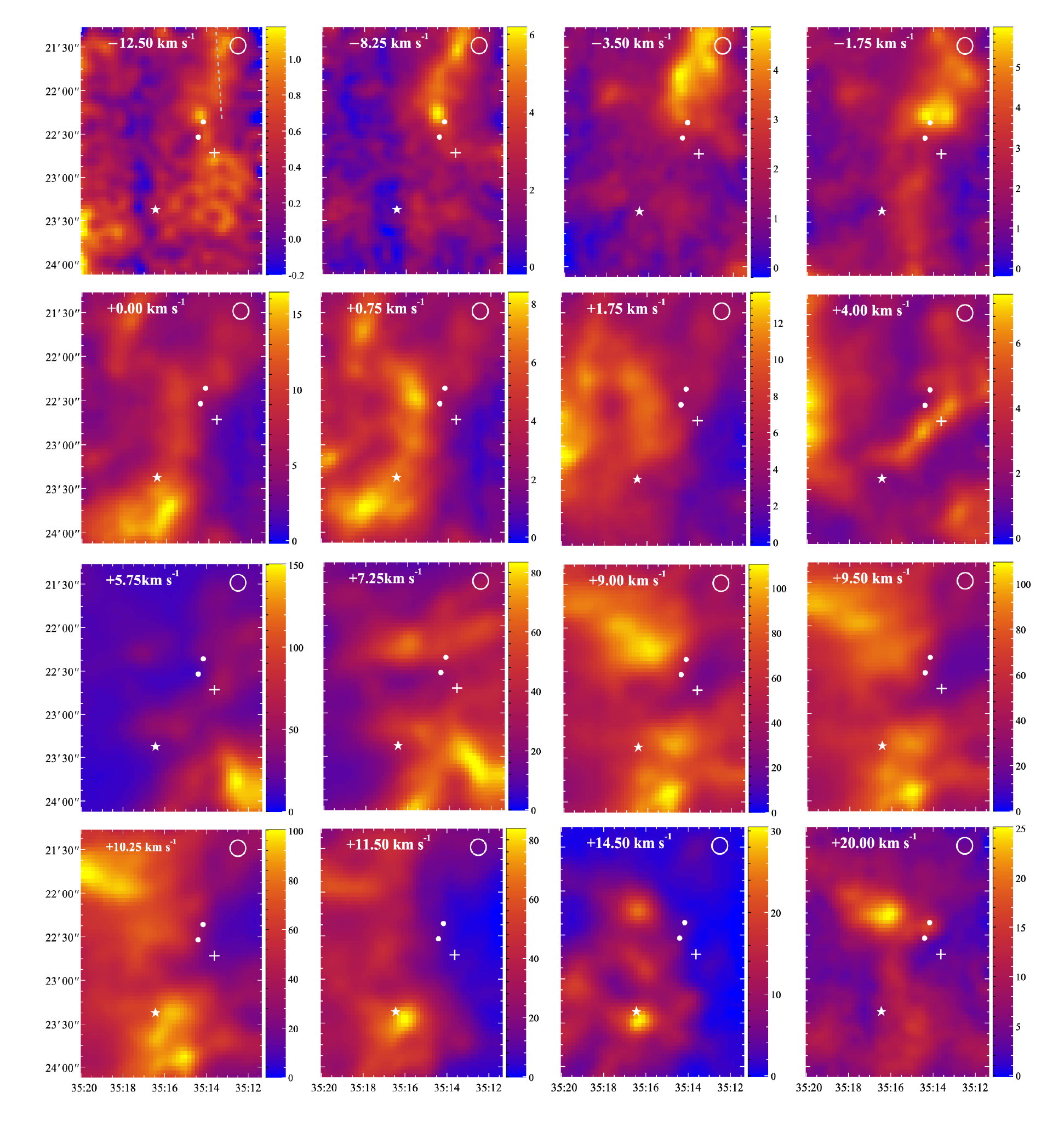} 
 \end{center}
  \caption{Velocity channels of $^{12}$C$^+$.  Each channel map is an integration over $\pm$0.5 \kms around the labeled velocity, on a linear intensity scale of K {\kms}.  Coordinates on horizontal and vertical axes are referenced to R.A. $5^h00^m00^s$ and decl. $-$5$^\circ00'00''$ (J2000.0), respectively. The HIFI 11$''$.4 beam size is indicated in all plots, and the location of the intensity profile cut in Fig.~\ref{fig:f7} is shown in the $-$12.5 \kms channel map.  The $+$20.0 \kms channel in the bottom right corner panel corresponds to the $^{13}$C$^+$ $F = 2-1$ fine structure line at \vlsr = $+$8.4 {\kms}.  The filled white circles indicate the locations of the BN (northern) and SMA~1 (southern) sources.  The cross and filled star indicate the locations of the Compact Ridge and {\thet1}, respectively. 
 \label{fig:f6}}
\end{figure*}

The dominant C$^+$ feature can be fitted by two Gaussian components with a principal component at $\langle {v_{1,\rm{lsr}}} \rangle$ = 9.3 {\kms} and a weaker secondary component at $\langle {v_{2,\rm{lsr}}}\rangle $ = 8.3 \kms, with line widths (FWHM) of $\langle {\Delta v} \rangle$ of 4.6 and 2.4 {\kms}, respectively.  The strongest emission from other locations can be similarly decomposed into two components, varying in LSR velocity of the primary component between 8.6 and 11.0 {\kms}.   Emission at these velocities is strongest in the plateau to the NE of the Hot Core and south of the Trapezium stars, as seen in the velocity channel maps in Figure~\ref{fig:f6}.  The dominant velocities and dispersions are consistent on average with early $^{12}$C$^+$ observations of Orion A with 3$'$ resolution by Balick et al.  (1974), who suggested a dynamically distinct and quiescent cloud interface between Orion~KL and the Trapezium stars, compared to the H~{\sc{ii}} emitting region.

\begin{deluxetable*}{lcccrccccccr}
\tabletypesize{\scriptsize}
\tablecaption{Optical depth and column density of $\mathrm{^{12}C}^{+}$ at selected positions \label{cplusline}}
\tablewidth{0pt}
\tablehead{
\colhead{Position} & \colhead{$\alpha_{2000}$} & \colhead{$\delta_{2000}$}     &  \colhead{$I_{12}(\mathrm{^{12}C}^{+})$\tablenotemark{a} } &  \colhead{$I(\mathrm{^{13}C}^{+})$\tablenotemark{b} } &   \colhead{$v_{\mathrm{1,lsr}}$\tablenotemark{c} } & \colhead{$\Delta v_1$\tablenotemark{c} }  &  \colhead{$v_{\mathrm{2,lsr}}$\tablenotemark{c}} & \colhead{$\Delta v_2$\tablenotemark{c} } & \colhead{$\langle {\tau}({\mathrm{^{12}C}^{+}}) \rangle $} & \colhead{$T_{\rm{ex}}$} & \colhead{$N({\mathrm{C}^{+}})$} \\
                             &  \colhead{(h) (m) (s)}&  \colhead{($^{\circ}$)($^{\prime}$)($^{\prime\prime}$)} & \colhead{(K~km~s$^{-1}$)} & \colhead{(K~km~s$^{-1}$)}    &    \colhead{(km~s$^{-1}$)}    &  \colhead{(km~s$^{-1}$)}  &   \colhead{(km~s$^{-1}$)}    &  \colhead{(km~s$^{-1}$)}  &  & (K) &  \colhead{($10^{18}$~cm$^{-2}$)}
}
\startdata
\multirow{ 2}{*}{Average\tablenotemark{d}}   & \multirow{ 2}{*}{05 35 14.5}    & \multirow{ 2}{*}{$-$05 21 59.5}  & \multirow{ 2}{*}{$621\pm21$}  &    $21.6\pm0.4$      & \multirow{ 2}{*}{7.07}  & \multirow{ 2}{*}{2.60}  & \multirow{ 2}{*}{9.38} & \multirow{ 2}{*}{4.50}  & $2.0^{+0.5}_{-0.4}$   &  \multirow{ 2}{*}{178.1} & $13^{+4}_{-3}$ \\
   &      &      &   &    $18.4\pm0.7$     &  &   & &   & $1.6^{+0.5}_{-0.4}$   & &  $10^{+4}_{-3}$  \\ \midrule

\multirow{ 2}{*}{A}     &  \multirow{ 2}{*}{05 35 16.1}      & \multirow{ 2}{*}{$-$05 22 15.2}     & \multirow{ 2}{*}{$617\pm20$}  &    $22.8\pm0.5$       &   \multirow{ 2}{*}{8.93}  & \multirow{ 2}{*}{2.49} & \multirow{ 2}{*}{10.95} & \multirow{ 2}{*}{6.61}  & $2.2^{+0.6}_{-0.4}$  &   \multirow{ 2}{*}{$257.2\pm$3.2} & $9.7^{+1.9}_{-1.5}$ \\
  &              &     &   &    $23.9\pm1.2$       &   &  &  &  & $2.3^{+0.6}_{-0.5}$  &  & $10^{+2}_{-2}$ \\ \midrule

\multirow{ 2}{*}{A$^\prime$}     &  \multirow{ 2}{*}{05 35 17.4}             & \multirow{ 2}{*}{$-$05 21 58.5}     & \multirow{ 2}{*}{$905\pm30$}  &    $38.4\pm0.5$       &  \multirow{ 2}{*}{8.93}  & \multirow{ 2}{*}{2.49} & \multirow{ 2}{*}{10.95} & \multirow{ 2}{*}{6.61}  & $2.6^{+0.6}_{-0.5}$  & \multirow{ 2}{*}{$274.5\pm1.5$}&  $18^{+3}_{-2}$ \\
    &   &     &   &      $38.0\pm2$  &  &  &  &  & $2.4^{+0.6}_{-0.5}$  & &  $17^{+3}_{-2}$ \\ \midrule

B     & 05 35 15.0              & $-$05 23 52.5     & $1160\pm38$  &    $44.7\pm0.7$      &   6.60  & 2.00 & 9.07  & 4.18  & $2.3^{+0.6}_{-0.4}$   & $298.7\pm2.1$ &  $24^{+4}_{-3}$ \\

C    & 05 35 16.5               & $-$05 23 37.3     & $1034\pm34$  &    $28.3\pm0.9$      &   8.09 & 2.27 & 10.22 & 4.90  & $1.4^{+0.5}_{-0.4}$   & $226.7\pm3.5$ & $10^{+3}_{-2}$ \\ \midrule

\multirow{ 2}{*}{D}    & \multirow{ 2}{*}{05 35 16.2}               & \multirow{ 2}{*}{$-$05 23 29.0}     & \multirow{ 2}{*}{$1002\pm33$}  &    $32.5\pm0.8$      &  \multirow{ 2}{*}{8.14} & \multirow{ 2}{*}{2.96} & \multirow{ 2}{*}{11.00} & \multirow{ 2}{*}{4.16} & $1.8^{+0.5}_{-0.4}$   & \multirow{ 2}{*}{$188.1\pm3.1$} & $11^{+3}_{-2}$ \\
    &              &     &  &    $33.4\pm1.6$      &    &  &  &  & $1.9^{+0.6}_{-0.4}$    & & $11^{+4}_{-3}$ \\ \midrule

E    & 05 35 12.5               & $-$05 23 42.0     & $1009\pm33$  &    $42.3\pm2.6$            &   6.07 & 2.79 & 8.78 & 3.69   & $2.6^{+0.7}_{-0.5}$    & $202.4\pm2.7$ & $24^{+6}_{-5}$ \\

F    & 05 35 15.4               & $-$05 24 08.1     & $932\pm31$  &    $38.5\pm0.4$      &   7.12 & 2.52 & 8.98 & 3.88  & $2.5^{+0.6}_{-0.5}$    & $250.5\pm1.6$ & $30^{+6}_{-4}$ \\

Hot Core & 05 35 14.4      & $-$05 22 30.0     & $423\pm14$  &    $39.4\pm1.9$      &  3.67 & 6.90 & 8.69 & 3.67 &  $6.2^{+1.3}_{-0.9}$ &$117.7\pm5.4$ & $35^{+9}_{-5}$ \\
Hot Core-S & 05 35 14.3      & $-$05 22 34.5     & $496\pm16$  &     $30.1\pm1.8$  &  3.67 & 6.90 & 8.69 & 3.67 &  $4.0^{+0.9}_{-0.7}$ & $116.9\pm5.9$ & $25^{+7}_{-5}$ \\
\enddata

\tablenotetext{a}{The $^{12}$C$^+$ emission $ I_{12}(\mathrm{^{12}C}^{+})$ is an integration of the two strongest components between $+$5 to $+$15 {\kms}.}
\tablenotetext{b}{Integrated $^{13}$C$^+$ intensities are estimated as a sum of all 3 hyperfine transitions (see Table~\ref{restfreq}), using a fit to the $F = 2-1$ component and the partition rule (see Sec.~\ref{CplusTau}).  Second values of $I(^{13}$C$^+$) for the Average and positions A, A$^\prime$, and D are from direct measurements of all 3 observed components.}
\tablenotetext{c}{LSR velocities and widths are measured from the two strongest $^{12}$C$^+$ emission components peaking between $+$5 and $+$15 \kms.}
\tablenotetext{d}{The ``average'' position refers to the area in common to all spectral maps in this study, excluding the inner 12$''$ around IRc~2; see Fig.~\ref{fig:f1}.}
\end{deluxetable*}

The spectrum centered on the Hot Core source SMA~1 is shown in Figure~{\ref{fig:f5}} on a wider velocity scale to show the spectral complexity along this sightline.  In this spectrum we find evidence of C$^+$ line absorption near 5.5 {\kms}; this will be treated in the next sub-section.  The $^{13}$C$^+$ triplet is also detected everywhere in our mapped region, although this is sometimes limited to the $F = 2-1$ transition, which is expected from the partition function to be strongest (see Sec.~\ref{13cplus}).

Very similar characteristics have been reported in HIFI observations of OH$^+$ and H$_2$O$^+$ absorption (Gupta et al.  2010) and HF P-Cygni profiles (Phillips et al.  2010) towards Orion BN/KL, and in these lines plus CH$^+$ absorption towards OMC-2 FIR4 roughly 13$'$ to the N-N-E of the BN object (L{\'o}pez-Sepulcre et al. 2013).  While there is a considerable amount of spatial and kinematic variation of the C$^+$ emission in our 12 arcmin$^2$ HIFI map around BN/KL (note the complexity of the profiles particularly at Position A and towards the Hot Core in Figure~\ref{fig:f4}, and the spatial variations in Figure~\ref{fig:f6}), the general kinematic similarities with CH$^+$ towards OMC-2 FIR4 adds support for the conclusion by L{\'o}pez-Sepulcre et al. (2013)  that the cloud observed further to the north is an extension of the C$^+$ interface between Orion~KL and the Trapezium, and thus interpreted to be part of the larger parent cloud of OMC~1.  At the higher resolution of our observations around BN/KL (HPBW $\approx 11''$), the C$^+$ gas exhibites a high level of velocity variation among multiple components.

\subsection{C$^+$ Absorption}

The Hot Core spectrum (Fig.~\ref{fig:f5}) can be decomposed into a predominant and relatively narrow emission component at $+$8.7 \kms, and two weaker flows at $+$5.8 and $-$7.4 \kms.  This compares with \vlsr $= 3 - 5$ km~s$^{-1}$ of molecular emission from the Hot Core (Blake et al.  1987), indicating that most of the detected C$^+$ emission is emitted from outside this warm and dense region.  The profile fit includes narrow ($\Delta v$ = 1.3 \kms) absorption against the continuum at \vlsr = $+$5.5 \kms, with an optical depth of $1.0-1.3$ at line center.  Absorption is expected when a significant column of low-density gas is observed towards a strong source of background emission at temperature $T_{\rm{bg}}$, and the excitation temperature $T_{\rm{ex}}$ is less than the brightness temperature of the background; i.e., 

\begin{equation}\label{eq:TexCplus}
T_{\rm{ex}}({\rm{C}}^+) < \frac{91.2}{{\rm{ln}}(1 + 91.2/T_{\rm{c}}(1900.5))} \simeq 25.0 \; \rm{K}
\end{equation}

\noindent where the continuum brightness temperature at 1900.5 GHz, $T_{\rm{c}}$(K)$ = 91.2/(e^{91.2/T_{\rm{bg}}} - 1)$.  The value 91.2 K is the equivalent temperature $T_*  \equiv h \nu/k$ at 1900.5 GHz.  At the position of SMA~1 where we observe the absorption, the continuum temperature peaks at around 25.0 K, and if we add 2.7 K for the cosmic microwave background temperature (thus $T_{\rm{bg}}$ = 27.7 K), then $T_c \simeq 3.5$ K and $T_{\rm{ex}} \lesssim$  28~K for the absorption component.   Goicoechea et al. (2015b) suggest that absorption may be indicated in C$^+$ profiles for some sight-lines in Orion's extended Veil region (a collection of absorbing H~{\sc{i}} layers with low H$_2$ densities) and predict the ranges of $T_{\rm{ex}}$ and H~{\sc{i}} densities where absorption by C$^+$ should occur.   Here we can at least confirm detection of C$^+$ absorption towards the Hot Core where $\tau$ is around unity.  At other positions in our map we cannot be as certain, as all profiles can be well represented by emission only.   We should note that the kinematics reported here are consistent with those by Goicoechea et al. (2015b) using HIFI observations of the larger OMC~1 region, however there are differences in line profiles and peak intensities, highlighted by the BN/KL spectrum, for example.  The spectrum shown in their Figure 3 lacks both the structure of the main C$^+$ emission components over $-$20 to $+$30 {\kms} as well as significant emission of molecular lines near $\pm$50 {\kms} relative to the main C$^+$ line.  This is probably a result of differences in sensitivities in the two sets of observations, as noted above.

\begin{figure}
 \begin{center}
 \includegraphics[width=\columnwidth]{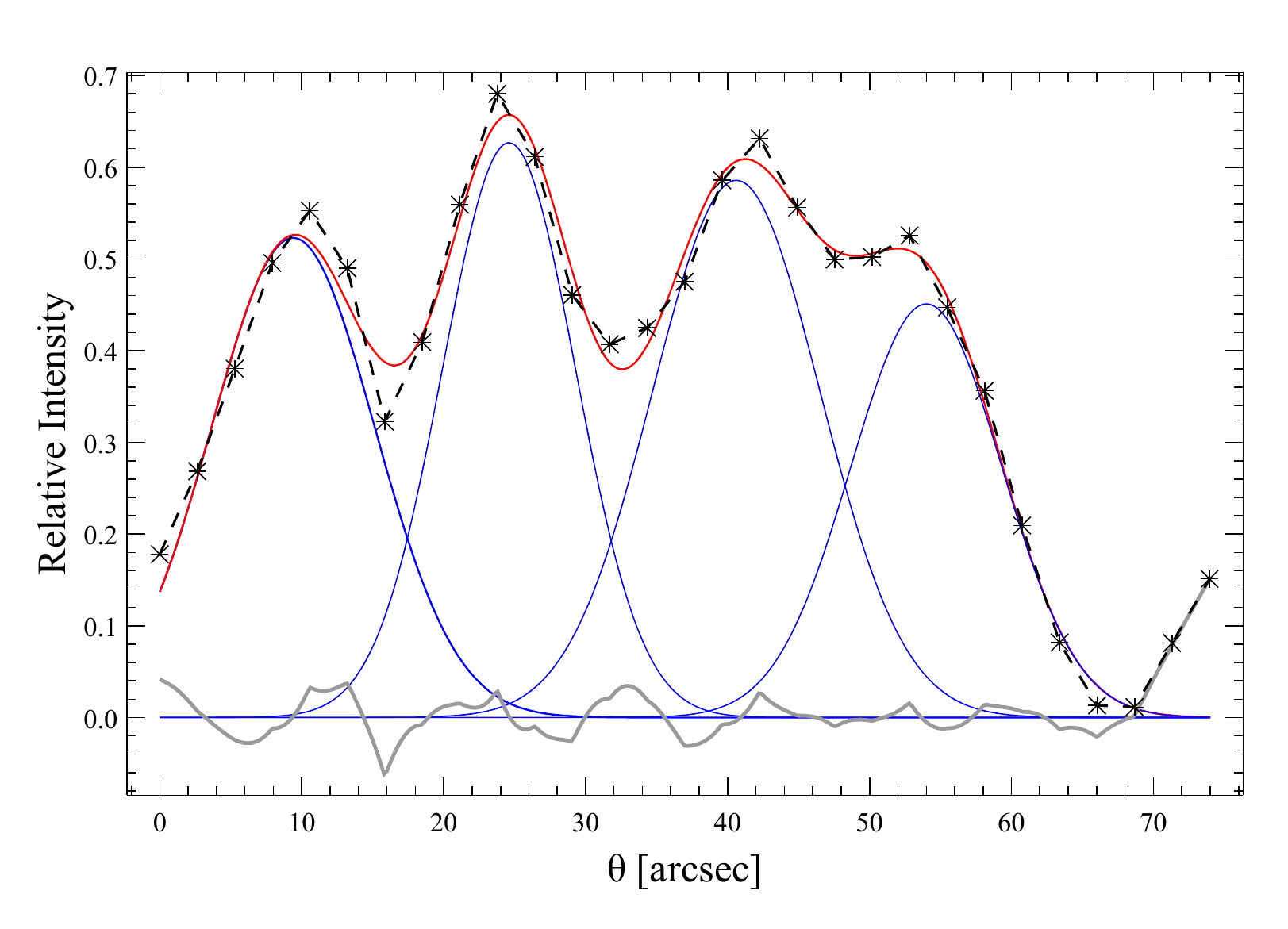}
 \end{center}
  \caption{Spatial profile of C$^+$ emission traced along a cut in the $-$12.5 {\kms} channel map as indicated in Fig~\ref{fig:f6} in the same units.  The trace is fitted by Gaussian profiles with FWHM between 11$''$.0 to 14$''$.4 with the residual shown in gray.
 \label{fig:f7}}
\end{figure}

Velocity channel maps are shown in Figure~\ref{fig:f6}.  Each channel is an integration of $\pm$0.25 {\kms} around the labeled velocity.  The structure is striking at velocities where emission peaks are spectrally resolved, owing to the high data quality of the baselines (particularly good corrections for the optical and electrical standing waves, as described in Sec.~\ref{obs_s}).  At \vlsr = $-$12.5 and $-$3.5 {\kms}, for example, the spectrally resolved emission can be measured to better than 5-$\sigma$ detection in ``velocity clumps'' (loosely speaking), which are spatially resolvable to as small as 11$''$.0-11$''$.5 FWHM in a representative profile tracing (see Figure~\ref{fig:f7}), compared to the 11$''$.2 HPBW of HIFI at these frequencies.  These sizes correspond to $\approx$0.02 pc (4.8$\times$10$^3$ AU) at a distance of 420 pc to Orion BN/KL (Hirota et al. 2007).  The C$^+$ structures are relatively smoothly distributed and do not appear to follow any particular correlation in distribution or intensity with the embedded sources in the core region, or with the hot UV source {\thet1} which is located some 0.3 pc in front of OMC~1 (Bally 2008).    

\begin{figure}
 \begin{center}
 \includegraphics[width=\columnwidth]{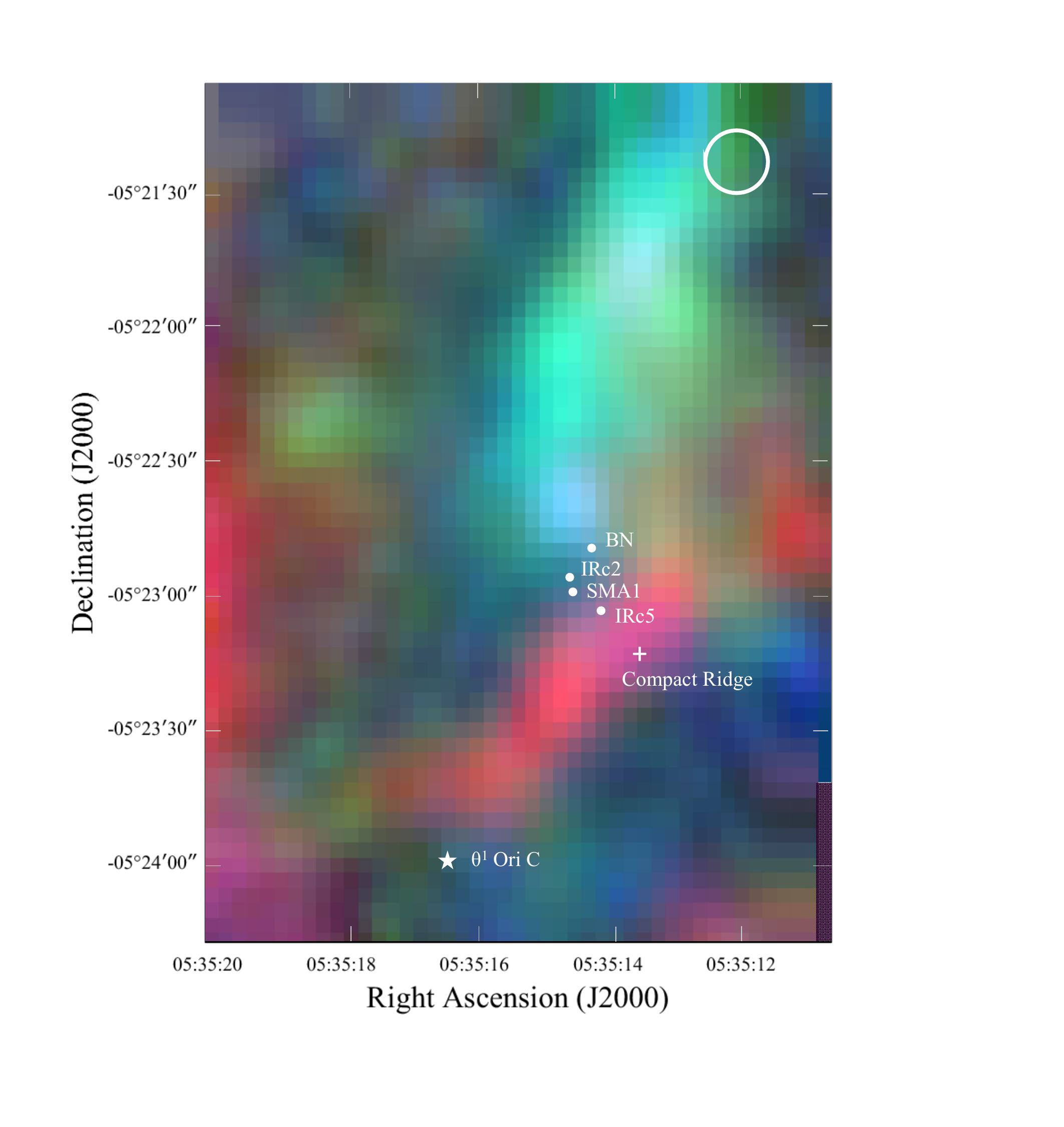}
 \end{center}
 \vspace{-2em}
  \caption{Distribution of the quiescent C$^+$ emission from $-$8.25 (blue), $-$3.5 (green), and $+$4.0 (red) {\kms} channels each normalized to its own peak integrated intensity.   
 \label{fig:f8}}
\end{figure}

The most blue-shifted emission from C$^+$ occurs at \vlsr = $-$12.5 {\kms}, shown in the upper left plot of Figure~{\ref{fig:f6}, and is concentrated very close (in projection) to the BN object with a clumpy appearance.  The intensities at this velocity are at the detection limit of these observations.  More blue-shifted emission from velocity channels at the spatial resolution of these maps cannot be discerned to better than 2-$\sigma_{\rm{RMS}}$.   Detection of C$^+$ at velocities $< -$10 {\kms} is an interesting question for its possible relation to the H~{\sc{i}} absorbing cloud observed in the south near the Orion Bar in the $-$20 to $-$10 {\kms} range by van der Werf et al. (2013), i.e., as part of the neutral atomic layers characterizing the Orion Veil, as discussed by Goicoechea et al. (2015b).  A portion of this blue-shifted C$^+$ emission should arise in the ionized region between the Veil and the Trapezium.  Spectral extractions around {\thet1} and at random locations throughout our mapped region do not reveal emission in that range of blue-shifted velocities.   By comparison, the red-shifted velocities of peak C$^+$ emission, $\approx$14.5 {\kms}, are similar to those of S~{\sc{iii}} and P~{\sc{iii}} absorptions observed in the UV by Abel et al. (2006).  The C$^+$ at this red-shifted velocity is part of a more generally distributed ionized cloud around BN/KL, not concentrated in way that can be associated with a foreground H~{\sc{ii}} region between the low density H~{\sc{i}} Veil and Trapezium stars as suggested by Goicoechea et al.   As we noted in the description of the data processing (Sec.~\ref{obs_s}), no C$^+$ contamination is present in the reference sky positions that might ``hide'' emission (by accidental subtraction) from the Veil in our observations.


The structure of the kinematically distributed gas, isolating the low intensity emission at velocities away from the peak emission,  is shown in Figure~\ref{fig:f8}, composed of $-$8.25 (blue), $-$3.5 (green), and $+$4.0 (red) {\kms} channels, each normalized to peak intensity.   The granular nature of the expanding and receding emission sources occur at scale sizes similar to that measured from Figure~\ref{fig:f7}, and possibly trace local enhancements in the electron density $n_e$.    Weilbacher et al. (2015) have recently estimated electron temperatures and densities over a $\sim$ 6$'$ $\times$ 5$'$ region that covers our observed area, using high quality spectroscopic observations of 
[N~{\sc{ii}}] and [S~{\sc{iii}}] (for electron temperature) and [S~{\sc{ii}}] and [Cl~{\sc{iii}}] (for electron density) doublets obtained with the MUSE optical integral-field spectrograph at the ESO VLT.   The values of $n_e$ based on the [S~{\sc{ii}}] 6731/6716 ratio range from $\sim$ 500 to 10$^4$ cm$^{-3}$, while $n_e$ based on the [Cl~{\sc{iii}}] 6731/6716 is much higher, ranging between 4000 to 2.5 $\times$ 10$^4$ cm$^{-3}$ (but are numerically noisier; see Weilbacher et al. 2015).   Both are broadly consistent in their spatial distribution, but neither exhibit a clumped morphology that correlates very well with our C$^+$ observations.  The MUSE observations were sampled at 0$''$.2 angular resolution, and applying a Gaussian filter to smooth the MUSE $n_e$ maps to the approximate beam width of our HIFI data does not reveal a morphology on any spatial scale that resembles the C$^+$ emission at any velocity shown in Figure~\ref{fig:f6}, a combination of channels as shown in Figure~\ref{fig:f8}, or integrated across the full range of C$^+$ velocities as in Figure~\ref{fig:f3}.   On the other hand, the distribution of the main C$^+$ emission component near {\vlsr} $\simeq \; +$9.0 {\kms} does broadly follow changes in excitation temperature and optical depth, which are discussed in Section~\ref{CplusTau}.

At velocities outside of the range shown in Figure~\ref{fig:f4} we do not detect any higher velocity (i.e., $\mid v_{\rm{LSR}} \mid > 20$ {\kms}) emission from C$^+$.   The emission features at $-$38.9 and $+$65.6 {\kms} in the Hot Core spectrum (Fig.~\ref{fig:f5}), which are relatively symmetric around the main emission component at 8.7 {\kms}, are at first suggestive of velocities in molecular lines such as HF $J = 1-0$ and H$_2$O$^+$  J = $3/2-1/2$ absorptions, or in the wings of broad CO $J=1-0$ emission that have been associated with the LVF (e.g., Zuckerman et al. 1976; Schulz et al. 1995; Wilson et al.  2001; Gupta et al. 2010; Phillips et al. 2010; Zapata et al.  2010; Peng et al.  2012a).  However, careful inspection shows that the blue-shifted feature can be attributed to H$_2^{18}$O while the red-shifted feature is dominated by H$_2$S, which have both been modeled by Crockett et al. (2014b).   The modeled profiles tend to under-estimate the strengths observed here, possibly a result of emitting source size uncertainties, but are sufficient to rule out any obvious contribution from C$^+$ at velocities that are typical of the molecular LVF.

\subsubsection{Hyperfine $^{13}$C$^+$ $^2P_{3/2} - ^2P_{1/2}$}\label{13cplus}

In the same range of velocities of $^{12}$C$^+$ we also detect the three hyperfine transitions of $^{13}$C$^+$, $F = 1-1, 2-1$, and $1-0$, as shown in the average spectrum in Figure~\ref{fig:f4}.  The  $F = 1-1$ and $1-0$ transitions do not meet our detection criterion everywhere over our mapped region, nor at every selected position above 3-$\sigma$ of the baseline noise, but the $F = 2-1$ line is detected everywhere and is sufficiently resolved (spectrally) from the main $^{12}$C$^+$ emission that we can interactively fit the individual components to provide estimates of the $^{13}$C$^+$ contribution at these positions.  It is not possible to construct a reliable $F=2-1$ integrated intensity map by profile decomposition over the entire region due to the very large variations in the number of $^{12}$C$^+$ components, LSR velocities, intensities and widths (as can be seen from Figure~\ref{fig:f6}), restricting reliable fits to the representative positions given in Table~\ref{cplusline}.  

The $+$19.5 {\kms} velocity channel shown in the bottom right panel of Figure~\ref{fig:f6} shows peak emission of the $F = 2-1$ transition corresponding to $v_{\rm{lsr}} = +$8.4 {\kms} coincident with the Peak A of $^{12}$C$^+$.  At that position, the {\vlsr} and width of $^{13}$C$^+$ are consistent with those of the dominant $+$8.6 \kms component of $^{12}$C$^+$.  The fact that $^{13}$C$^+$ does not reach peak intensity at the same locations B and D as $^{12}$C$^+$ could be due to isotopic abundance variations, but is more likely due to differences in $^{12}$C$^+$ optical depth, discussed next.

\begin{figure}
 \includegraphics[width=\columnwidth]{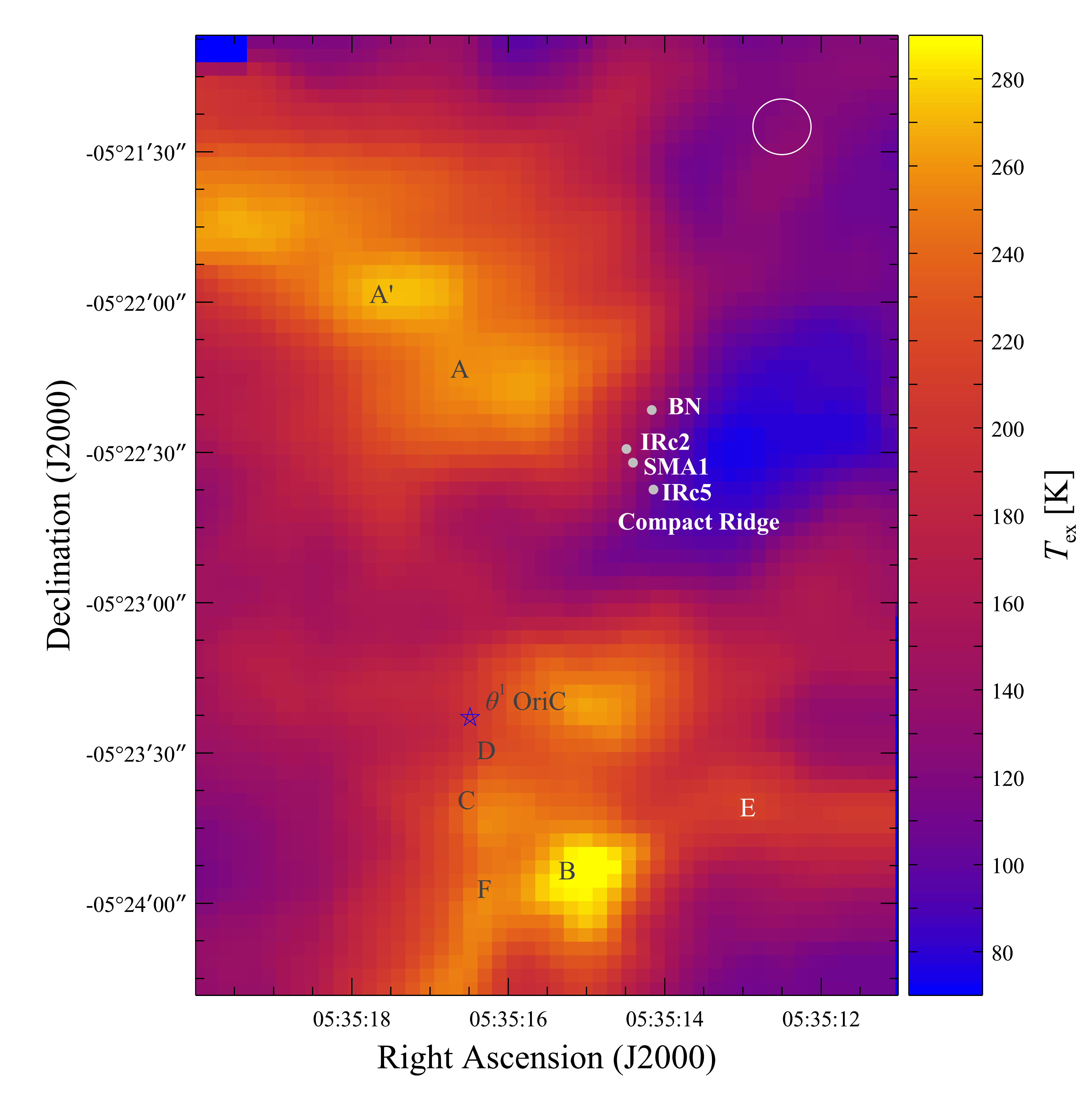}
 \caption{Excitation temperatures for main emission component of $^{12}$C$^+$ estimated from Eq.~\ref{eq:TexCplus}. 
 \label{fig:f9}}
\end{figure}

\subsubsection{The C$^+$ optical depth, excitation temperature, and column density}
\label{CplusTau}


A determination of the C$^+$ gas column densities depends on line opacities and excitation temperatures through the equation (e.g. Ossenkopf et al. 2013)

\begin{equation}\label{eq:CplusColDenEq}
\begin{aligned}
  N{\rm{(^{12}C^+)}} &  = 1.40 \times 10^{17} \frac{ 1 + 2 \exp (-91.2 / {\rm{T_{ex}}}) }{ 1 - \exp (-91.2 / {\rm{T_{ex}}}) } \int{\tau dv}  \\
                                  & \simeq 1.49 \times 10^{17} \frac{ 1 + 2 \exp (-91.2 / {\rm{T_{ex}}}) }{ 1 - \exp (-91.2 / {\rm{T_{ex}}}) } \Delta\nu \times {\tau}_{12} \; {\rm{(cm^{-2})}}
\end{aligned}
\end{equation}

\noindent where $\tau_{12}$ is the $^{12}$C$^+$ optical depth at line center and $\Delta \nu$ is the FWHM Gaussian line width in {\kms}.  In order to estimate $\tau_{12}$, we recognize that the $^{12}$C$^+/^{13}$C$^+$ integrated intensity ratio reflects the $^{12}$C$/^{13}$C abundance ratio of the gas, provided that the $^{13}$C$^+$ emission can be treated as optically thin.   Following Boreiko \& Betz (1996), the total isotopic abundance ratio $R \equiv ^{12}$C/$^{13}$C is related to the ionic line intensity ratio $I_{\rm{13}}/I_{\rm{12}}$ and $\tau_{12}$ by 

\begin{equation}
R = \frac{- \tau_{\rm{12}}}{{\rm{ln}}\Big[ 1 - \frac{I_{\rm{13}}}{I_{\rm{12}}} ( 1 - e^{-\tau_{\rm{12}}} ) \Big] }
\end{equation}

 \noindent There is no dependence on the excitation temperature, but it is implicitly assumed that both isotopes are emitting from the same gas.  The total integrated intensities are tabulated in Table~\ref{cplusline} at the positions indicated in Figure~\ref{fig:f3}.  The $^{13}$C$^+$ intensities have been obtained by summing over all three measured hyperfine components following our detection criterion (Sec.~\ref{results}) or otherwise by measurement of the strongest $F = 2-1$ component and applying the theoretical 0.25:0.125:0.625 partitioning of $F = 1-0:1-1:2-1$, when either or both of the two weaker components are below the detection limit.  We find that this partitioning holds up very well where we can measure all three components, as can be seen in Table~\ref{cplusline} where we give the $I_{\rm{13}}$ integrated sum using both methods, thus supporting the assumption of a low $^{13}$C$^+$ optical depth.  

If $^{12}$C$^+$ line emission is also treated as optically thin, then $R$ matches the tabulated values of $I_{\rm{12}}/I_{\rm{13}}$.  However these would yield abundance ratios that are a factor of 2 to 3 lower than the total  ratio of 67 based on local $^{12}$C$^{18}$O and $^{13}$C$^{18}$O observations in Orion by Langer \& Penzias (1990, 1993).   Boreiko \& Betz (1996) derived $\tau_{\rm{12}} = 1.2$ towards {\thet1}, based on a kinetic temperature $T_{\rm{kin}}$ = 185$\pm$15 K calculated from O~{\sc{i}} and C~{\sc{ii}} observations along that sightline.  If we set $\tau_{\rm{12}} = 1.2$ as an average over our mapped region around KL, then we obtain an abundance ratio $R$ = 53 for the diffuse medium outside the Hot Core.  This is very close to the value of 57 given by the relationship of $^{12}$C/$^{13}$C with Galactocentric radius obtained by Langer \& Penzias (1990), and the value $58^{+6}_{-5}$ for \thet1 by Boreiko \& Betz (1996) based on their measured ionic intensity ratio and independently derived value of $\tau_{\rm{12}}$.

\begin{figure*}
\centering
  \includegraphics[width=14.0cm]{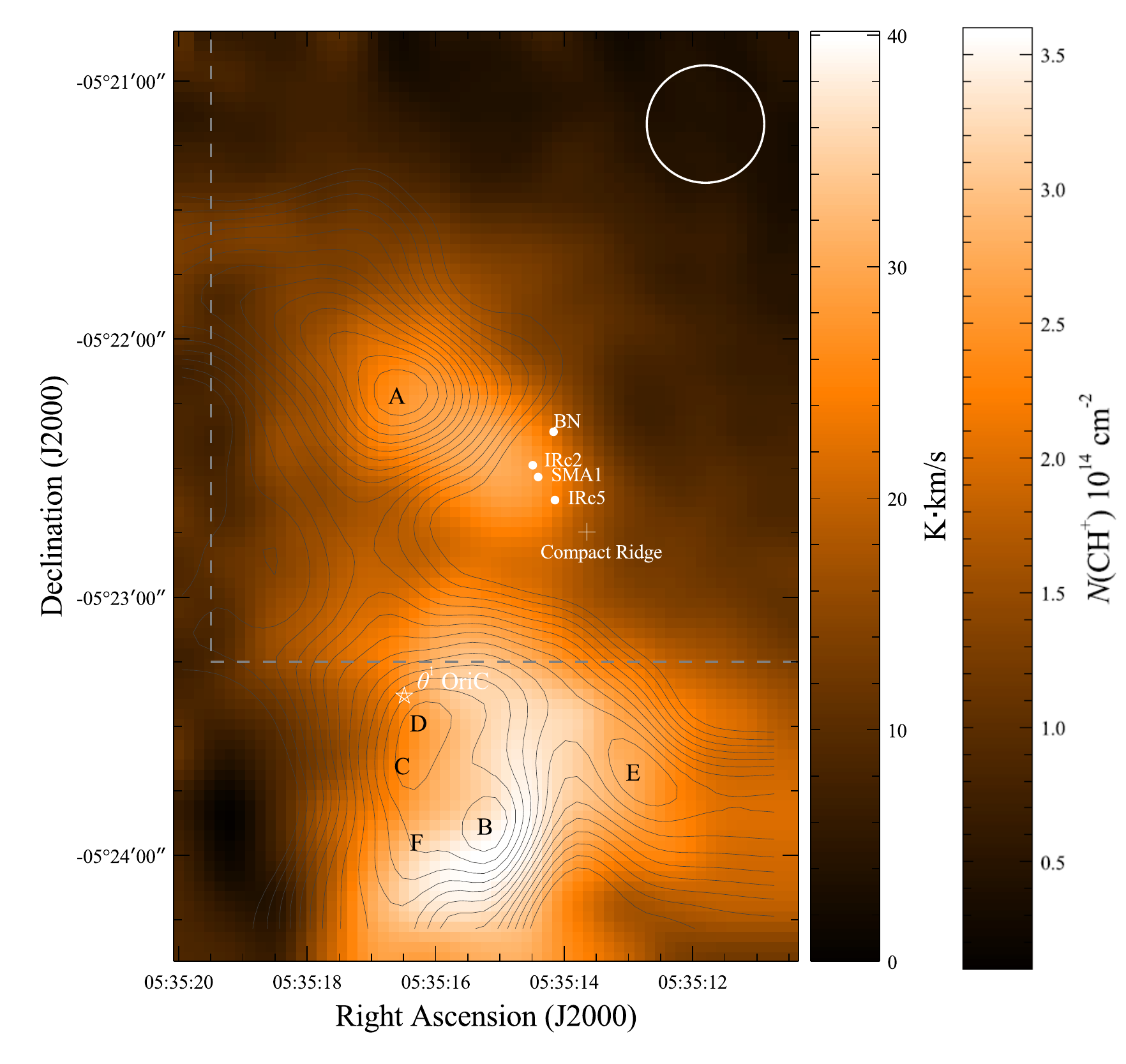}
  \caption{CH$^+$ $J = 1-0$ intensity integrated over $-$5 to $+$20 {\kms} and a column density scale calculated at $T_{\rm{ex}}$ = 30~K, with contours of $^{12}$C$^+$ integrated over $-$10  to $+$15 km~s$^{-1}$ in gray on a linear scale between 450 and 850  K~ km~s$^{-1}$ (approximating the upper 65\% of the total power) at intervals of 25  K~ km~s$^{-1}$.  The dashed box represents the area mapped in the CH$^+$ $J = 2-1$ line (see Fig.~ \ref{fig:f11}).
\label{fig:f10}}
\end{figure*}

This demonstrates that our result for the average isotopic abundance ratio based on a fixed value of $\tau_{\rm{12}}$ is consistent with previous results along the specific sightline of {\thet1}, and also according to the $^{12}$C/$^{13}$C ratio with Galactocentric radius. We could assume that the isotopic abundance ratio $R$ is constant within the mapped region, and take all variations of the intensity ratio to reflect changes in the column density. Thus we can switch to exploring the spatial variations in the $^{12}$C$^+$ optical depth more quantitatively at a fixed value of $R$. For $R$ we adopt a value of $67^{+6}_{-5}$.    This is high (but within the uncertainty) compared to the value calculated by Boreiko \& Betz (1996) which we used as a consistency check using their value of $\tau_{\rm{12}}$ they derived towards {\thet1}.   Here we adopt the estimate by Langer \& Penzias (1993) since we are re-deriving $\tau_{12}$ based on our HIFI observations, and this value of $R$ has been adopted by Ossenkopf et al. (2013) for the Orion Bar to which we will compare our results.  The assumptions that the $^{13}$C$^+$ emission is optically thin and that the excitation temperature is the same for $^{12}$C$^+$ and $^{13}$C$^+$ are included in the $15\%$ uncertainty.

Our results for the $^{12}$C$^+$ optical depths,  given in Table~\ref{cplusline}, show large variations around the average value, because the $^{12}$C$^{+}/^{13}$C$^+$ intensity ratio varies by position. Generally, the $^{12}$C$^+$ line optical depth ranges from 1.4 to 2.6 outside the Hot Core, and from 4.0 to 6.2  for the Hot Core sightline.  The tabulated values are consistent overall with results by Goicoechea et al. (2015b) for the whole of OMC~1 at the same value of $R$ (although they do not quote an estimate for the Hot Core region where we find very high values of $\tau_{12}$), considering the sensitivity differences of the observations. Our estimates are based on measurements of the two strongest emission components of $^{12}$C$^+$ with {\vlsr} and widths similar to observed $^{13}$C$^+$ (and CH$^+$ and CH presented below).  If we had restricted our estimates to the main emission component, as done by Goicoechea et al., our values of $\tau_{12}$ would increase by $\approx$20\% (on average).  Conversely, if we include the full range of observed $^{12}$C$^+$  velocity components $-$10 to $+$15 \kms, the optical depths will decrease by about the same amount.  

\begin{figure}
\centering
  \includegraphics[width=\columnwidth]{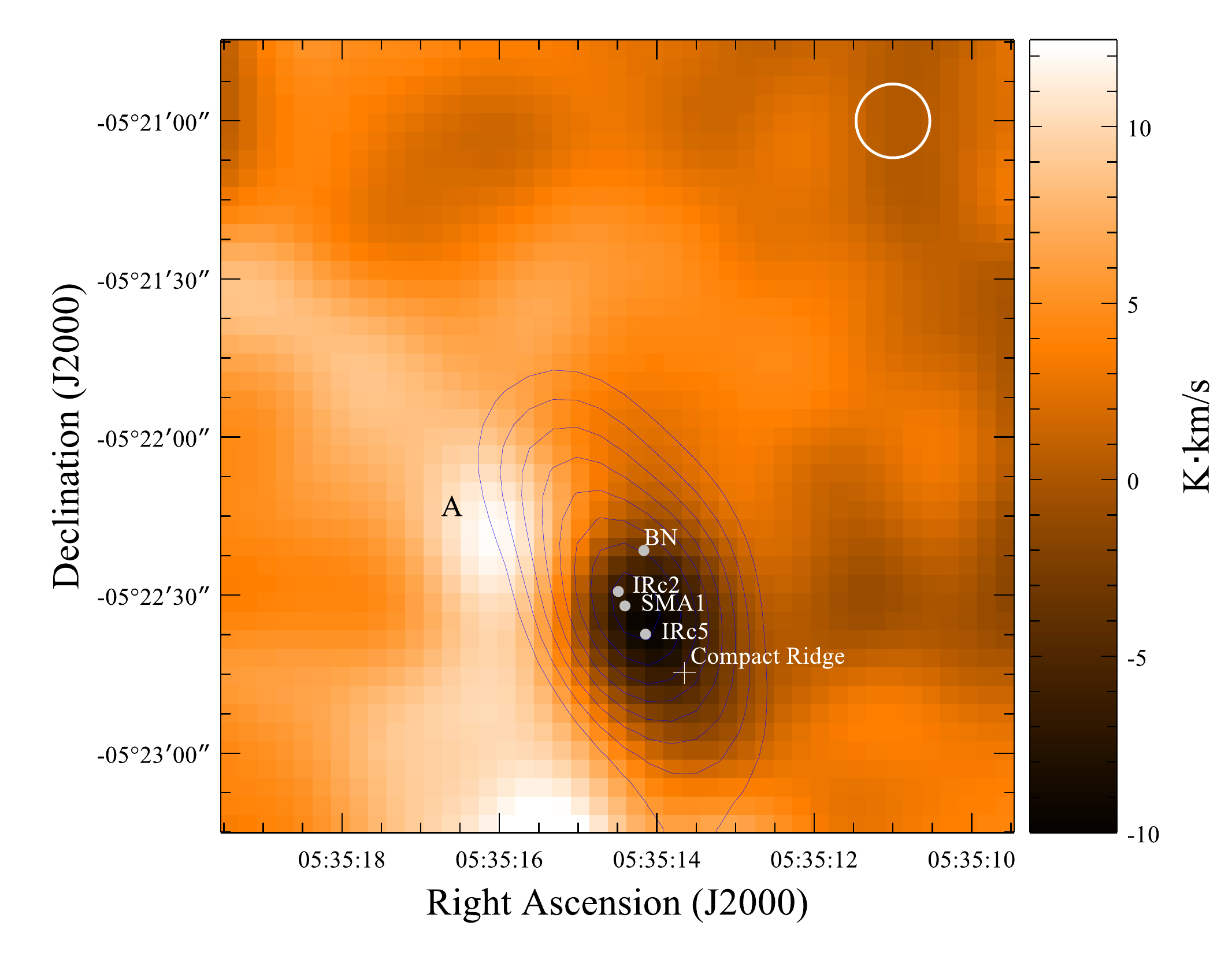}
  \caption{Integrated CH$^+$ $J= 2-1$ integrated intensity with contours of the 1900 GHz continuum as in Fig.~\ref{fig:f3}. 
\label{fig:f11}}
\end{figure}

The variation of $^{12}$C$^+$ optical depths around Orion BN/KL is similar to that observed in the Orion Bar PDR (Ossenkopf et al. 2013).  There, as in Orion BN/KL,  the role of chemical fractionation of carbon driven by the reaction $^{13}$C$^+$ + CO $\rightleftharpoons$  $^{12}$C$^+$ + $^{13}$CO + 34.8 K (Langer et al. 1984) is difficult to quantify due to the variation in optical depth of the main isotopologues.  A significant fractionation into $^{13}$CO would enhance the $^{12}$C$^{+}/^{13}$C$^+$ abundance ratio, yielding $I_{\rm{12}}/I_{\rm{13}}$ that is equal to or higher than the elemental abundance ratio.  However, our measured $^{12}$C$^{+}/^{13}$C$^+$ intensity ratios at all positions in Orion BN/KL are {\em lower} than the adopted $^{12}$C$/^{13}$C isotopic abundance ratio of $67^{+6}_{-5}$, indicating that moderate to high optical depths of C$^+$ lead to a reduction in the $^{12}$C$^{+}/^{13}$C$^+$ intensity ratio, thereby offsetting the effect of a $^{12}$C$^{+}/^{13}$C$^+$ abundance ratio enhanced by chemical fractionation or isotope-selective photo-dissociation.  A similar conclusion was reached by Keene et al.  (1998) with $^{12}$C {\sc{i}}, $^{13}$C {\sc{i}}, $^{12}$C$^{18}$O, and  $^{13}$C$^{18}$O observations at a position near Orion S.  Their derived $\langle\tau\rangle$ for C {\sc{i}} $4\arcmin$ south of IRc~2 is comparable to what we estimate for C$^+$ in Orion~KL, while isotopic abundance ratios in Orion S are also close to that of the ISM.  This indicates that either carbon fractionation is negligible, or that the $^{13}$C$^+$ abundance is enhanced by photo-dissociation of $^{13}$CO in the strong FUV field. Ossenkopf et al.  (2013) have concluded that isotope-selective photo-dissociation plays a very minor role in their PDR models for the Orion Bar region.     

The excitation temperature $T_{\rm{ex}}$ is related to the measured radiation temperature $T_{\rm{R}}$ through the equation (e.g. Goldsmith \& Langer 1999)

\begin{equation}\label{eq:TRad}
T_{\rm{R}} = \left( \frac{T_*}{\exp (T_*/T_{\rm{{ex}}}) - 1} - \frac{T_*}{\exp (T_*/T_{\rm{{bg}}}) - 1} \right) \left[ 1 - e^{-\tau} \right] (\rm{K}).
\end{equation}

\noindent For optically thick C$^+$ line emission that fills the HIFI beams, 

\begin{equation}\label{eq:CplusTex}
T_{\rm{ex}}({\rm{C}}^+) = \frac{91.2}{ {\rm{ln}} \left[ 1 + 91.2/ \left( T_{\rm{MB}}(^{12}{\rm{C}}^+) + T_{\rm{c}}(1900.5)  \right) \right] } \; ({\rm{K}})
\end{equation}

\noindent where $T_{\rm{MB}}(^{12}{\rm{C}}^+)$ is the line peak on a main beam temperature scale and $T_{\rm{c}}(1900.5)$ is the continuum brightness temperature (as defined in Sec.~\ref{cplusproperties}) in K, including source continuum emission and a cosmic microwave background component of 2.74~K.  Goicoechea et al. (2015) have noted the rather low brightness temperatures compared to the peak temperatures of the main spectral component across OMC~1.  Over our mapped region, $T_c$/$T_{\rm{MB}}$ is at most $\approx$5\%, occurring at the location of peak continuum emission is strongest, which is where C$^+$ line emission is weak in the direction of SMA~1.   Equation~\ref{eq:TRad} assumes that the HIFI beam is filled, while Figures~\ref{fig:f7} and \ref{fig:f8} prove that some of the emission is very structured, with spatial variations below the beam widths. For this weak, structred emission the assumption of a filled beam is not well justified, but for the peak emission, the spatial profile is much smoother.

The distribution of C$^+$ excitation temperatures estimated using Equation~\ref{eq:CplusTex} is shown Figure~\ref{fig:f9}.  Over this region $T_{\rm{ex}}$(C$^+$) ranges from a minimum of 73 K to the west of SMA~1, to a maximum of 300 K at Position B, with an average $\langle T_{\rm{ex}} \rangle$ = 178 K.  This value is comparable to that estimated towards {\thet1} by Boreiko \& Betz (1996) and adopted for the Orion Bar PDR by Ossenkopf et al. (2013), but almost a factor 2 higher than the average $T_{\rm{ex}}$(C$^+$) reported over the larger mapped 75 arcmin$^2$ area of OMC~1 by  Goicoechea et al. (2015b), who also assumed that all C$^+$ emission is optically thick.   The higher Orion BN/KL average is probably explained by the closer proximity of the mapped gas to the Trapezium stars, although there is no strict trend of temperature decreasing with projected distance from the Trapezium within this region, as can be seen in Table~\ref{cplusline} values of $T_{\rm{ex}}$ at the different locations.  For the values of $T_{\rm{ex}}$ given in Table~\ref{cplusline}, we used the form of Eq.~\ref{eq:CplusTex} without taking a limit on $\tau$, instead using the optical depths derived specifically for each location.

The column densities of C$^+$ are given in Table~\ref{cplusline}, calculated using the derived optical depths and excitation temperatures.   Outside of the Hot Core region, the average column density $\langle N({\rm{C}}^+) \rangle \simeq 1.0 \times 10^{19}$ cm$^{-2}$ agrees almost exactly with the value estimated by Ossenkopf et al. (2013) for the Orion Bar where C$^+$ intensity peaks  and a factor of 3 higher than the OMC~1 average quoted by Goicoechea et al. (2015b), while the column densities at the positions marked in Figure~\ref{fig:f3} around BN/KL are a factor 2 to 3 higher than at the Orion Bar peak. 

\begin{deluxetable*}{lccccccccccccc}
\tabletypesize{\small}
\tablecaption{Line measurements of CH$^+$ at selected positions \label{chplusline}}
\tablewidth{0pt}
\tablehead{
\colhead{Position}   &  \colhead{$I_1$}  &  \colhead{$v_{\mathrm{lsr,1}}$} & \colhead{$\Delta v_1$} & 
                      \colhead{$I_2$}  &  \colhead{$v_{\mathrm{lsr,2}}$} & \colhead{$\Delta v_2$} & 
                      \colhead{$I_1$}  &  \colhead{$v_{\mathrm{lsr,1}}$} & \colhead{$\Delta v_1$}&  
                      \colhead{$I_2$}  &  \colhead{$v_{\mathrm{lsr,2}}$} & \colhead{$\Delta v_2$} &  \colhead{$N$(CH$^+$)} \\
 & \multicolumn{3}{c}{($J=1-0$ comp. 1)} &  \multicolumn{3}{c}{($J=1-0$ comp. 2)} &  \multicolumn{3}{c}{($J=2-1$ comp. 1)} &  \multicolumn{3}{c}{($J=2-1$ comp. 2)} & \colhead{($10^{14}$~cm$^{-2}$)} 
}
\startdata
Average & 10.1 &$\;$ 9.4 & 5.4 & $\;$ 1.1 & 12.9 & 14.7 & 1.9 & $\;$ 8.7 & 3.9 & \dotfill & \dotfill & \dotfill &  0.8 \\
A & 16.4 & $\;$ 9.3 & 4.7 & 12.2 & 12.8 & 12.1 &  5.5 & 10.0 & 4.7 & 1.7 & 14.8 & 3.3 &  2.6 \\
A$'$ & 12.8 & $\;$ 9.3 & 5.1 & 13.6 & 12.6 & 13.5 & 4.9 & 10.3 & 4.1 & \dotfill & \dotfill & \dotfill & 1.4 \\
B & 36.8 & $\;$ 9.0 & 5.4 & \dotfill & \dotfill & \dotfill & nc & nc & nc & nc  & nc & nc & 3.3 \\
C & 24.6 & 10.1 & 6.1 & \dotfill & \dotfill & \dotfill & nc & nc & nc & nc  & nc & nc & 2.3 \\
D & 30.2 & 10.2 & 6.4 & \dotfill & \dotfill & \dotfill &  nc & nc & nc & nc  & nc & nc & 2.5 \\
E & 27.5 & $\;$ 7.3  & 6.1 & \dotfill & \dotfill & \dotfill &  nc & nc & nc & nc  & nc & nc & 2.4 \\
F & 36.9 & $\;$ 8.8 & 4.6 & \dotfill & \dotfill & \dotfill &  nc & nc & nc & nc  & nc & nc & 3.1 \\
Hot Core  & $\;$ 7.2 & $\;$ 8.6 & 5.4 & $\;$ 2.4 & 11.8 & 15.3 & 6.5 & 12.6 & 4.8 & $-$28.9 & 8.8 & 9.3 & 3.1 \\
\enddata

\tablecomments{Column 1 positions are the same as in Table~\ref{cplusline}.  Measurement uncertainties at all positions are 5\%, except towards the Hot Core where the $J=2-1$ intensity uncertainty is 20\% due to strong blending with LOS absorption. Line integrated intensities $I$ are in units of K \kms, peak velocities $v_\mathrm{lsr}$ and velocity widths $\Delta v$ are in \kms.  Subscripts '1' and '2' denote the strongest and secondary components, where detected.  Dotted entries indicate non-detections, 'nc' at Positions B through F for the $J=2-1$ line indicate no observational coverage. Column densities are computed with $T_{\rm{ex}}({\rm{CH}}^+)$ = 30~K, and are based on averaging weighted by SNRs of the emission measured from the ground and first excited level transitions at Position A, A$'$, the Hot Core  (the fitted line emission component only) and the Average.  All other positions are based on the $J=1-0$ line only.  The Hot Core South is not included due strong overlap of signal with the Hot Core (due to the beam size at 835 GHz).}

\end{deluxetable*}


\subsection{Morphology of CH$^+$ $J = 1-0, 2-1$}
\label{CHplus}

The intensity distribution of emission around Orion BN/KL from transitions to the ground and first excited levels of CH$^+$ are shown in Figures~\ref{fig:f10} and \ref{fig:f11}, integrated over \vlsr \ between 0 and $+$20 \kms.  Column densities $N$(CH$^+$) based on the $J=1-0$ measurements, where we have largest spatial coverage, are included with Figure~\ref{fig:f10}, calculated at an excitation temperature $T_{\rm{ex}}$(CH$^+$) = 30~K, as discussed below.  For comparison, the overall distribution of the CH$^+$ $J=1-0$ emission in OMC~1 (right panel of Fig.~\ref{fig:f1}) appears similar to that of the 350 $\mu$m (857 GHz) emission mapped with SPIRE.  The CH$^+$ $J=1-0$ line falls within the SPIRE filter bandpass, but  CO $J_{\rm{upper}}$ = 5 through 8 probably dominates the extended emission, especially towards the BN/KL complex.  Continuum emission extending from Orion South to Orion North, peaking in the Hot Core region, also contributes.  The extended CH$^+$ emission from the larger OMC~1 map (see Fig.~\ref{fig:f1}, right) roughly encircles a  3$'$.1 diameter region of cloud surfaces bound by strongest emission from the Orion Bar PDR, the Orion South star formation site, the Hot Core and BN/KL complex,  and a diffuse arc of emission extending to a bright region of CH$^+$ emission just to the north of the Orion Bar at its eastern extent.  

The CH$^+$ emission is somewhat more smoothly distributed than that of C$^+$, exhibiting lower variations in peak intensity overall. There are no convincing indications of CH$^+$ $J=1-0$ absorption along any line of sight.  Towards the BN/KL complex, the $J=1-0$ intensity is $\approx$30\% lower whereas the C$^+$ emission is more than a factor 2 lower (see Tables \ref{cplusline} and \ref{chplusline}) compared to the respective averages.   In this line of sight, the baseline around the CH$^+$ line is partially blended with other molecules such as SO$_2$ and CH$_3$OH and their isotopologues emitting at nearby frequencies, so that we cannot be certain about possible CH$^+$ $J=1-0$ absorption.   

The CH$^+$ emission peaks exhibit a range of {\vlsr} between 7.0 to 11.0 {\kms}, while the distribution of large bright features is similar to that of C$^+$, i.e., noting similar locations of the C$^+$ Peak A and a CH$^+$ peak emission near the Hot Core, as well as in the peaks to the south of the Compact Ridge.   The highest CH$^+$ $J=1-0$ emission peak is similarly located in an arc of material projected to the south of \thet1.  There is a tendency for the strongest CH$^+$ emission to occur where C$^+$ intensity gradients are strongest, but a notable distinction of the CH$^+$ emission is that LSR velocities are red-shifted $\approx$ $+$2 {\kms} and are almost twice as broad on average {\em{versus}} the main emission component of C$^+$.  At C$^+$ velocities below 8 {\kms} and above 10 {\kms} there is far less correspondence with the distribution of CH$^+$ emission (comparing Fig.~\ref{fig:f6} with Fig.~\ref{fig:f10}).    The coarse radial layering of C$^+$ and CH$^+$  around \thet1, in which the C$^+$ emission peaks closer to the Trapezium stars, is compatible with photo-dissociation of the molecular gas in the hot stars' strong UV field.  By contrast, C$^+$ peaks in projection further to the N-E of BN/KL than CH$^+$, which does not match a PDR scenario for a source of UV radiation from the BN/KL complex or the Trapezium on the arcminute scale.  Goicoechea et al. (2015a,b) showed that the C$^+$ emitting region is widespread in OMC~1, and although it is not exactly following the distribution of CO which has its own velocity-dependent morphology, the overall ionized/PDR/molecular gas interfaces are revealed on the {\it{larger}} scales across OMC~1, through comparison of hydrogen recombination, CO $J=2-1$, and C$^+$ integrated intensity distributions.  

\subsubsection{CH$^+$ absorption}\label{sec:chpabs}
  
The distribution of CH$^+$ $J=2-1$ emission is consistent with that of $J=1-0$ over region mapped in common, comparing Figures~\ref{fig:f10} and \ref{fig:f11}.  The only difference is towards the dense core region where strong $J=2-1$ absorption is detected.   As noted above, there is no indication of absorption in the $J=1-0$ line, though we cannot rule it out due to strong emission and crowding by other emitting molecules.  We can only conclude that the relative strengths of the absorption lines $I_{\rm{abs}}(1-0)/I_{\rm{abs}}(2-1) < 1$.   Profiles are compared in Figure~\ref{fig:f12} at Position A and the Hot Core (centered on SMA~1) with a 30$''$ extraction aperture, and the overall averages.  The comparison shows that the $J=1-0$ line fluxes extracted over the same aperture size (which gives the same result if we convolve the 1670 GHz observations to the beam size at 835 GHz) are below average strength towards the Hot Core (see also columns 1 and 2 of Table~\ref{chplusline}).  A low absorption ratio of $I_{\rm{abs}}(J_u=1)$/$I_{\rm{abs}}(J_u > 1)$ is thought to be an excitation signature of a diffuse columns of material undergoing dissipative turbulence processes, in contrast to regions in which the level populations are governed by formation pumping (Godard \& Cernicharo 2013).   On an optical depth scale, Figure~\ref{fig:f13}, the $J = 2 - 1$ line  is a 25\% absorber of the 14~K continuum towards BN/KL, using a spectral extraction aperture of 30$''$ diameter.   The absorption implies an excitation temperature of the absorbing material

\begin{equation}\label{TexCHp21}
T_{\rm{ex}}({\rm{CH}}^+) < \frac{80.1}{{\rm{ln}}[1 + 80.1/T_{\rm{c}}(1669)]} \simeq 25.0 \; \rm{K}
\end{equation}

\noindent where the continuum brightness temperature at 1669 GHz, $T_{\rm{c}} = 80.1/(e^{80.1/T_{\rm{bg}}} - 1) \simeq 3.4$ K.  The continuum levels are much lower at 850 GHz, $\approx$5.5~K, yielding $T_{\rm{c}}$ = 0.7~K and an absorption condition $T_{\rm{ex}} <$ 10~K on the $J=1-0$ line.  While we cannot claim that $J=1-0$ is not in absorption due to the mentioned emission line crowding, we can assert that it is {\it not detected} either directly or by profile decomposition, thus $10 \; < T_{\rm{ex}}({\rm{CH}}^+) < \; 25$~K.

\begin{figure}
\centering
  \includegraphics[width=\columnwidth]{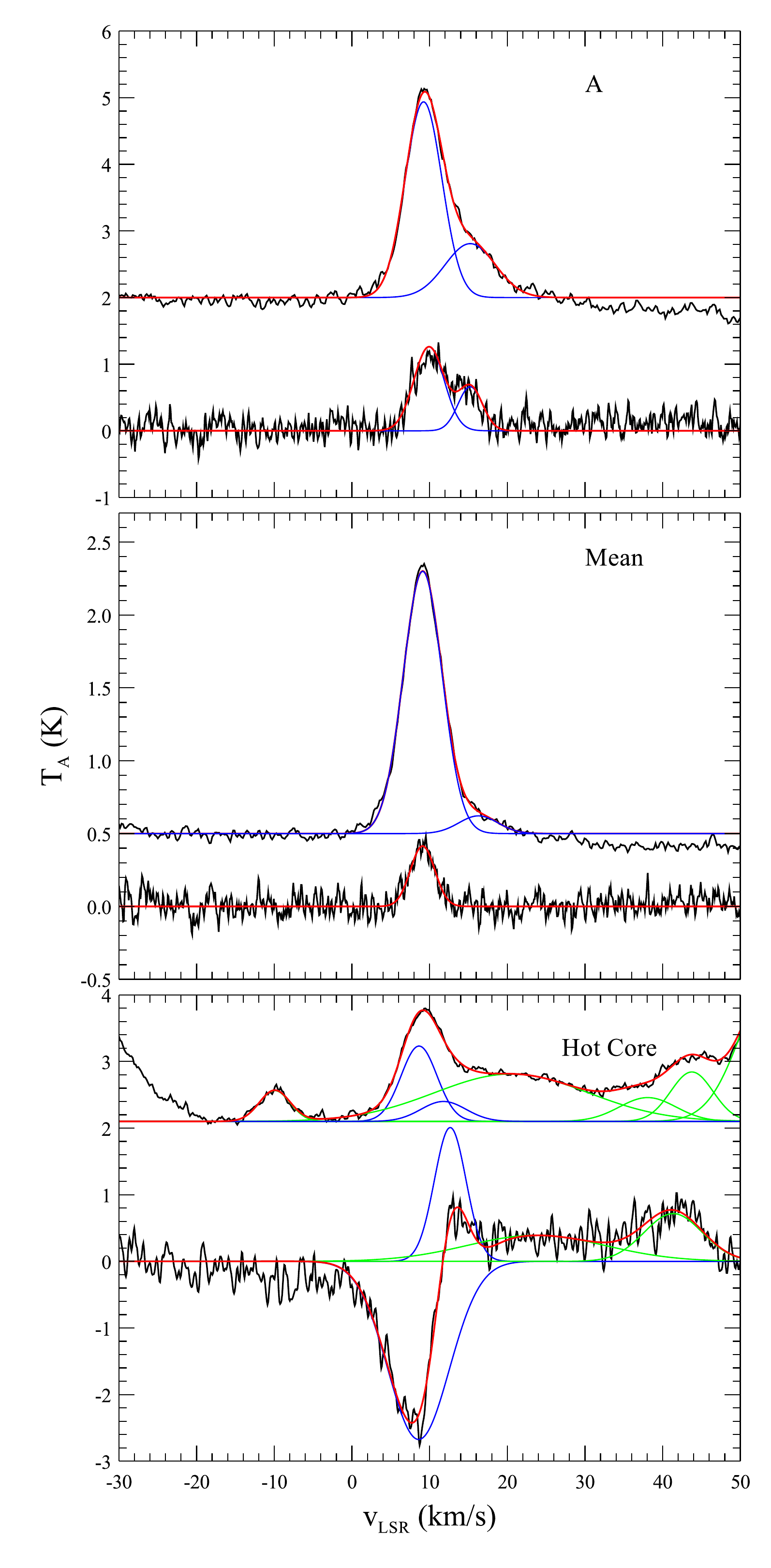}
  \caption{CH$^+$ $J = 1-0$ and $2-1$ spectra at Position A (top panel), averaged over the common mapped area around BN/KL (middle panel), and towards the Hot Core region (bottom panel).
 Continuum levels have been subtracted, and in each panel the $J=1-0$ spectrum is plotted on top, offset for clarity.  Spectra at C$^+$ Peak A and the Core regions have been extracted over a circular aperture of 30$''$ diameter suited to the HIFI beam at 835 GHz.  The mean spectrum corresponds to the region marked by the dashed box in Fig.~\ref{fig:f10}, excluding a 30$''$-diameter region centered on IRc~2 where CH$^+$ $J=2-1$ is in absorption.  Gaussian profiles obtained by Levenberg-Marquardt minimization are shown in blue for each attributed CH$^+$ component, in green for nearby features from other species contributing to the overall fit, and in red for the total model.  Best fit parameters for the CH$^+$ components are given in Table~\ref{chplusline}. 
\label{fig:f12}}
 \vspace{-2em}
\end{figure}

The Gaussian decompositions to the Hot Core CH$^+$ spectra are numerical best fits, but are not unique due to blending with nearby emission features.  For example, the best fit to the $J=2-1$ profile yields an emission component at {\vlsr} = 12.6 {\kms}, considerably red-shifted from the average of 9.5 {\kms}.  If we were to fix the emission component to the average LSR velocity, the next best fit at nearly equal confidence level yields 30\% weaker emission, and the absorption profile must be fit by two Gaussian profiles at LSR velocities {\vlsr} of 6.1 and 8.4 \kms, and widths $\Delta v$  of 5.9 and 3.8 \kms.  Hence, the column densities tabulated in Table~{\ref{chplusline}}, which are based on an average of measurements of both $J=1-0$ and $2-1$ (weighted by SNRs), may be somewhat overestimated towards the Hot Core due to this fitting uncertainty.  If we rely only the $J=1-0$ line in the Hot Core spectrum, which is how Figure~\ref{fig:f10} has been produced, then the value of $N$(CH$^+$) toward the Hot Core decreases by around 35\%.  The uncertainty is entirely attributable to blending with nearby species and the mixed gas temperatures in this line of sight.  

\subsubsection{CH$^+$ emission}\label{sec:chpem} 

Outside of the Hot Core line of sight, the ground and first excited CH$^+$ transitions share a similar kinematic structure; see Table~\ref{chplusline} for {\vlsr} and line widths.  Both $J=1-0$ and $2-1$ emission lines are well fit by a single Gaussian profile over most of the mapped regions with {\vlsr}  in the range of 9.3 to 10.0 {\kms} (with the exception of Position E with $J=1-0$ emission at 7.3 \kms), and widths $\Delta v_{\mathrm{LSR}}$ in the range of $4.6-6.4$ \kms.   At Peaks A, A$'$, and towards the nearby Hot Core we detect a secondary component of $J=1-0$ emission red-shifted to $13 - 15$ {\kms}, which is strong enough to be observed in the overall average spectrum.  Only at Position A can we detect the second component in the $J=2-1$ emission profile, consistently red-shifted but at a lower intensity ratio with respect to the main component compared to the corresponding $J = 1-0$ ratio of the two velocity components at the same position.    

\begin{figure}
  \begin{center}
  \includegraphics[width=\columnwidth]{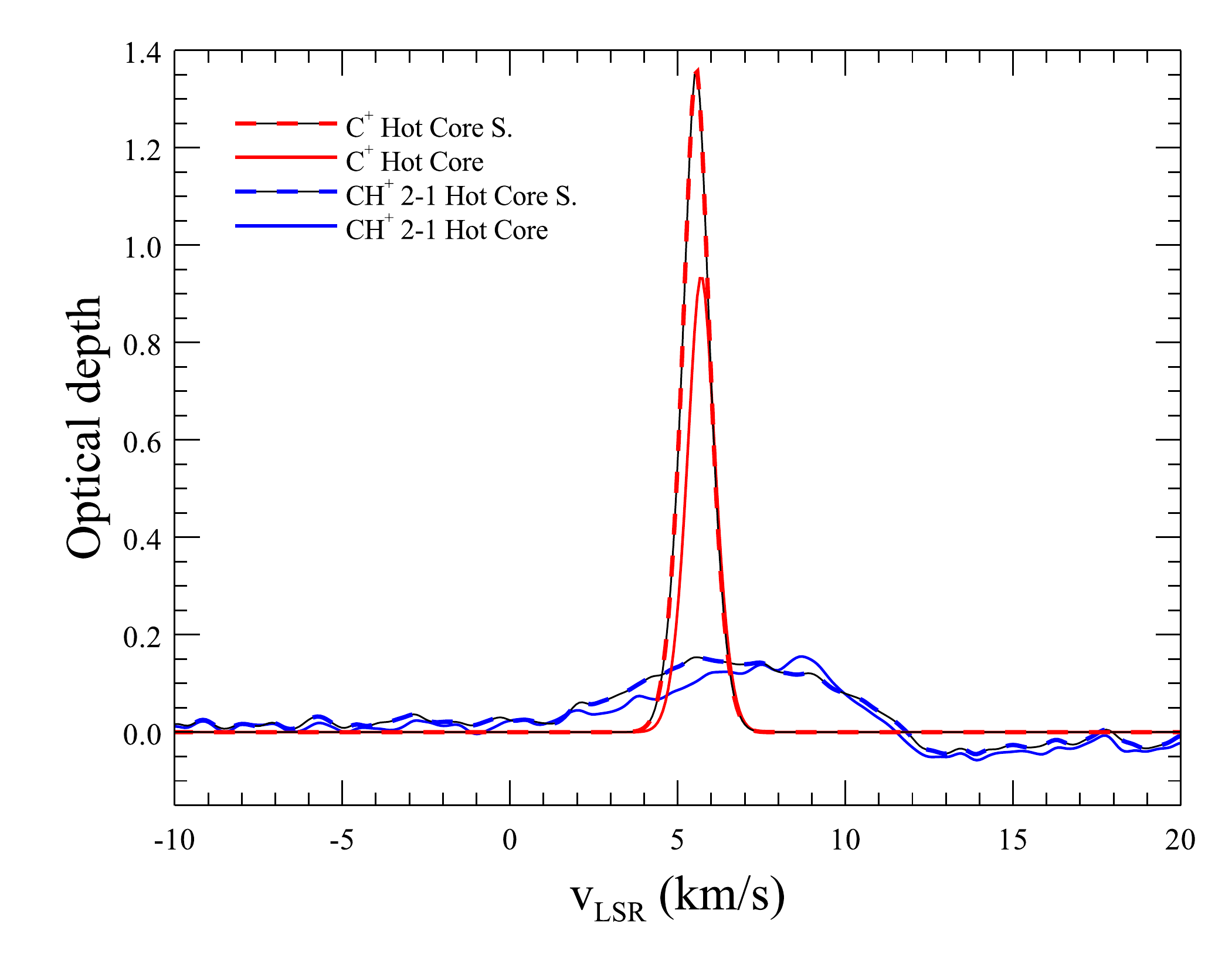}
  \end{center}
  \caption{Optical depths of the C$^+$ and CH$^+$ $J = 2-1$ lines towards the Orion Hot Core and Hot Core South as indicated, each extracted in a 30$''$ diameter aperture.  No CH$^+$ $J=1-0$ absorption is observed (see text). 
\label{fig:f13} }
\end{figure}

\begin{figure}
  \begin{center}
  \includegraphics[width=\columnwidth]{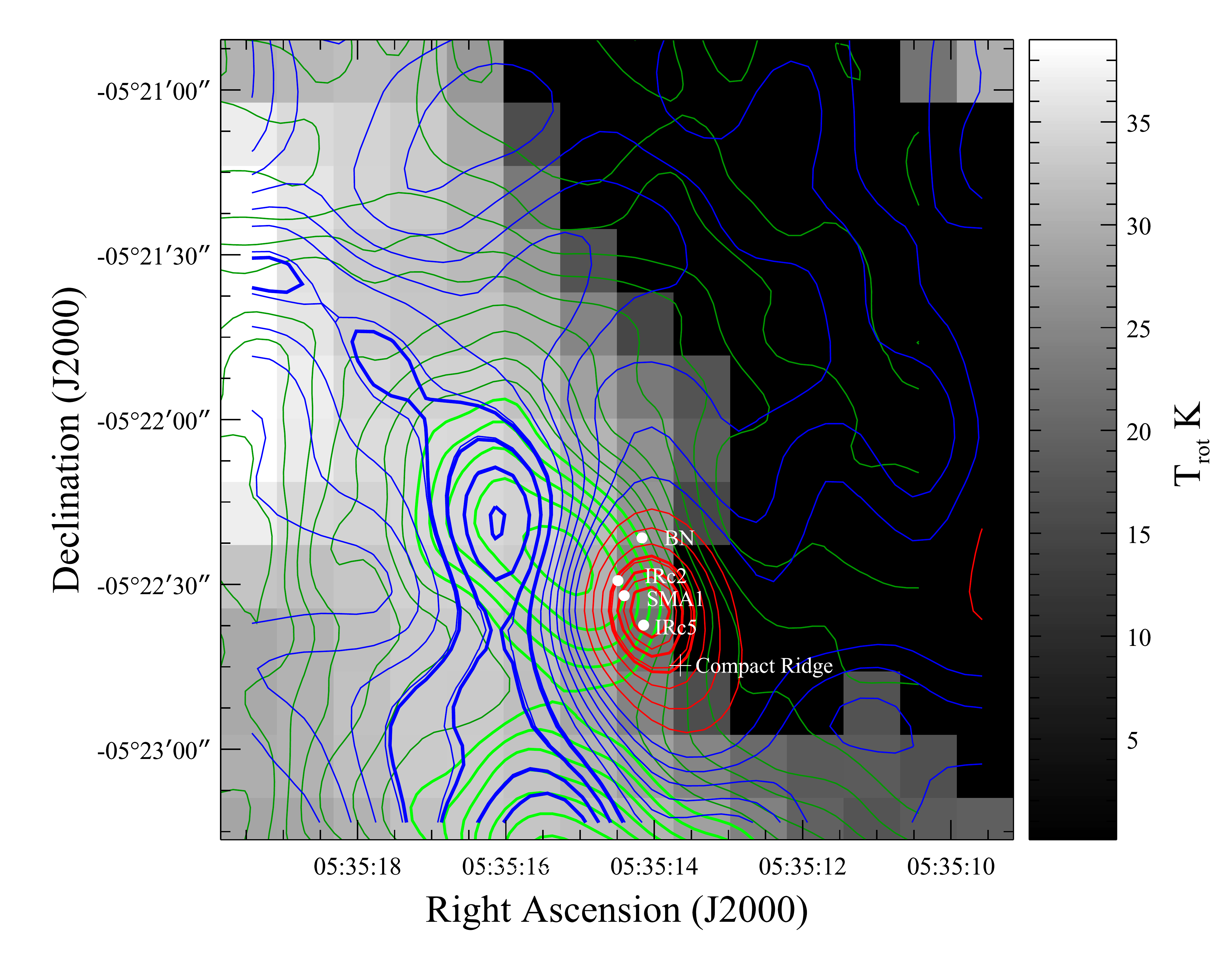}
  \end{center}
  \caption{Rotational temperature map of CH$^+$ derived from the $J=1-0$ and $2-1$ emission lines, under the assumption of the emitting gas in LTE at a single temperature, i.e., a linear slope ($-1/kT$) in the Boltzmann excitation representation.  Overlaid are contours of integrated intensities: the $J=1-0$ emission (green) between 0 and 34 K~{\kms} on intervals of 2 K~{\kms}; $J=2-1$ emission (blue) between 0 and 10 K~{\kms} on intervals of 1 K~{\kms}; and $J=2-1$ absorption (red) between $-$0.5 and $-$12.5 K~{\kms} on intervals of 2 K~{\kms}.  Thick contours indicate the upper 60~\% power of integrated intensities.  The two pixels in the upper right (NW) corner of the map are spurious, where detection of $J=2-1$ emission is at the noise limit. 
\label{fig:f14} }
\end{figure}

A minimum of two rotational transitions of CH$^+$ can be used to derive rotational temperatures $T_{\rm{rot}}$ following the analytical expressions by Goldsmith \& Langer (1999), using a linear fit to a Boltzmann excitation diagram of the emission lines at each map point.   With this approach, plotting the logarithm of the column density per rotational level, ln($N_J /g_J$), versus energy, $E_J$, and fitting a straight line between the two CH$^+$ transitions yields rotational temperatures in the range of 25 to 35 K (Fig.~\ref{fig:f14}), which is clearly sub-thermal.  The column densities are a few 10$^{13}$ cm$^{-2}$.

The assumption of a single temperature (or of strict LTE) of the emitting gas is required along all sightlines in order to derive the column density from a linear slope ($-1/kT$) in the Boltzmann diagram.   If this condition is not met, the shape of the curve can deviate from linearity, as Neufeld (2012) has shown with {\em{Herschel}}/PACS observations of the CO ladder along various warm ISM sightlines.  The sub-thermal value of $T_{\mathrm{rot}}$ derived from these Orion BN/KL observations of the two lowest rotational lines of CH$^+$ could be an indicator of negative curvature of Boltzmann excitation curve that would be revealed when higher-J transitions are included.  A non-linear curve does not necessarily mean that non-LTE conditions prevail, but to get a reliable column density estimate, the partition function must be known and for this we need to know $T_{\rm{ex}}$ for all transitions. This is the same number only in LTE.  Otherwise the column densities based on only two lines may be incorrect.   Nonetheless, the excitation (i.e., rotation) temperature of the absorbing material revealed by the CH$^+$ $J=2-1$ Hot Core profile (Fig.~\ref{fig:f12}) we estimated using Eq.~\ref{TexCHp21} is in very close agreement with $T_{\rm{rot}}$ along that sightline.  Similarly, the range of excitation temperatures based on the {\it{emission}} lines is 25 K $\leq \; T_{\rm{ex}}({\rm{CH}}^+) \; \leq$ 32 K at Peaks A, A$'$, and the Hot Core, using the appropriate form of Eq.~\ref{eq:TRad} with $T_* = 80.1$ K, again in very good agreement with the simple rotation diagram temperatures.  Hence a consistent value of $T_{ex}$ is reached by different analytical approaches, further implying that a single CH$^+$ gas temperature is a good approximation. 

In order to further explore the formation and excitation of CH$^+$ in this region, we next compare the observed line intensities to predictions by PDR and full radiative transfer modeling, proceeding similar to the recent study by Nagy et al. (2013) of reactive ions CH$^+$, SH$^+$, and CF$^+$ observed in the Orion Bar.   To do so, we must include higher-$J$ transitions which were observed as part of the HEXOS program, using the {\em{Herschel}}/PACS integral field unit spectrometer.   The coverage of the PACS data is similar to that of our HIFI map of the $J=2-1$ transition, represented as the gray box in Figure~\ref{fig:f1}.  From these data, we can measure only the $J = 4-3$ and $5-4$ lines towards C$^+$ Peak A, and in overall averages that exclude the saturated BN/KL complex.    We obtain estimates of 1.05 and 0.21 K {\kms} for $J=4-3$ emission at Peak A and for the average, and similarly 0.33 and 0.13 K {\kms} (both upper limits accounting for uncertainty in surrounding baseline levels) for $J=5-4$.  Unfortunately, non-linear photometric responses caused by high brightnesses of the BN/KL region and spectral line confusion create difficulties to extract reliable $J = 3-2$ through $6-5$ line fluxes from these data.   A full description of the observations and reductions of the PACS observations will be provided by Goicoechea et al.  (in preparation).

\begin{figure}
\centering
\includegraphics[width=\columnwidth]{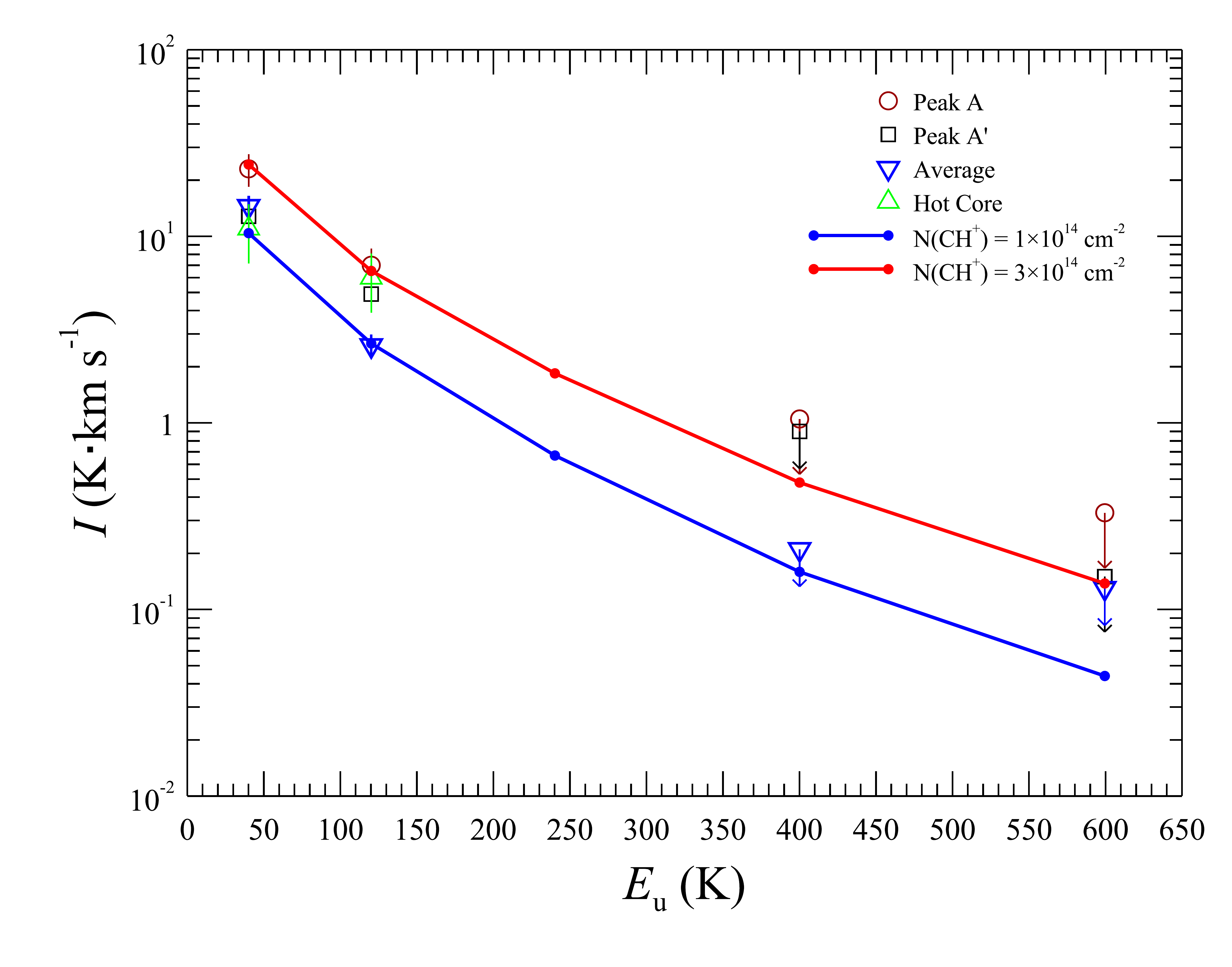}
\caption{The CH$^+$ line intensities measured toward Orion BN/KL peak positions A and A$'$, the Average, and the Hot Core (towards SMA~1), with symbols given in the legend.  Plotted values for the $J=4-3$ and $5-4$ integrated line fluxes are upper limits.  The over-plotted RADEX models in red and blue dots correspond to a kinetic temperature of 500 K, an H$_2$ volume density of 2$\times$10$^5$ cm$^{-3}$, an electron density of 10 cm$^{-3}$, and CH$^+$ column densities as indicated.
\label{fig:f15} }
\vspace{-2em}
\end{figure}

\subsubsection{``RADEX'' radiative transfer models}
\label{radex}

We first analyze the observed CH$^+$ line intensities using the non-LTE radiative transfer code RADEX \citep{vandertak2007}, by applying physical parameters that are expected for the CH$^+$ emitting gas. We include a continuum model based on measurements with the Short and Long Wavelength Spectrometers which flew on the Infrared Space Observatory, and with \textit{Herschel}/HIFI as described in \citet{crockett2014} and references therein. The collisional rates for CH$^+$-He collisions were taken from \citet{turpin2010} and were scaled to represent CH$^+$-H$_2$ collisions based on \citet{schoier2005}. Collisional rates for CH$^+$-e$^-$ collisions are taken from \citet{lim1999}; see also Hamilton et al. (2015).
As CH$^+$ is very reactive, inelastic collision rates with H$_2$ and electrons are similar to the chemical reaction rates with these species (e.g. \citealp{stauberbruderer2009}). 
Therefore, we consider the chemical formation and destruction rates in the statistical equilibrium calculation as described in \citet{nagy2013}. However in Nagy et al., due to the lack of knowledge of the state-to-state formation rates of CH$^+$, the formation rate into level $i$ was expressed as a Boltzmann distribution over all states at an effective formation temperature $T_{\rm{f}}$. In this paper we use the state-to-state formation rates for the reaction H$_2$($v$=1) + C$^+$ $\rightarrow$ CH$^+$ + H computed by \citet{zanchet2013}.

\begin{figure}
 \centering
  \includegraphics[width=\columnwidth]{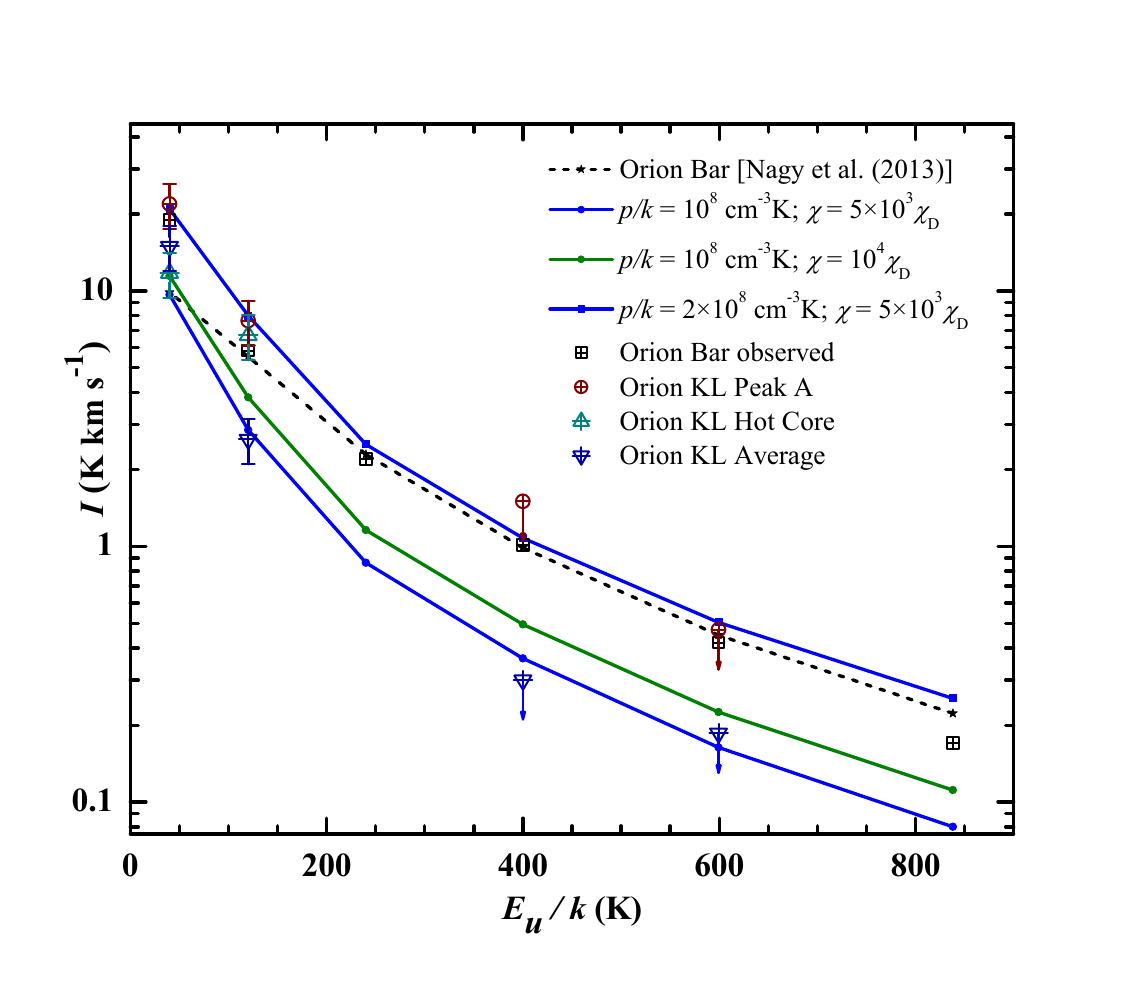}
  \caption{Observed intensities of CH$^+$ compared with those calculated from isobaric PDR models.  The observed and calculated values for the Orion Bar PDR (Nagy et al. 2013) are shown for comparison. 
\label{fig:f16} }
\vspace{-2em}
\end{figure}

Line intensities or upper limits are available for four transitions toward the C$^+$ Peak A position and in the average spectrum toward the measured area. RADEX models for CH$^+$ line intensities toward the Orion Bar suggested high kinetic temperatures, such as 500 K. 
Assuming a kinetic temperature of 500 K and an H$_2$ volume density of 5$\times$10$^5$ cm$^{-3}$ the observed line intensities are consistent with a CH$^+$ column density of 1 $\times$10$^{14}$ cm$^{-2}$ for the Orion BN/KL Average, and 3 $\times$10$^{14}$ cm$^{-2}$ at Peak A; see Figure~\ref{fig:f15}. 
The excitation temperatures are in the range between 23~K and 40~K from low- to high-$J$ for both models.
For these results we assumed an electron density of 10 cm$^{-3}$. 
Increasing the electron density to 100 cm$^{-3}$ increases the predicted integrated intensities by factors of $1.3-1.4$. 
Assuming an electron density of 100 cm$^{-3}$ and applying the H$_2$ volume density and kinetic temperature quoted above, the CH$^+$ column densities decrease to 8$\times$10$^{13}$ cm$^{-2}$ and to 2 $\times$10$^{14}$ cm$^{-2}$ toward the Orion BN/KL average and Peak A, respectively.  The value of 100 cm$^{-3}$ is where approximately where $n_e$ peaks at maximum cloud penetration depth, as will be shown in the next section.  Furthermore, since CH$^+$ is produced where C$^+$ is the dominant ion, $n_e$ scales with $n_{\rm{H_2}}$, and at the H$_2$ density of $5 \times 10^5$ cm$^{-3}$ assumed above, C/H= $1.4 \times 10^{-4}$ and all carbon ionized, $n_e \approx$ 140 cm$^{-3}$.

To investigate the role of the background continuum in the excitation of CH$^+$, we ran models where we ignored the background continuum measured toward Orion BN/KL (which has a mean dust temperature of $\approx$30 K; Goicoechea et al. 2015b)  and assumed only a cosmic microwave background temperature of 2.76 K. 
In this case, the excitation temperatures drop by factors of 1.2-2.9. 
The integrated intensities for the three lowest transitions increase from 1-50\%, and then decrease by 8-21\% for the $J=4-3$ and $5-4$ transitions.  

In conclusion, the CH$^+$ column density toward the Orion BN/KL average is (0.8-1.0)$\times$10$^{14}$ cm$^{-2}$, and it is (2-3)$\times$10$^{14}$ cm$^{-2}$ toward the C$^+$ Peak A. The excitation of CH$^+$ is affected by its formation and destruction, the continuum radiation field (such as for other reactive ions like OH$^+$, \citealp{vandertak2013}), and the electron density. 

\begin{figure}
\centering
\includegraphics[width=\columnwidth, angle=0]{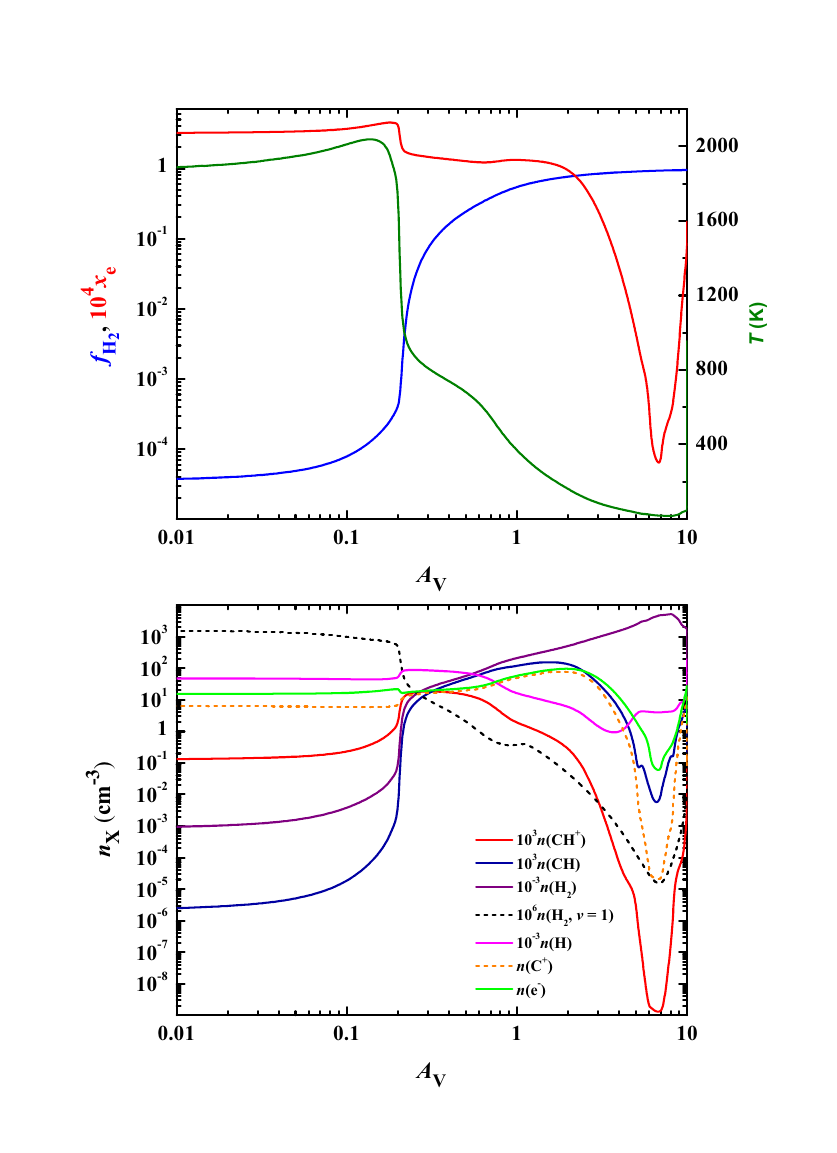}
\caption{Depth dependence of parameters in the PDR model. {\em Bottom}: Number densities of key species involved in the production of CH and CH$^+$; the values of all but C$^+$ and $e$ are scaled for clarity of the comparisons.  {\em Top}: Profiles of the gas kinetic temperature, molecular fraction, and ionization fraction (scaled by a factor of $10^4$).  The results shown are for an isobaric PDR at a pressure $p/k = 10^8$ cm$^{-3}$K, and illuminated by a radiation field $\chi_{front} = 5\times 10^{3} \chi_{D} = 10^{3}\chi_{back}$, where $\chi_{D}$ is the field strength in Draine units.  Adopted initial atomic and molecular abundances and extinction parameters are described in the text.
}
\label{fig:f17}
\vspace{-2em}
\end{figure}
  
\subsubsection{PDR models}
\label{PDRmodels}

Intensities of the CH$^+$ emission lines in Orion~KL have been further modeled using the Meudon PDR code v~1.4.4 \citep{lep06,goi07,leb12}, following the approach of \citet{nagy2013}.  Isobaric models were run for a grid of pressures and radiation fields around the best fit parameters for the Orion Bar PDR through a depth equivalent of $A_V = 10$~mag (comparable to the beam size of the $J=1-0$ observations, or $\approx 5.0 \times 10^3$ AU, at the distance of 420 pc to Orion BN/KL), with the gas illuminated from two sides by a radiation field $\chi=\chi_\mathrm{front}=1000\chi_\mathrm{back}$ in Draine units, and a cosmic-ray ionization rate $\zeta=2\times10^{-16}$~s$^{-1}$.    The initial elemental abundances of He, C, N, O, S, and Fe that we adopt in our models are the values in the Meudon PDR code suited to diffuse clouds (see Table 6 in Le Petit et al. 2006), which are relative to the total number density of H nucleons.  The atomic and molecular hydrogen number densities are given by $n_{\rm{H}} = n$(H) + 2$n$(H$_2$) $= 2 \times 10^5$ cm$^{-3}$.  We also adopt an extinction to color excess ratio $R_V \equiv A_V / E(B-V) = 5.6$, a gas to dust ratio of 100, a grain size distribution index of 3.5 (Mathis, Rumpl, \& Nordsieck 1977; for a range of grain radii $0.003 - 0.3 \mu$m), and $A_V / N_{\rm{H}} = 5.3\times10^{-22}$ mag cm$^{2}$ H$^{-1}$.  

Collisional excitation and de-excitation as well as the the chemical pumping effect of destruction and formation of CH$^+$ on the  level populations are treated in the Meudon code.  The destruction of CH$^+$ occurs in 
collisions with H, H$_2$, and electrons, and by photo-ionization at rates tabulated by Nagy et al. (2013), based on Woodall et al. (2007).  The formation of CH$^+$ includes rotational and vibrational levels of H$_2$ to react with C$^+$, as

\begin{equation}
{\rm{H}}_2(J = 0 - 7)  + {\rm{C}}^+ \xrightarrow[]{k_1} {\rm{CH}}^+ + {\rm{H}} 
\end{equation}
\begin{equation}
{\rm{H}}_2(\nu = 1)  + {\rm{C}}^+ \xrightarrow[]{k_2} {\rm{CH}}^+ + {\rm{H}}
\end{equation}

\noindent (Hierl et al. 1997; Ag{\'u}ndez et al. 2010; Nagy et al. 2013), where rotational levels up to $J = 7$ (with an energy $E_{J=7}$ = 4586.4 K, close to the CH$^+$ activation barrier) are included for reaction, and the rate of formation through rotationally excited H$_2$ is $k_1 = 1.58 \times 10^{-10} {\rm{exp}}(-[4827-E_J/k]/T)$ cm s$^{-1}$ (Gerlich et al. 1987).  In the reaction with vibrationally excited H$_2$, only the $\nu = 1$ level is needed since its energy $E_{\nu = 1}$ = 5987 K is sufficient to overcome the activation barrier, forming CH$^+$ at a rate $k_2 = 1.6 \times 10^{-9}$ cm s$^{-1}$ (Hierl et al. 1997).  

Chemical equilibrium is assumed, though we must acknowledge some risk in the assumption due the dynamic nature of the relatively young BN/KL outflow (e.g., Bally et al. 2011).  The assumption can be justified for most of the mapped region by lack of direct evidence of recent changes in gas dynamics that would influence the CH and CH$^+$ emission, for example as variations in line profiles.  The profiles show very little variation across the region including sightlines to the northern and southern lobes of H$_2$ emission.   We cannot assert this to be true within $\approx$15$''$ of the Hot Core itself, near the origin of the explosion some 500 to 1000 years ago, due to spectral confusion with methanol and SO$_2$ transitions.  Therefore, the assumption of chemical equilibrium could break down for some portion of the gas, but this remains indiscernible in the current observations.   In the Discussion section we will also address formation timescales compared to the age of the outflow, and other chemical factors which may influence the CH$^+$ production there. 

Figure~\ref{fig:f16} shows the results of the PDR models: good agreement with the observed intensities is obtained using a radiation field $\chi = 5\times 10^{3} \chi_{D}$ where $\chi_D$ denotes the field strength in Draine units, using a pressure $p/k = 1\times10^{8}$~cm$^{-3}$~K for the average line intensities, and $p/k = 2\times10^{8}$~cm$^{-3}$~K for those at Peak A.  The applied radiation field is a factor of 2 lower than that required to reproduce the CH$^+$ emission in the Orion Bar, and the pressures comparable to or twice the best fit value for the Bar.  Under these conditions, the total column density of CH$^+$ is $1-2\times10^{14}$~cm$^{-2}$---in excellent agreement with estimates from non-LTE radiative transfer models described in section~\ref{radex}, which is reassuring since both sets of models contain all of the important physical ingredients.

Figure~\ref{fig:f17} shows the calculated abundance profiles of CH$^+$, CH, and the key species involved in their production, as well as the molecular fraction, ionization fraction, and gas temperature as a function of the visual extinction.  The CH$^+$ abundance peaks in the outer layers of the cloud between 0.2 and 0.6 $A_V$, corresponding to low to intermediate values of the molecular fraction ($0.01<f_{\mathrm{H_2}}<0.5$), an ionization fraction that is two-fold lower than in the outermost layers of the cloud, and gas temperatures of $400 - 1000$~K.  By contrast, the CH abundance shows a broad peak from about 0.2 to 5 $A_V$ over a range of ionization fractions, and maximizes at $A_V=2$, $f_{\mathrm{H_2}}=0.8$ and $T<200$~K.  The C$^+$ abundance increases and the vibrationally excited H$_{2}(v=1)$ abundance decreases over most of the region in which the CH$^+$ abundance is at its maximum.  In addition, the H$_{2}(v=1)$ abundance shows no correspondence with the CH abundance.  Instead, the CH abundance corresponds well with the H$_2$ abundance through $A_V=2$ mag.
  
The calculated depth dependence may be understood qualitatively in terms of the chemistry of CH$^+$ and CH.  The C$^+$ ion reacts with H$_{2}(v=1)$---produced efficiently only in the outer layers of the molecular gas---to yield CH$^+$.  The CH$^+$ thus formed can undergo dissociative recombination to yield C and H; the photo-ionization of C formed in this, as well as other processes, likely contributes to the increased C$^+$ abundance deeper into the cloud.  Additionally, in gas of substantial molecular fraction, CH$^+$ can react exothermically with H$_2$ to yield CH$_2^+$, which in turn undergoes dissociative recombination to yield CH.  This may explain the higher calculated CH abundance deeper into the cloud, as well as the observed CH intensity peak closest to IRc~2, the region with the densest molecular gas.  

We also note that while photo-ionization of neutral carbon provides the main source of free electrons in PDRs, additional contributing sources exist to raise the $n_e$/C$^+$ ratio above unity to depths of around $A_V$ = 3.0 mag.   Other elements with low ionization potential (Mg, Si, S, etc.) contribute between 20 and 30\% that of carbon.  Also, H can be ionized by cosmic rays and soft X rays and subsequently yields O$^+$ by charge transfer to O.  The combination of these two mechanisms results in a higher electron abundance in the low-$A_V$ region.  In the higher-$A_V$  region where carbon recombines, sulfur and other metals continue to provide electrons, until around $A_V$ = 3.0 mag in our model. Indeed, the adopted S abundance is 1.9$\times 10^{-5}$, which alone would contribute about 15\% of the electrons relative to those from carbon.  The electron density $n_e$ peaks at around the same depths as C$^+$, which may result in an increase in CH production at such depths as CH$_2^+$ recombines.  

\begin{figure}
\centering
\includegraphics[width=1.0\columnwidth, angle=0]{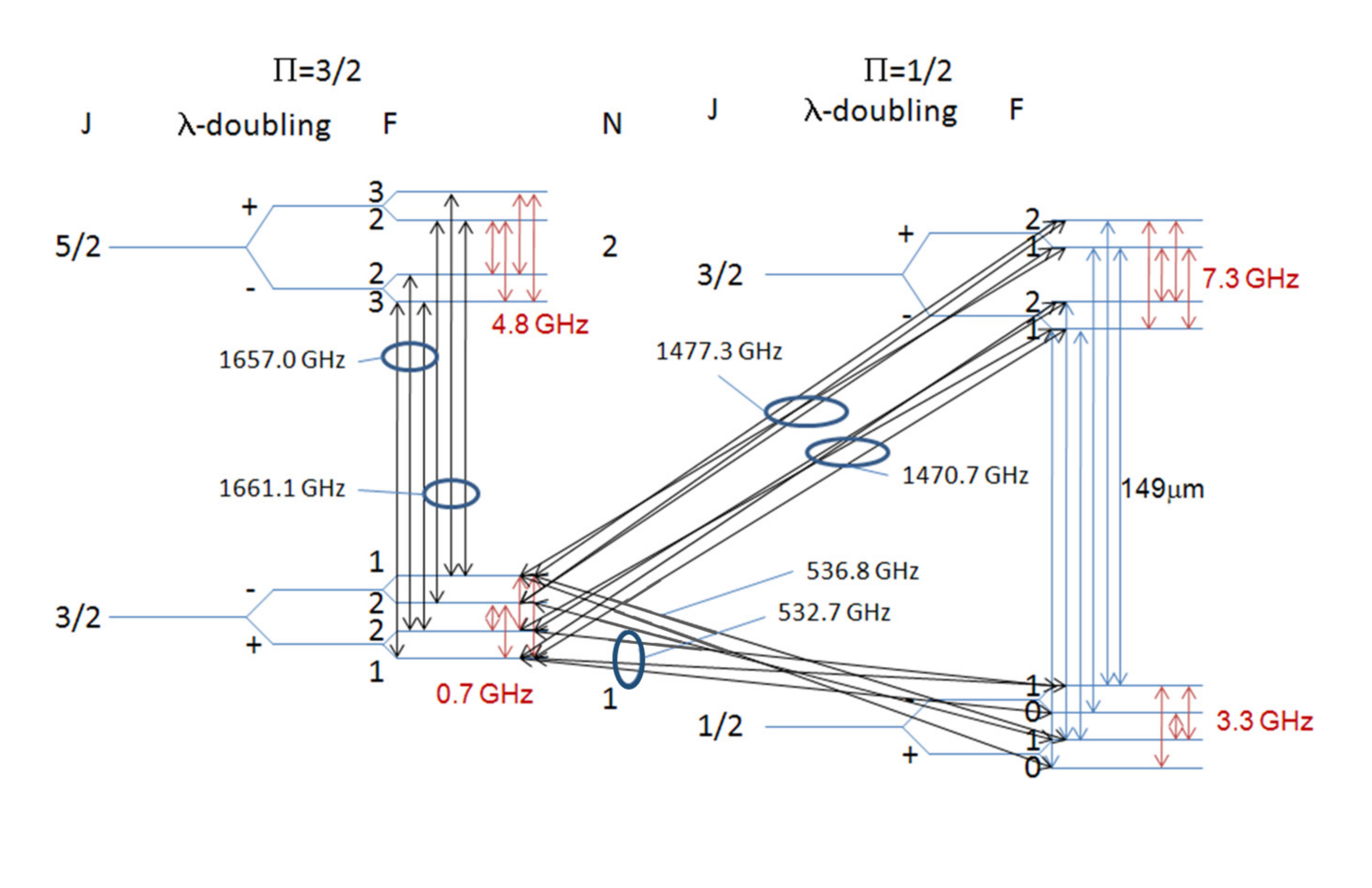}
\caption{Level diagram for the lowest four rotational levels of CH.  The $\Lambda$-doubling and hyperfine splitting are exaggerated for clarity.  The red arrows represent microwave frequency $\lambda$-doubling transitions.  Black arrows are transitions which were within the HIFI frequency range, blue ellipses indicate those observed in this study and the blue lines represent the 2.0 THz transitions, which lie above the Herschel HIFI frequency range.}
\label{fig:f18}
\vspace{-2em}
\end{figure}

\begin{figure}
\centering
\includegraphics[width=\columnwidth, angle=0]{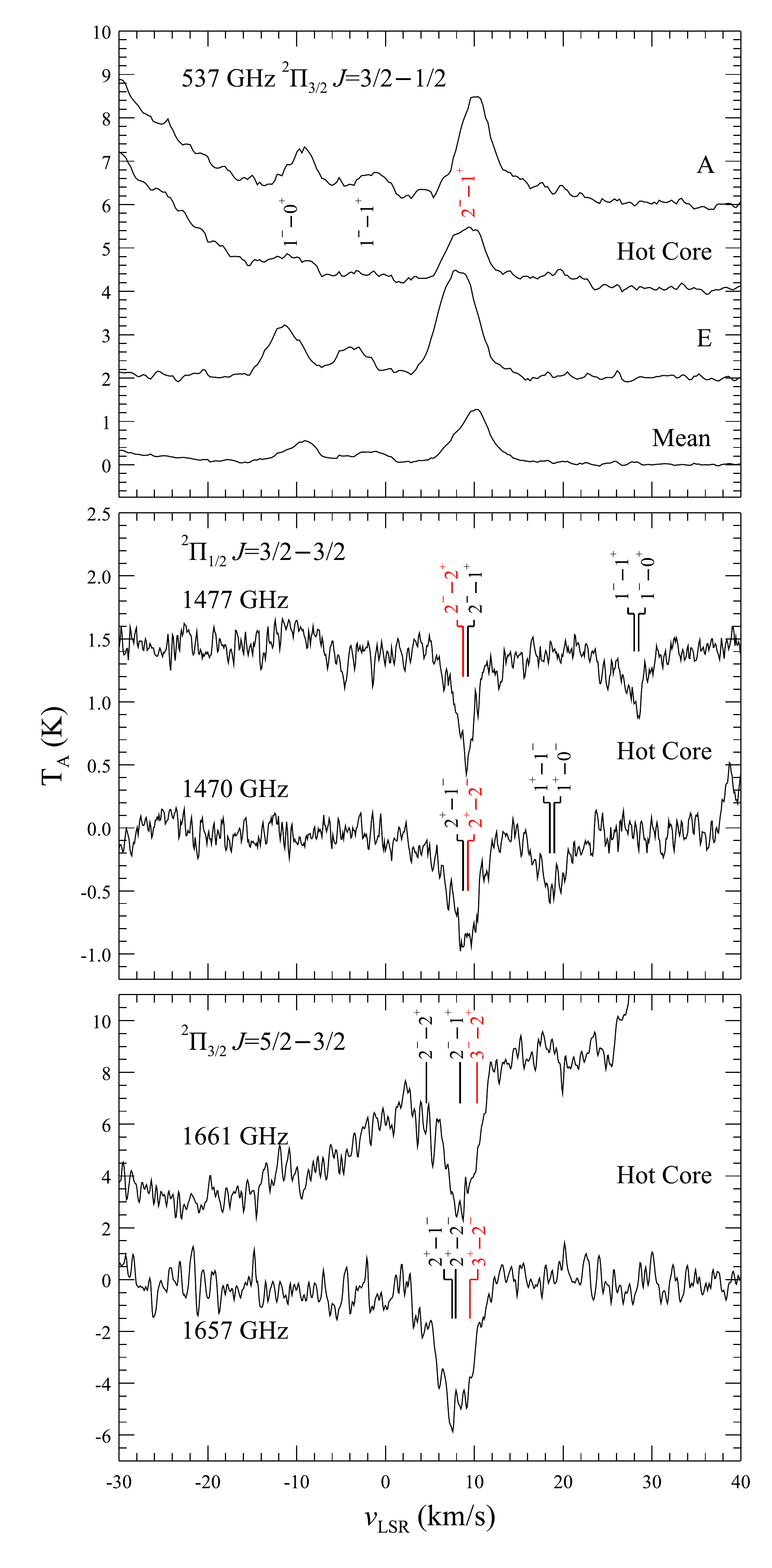}
\caption{Observed CH $\Lambda$-doubled lines of the $^2{\Pi}_{1/2}$ (upper panel) and $^2{\Pi}_{3/2}$ (middle and lower panel) states.  The 1471/1477 and 1661/1657 pairs have been observed in pointed mode towards the BN/KL complex, while the 557 GHz triplet has been spectrally mapped over an area similar to the CH$^+$ $J = 2-1$ line (see Table~\ref{obs_t}) and is plotted as the average.  Continuum levels have been subtracted and spectra offset for clarity.}
\label{fig:f19}
\vspace{-2em}
\end{figure}

Because the Meudon code applies to steady-state physical and chemical conditions, variations in the adopted atomic and molecular abundances have almost no effect on the output ionization abundances {\em{versus}} extinction profiles.  However, a more discernible effect can result from different extinction parameters.   We have adopted a total to selective extinction ratio $R_V \equiv A_V / E(B-V)$ = 5.5 and $A_V/N$(H) = $3.5 \times 10^{-22}$ mag cm$^{2}$ H$^{-1}$ (as have Nagy et al. 2013 for the Orion Bar, and Goicoechea et al. 2015b for larger OMC1 region) that are consistent with larger grains, leading to deeper penetration of the UV field, and producing a larger C$^+$ zone and higher values of $N$(C$^+$).  Similarly, a variation in the adopted UV field strength will have obvious impact on the atomic and molecular number densities at the cloud boundaries.  The field strength and radiation pressures which we adopt give a best fit to the CH$^+$ line intensities and yield a C$^+$ layer to A$_V \approx 3$ mag, while Nagy et al. (2013) use a UV field strength for the Orion Bar that is a factor $\sim$2 higher (but no continuum radiation field), producing a C$^+$ layer to A$_V \approx 4$ mag.  These effects are still minor, and do not affect our conclusions which demonstrate the role of UV chemistry around Orion BN/KL.


\subsection{CH $\Lambda$-doubled lines $^2{\Pi}_{3/2}$, $^2{\Pi}_{1/2}$}

The ground state of CH is $X^2\Pi$ regular (i.e., $\Pi_{1/2}$ lower than $\Pi_{3/2}$), with a spin-orbit coupling constant of 843818.4 MHz.   Figure~\ref{fig:f18} shows the lowest four rotational levels of CH.   The unpaired electron couples with rotation resulting in $\Lambda$ doubling and with the hydrogen spin, leading to clearly resolved hyperfine structure.  We have spectrally mapped emission from one of the ground state cross ladder pairs at 537 GHz (the 533 GHz hyperfine lines were not mapped).  Continuum levels at this frequency peak at $T_{\rm{c}}(557) = 2.1$ K towards SMA~1.  The pairs at 1471/1477 and 1657/1661 GHz were observed only towards the Hot Core, where the continuum is much stronger, $T_{\rm{c}}(1473) = 12.0$ K and $T_{\rm{c}}(1658) \approx 13.0$ K.   These pairs both originate from the lowest level in the excited $^2\Pi_{3/2}$ state, and are lines are in absorption; see Figure~\ref{fig:f19}. The profile measurements are summarized in Table~\ref{chline}.  The  velocities and line widths are very similar to those of the CH$^+$ rotational lines, and likewise kinematically consistent with observations towards the Compact Ridge (e.g. Crockett et al.  2014).  


All of the C$^+$, CH$^+$ $J=1-0$ and $2-1$, and CH $^2\Pi_{3/2} - ^2\Pi_{1/2}$ profiles extracted from the Hot Core, Position A, as well as averages are compared in Figure~\ref{fig:f22} on a normalized intensity scale.  Outside of the Hot Core, the profile shape similarities are very strong, especially in the averages where LSR velocities and line widths would be indistinguishable to within $\sim$2 {\kms}.   Variance in the strength of the red-shifted component can be seen in the C$^+$ peak Position A, but all species emit over the same red-shifted velocity range.

\begin{deluxetable}{lcccc}
\tabletypesize{\small}
\tablecaption{Line measurements of CH $\Pi_{3/2}$ $J = 3/2-1/2$ (537 GHz) at selected positions \label{chline}}
\tablewidth{0pt}
\tablehead{
\colhead{Position}  &  \colhead{$I$}  &  \colhead{$v_{\mathrm{lsr}}$} & \colhead{$\Delta v$} &  $N$(CH) \\
    \colhead{} & \colhead{(K {\kms})}  &  \colhead{{\kms}} & \colhead{{\kms}} & \colhead{10$^{14}$ cm$^{-2}$}
}
\startdata
Average & 10.3 & 9.7 & 5.7 & 0.7  \\
A & 20.0 & 10.4 & 4.7 & 1.5 \\
A$'$ & 21.2 & 10.2 & 5.4 &  1.0 \\ 
C & 14.4 & 8.6 & 4.6 & 1.3  \\
D & 16.8 & 8.5 & 4.6 & 1.4  \\
E & 23.1 & 8.2 & 5.1 & 1.5 \\
Hot Core & 18.4 & 8.2 & 4.8 & 1.7 \\
\enddata
\tablecomments{Column 1 positions are the same as in Table~\ref{cplusline}.  Integrated intensities $I$ are summed over all three hyperfine components, while \vlsr and $\Delta v$ are measured from the strongest $F = 2^+ - 1^-$ .  Measurement uncertainties at all positions are 5\%, except towards the Hot Core and nearby Position A where the  $F = 1^+ - 0^-$ and $1^+ - 1^-$ lines are weakly blended with SO$_2 10_{4,6} - 9_{3,7}$ 549.30 GHz emission observed in the opposite sideband, and contribute 8-10\% uncertainty to the total integrated line intensity.  Column densities $N$(CH) are calculated at $T_{\rm{ex}}$(CH) = 35~K.}

\end{deluxetable}

\begin{figure*}
\centering
\includegraphics[width=\textwidth, angle=0]{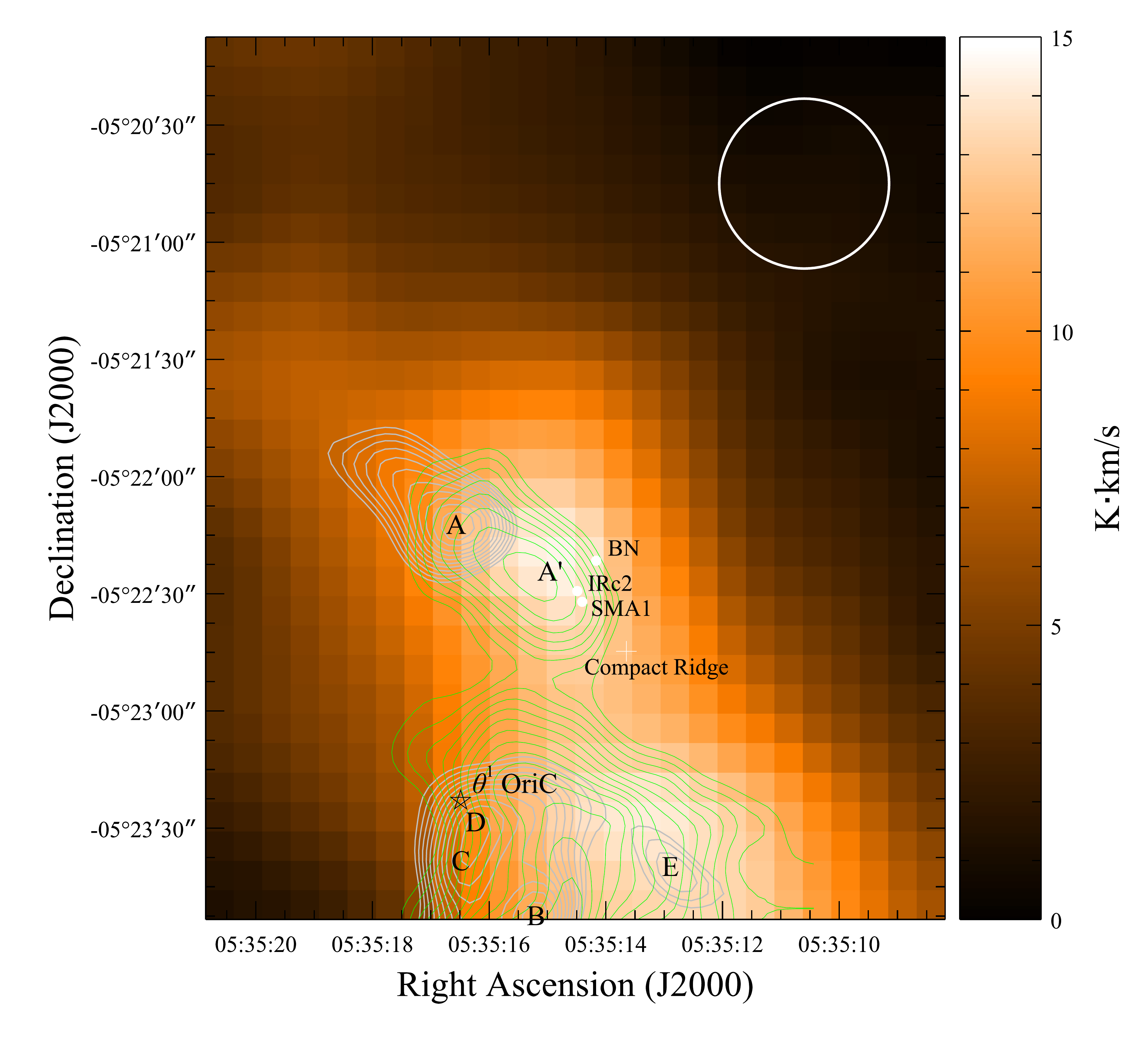}
\caption{CH intensity integrated over the unblended $\Pi_{3/2}$ $J= 3/2-1/2$ hyperfine line F $= 2^- - 1^+$.  Blending of the other two hyperfine transitions occurs with SO$_2$.  Over the plotted range of intensities, the total integrated emission would scale by a factor $\approx$1.6 for a theoretical 5:1:2 ratio for $F = 2-1:1-1:1-0$. Light gray contours indicate the upper 50\% emitted power of C$^+$ 650 to 850 K \kms, green contours are of CH$^+$ $J=1-0$ 20 to 40 K \kms.  Labeled positions are the same as in Fig.~\ref{fig:f3}, with A$'$ added to indicate a local intensity maximum; see also Table~\ref{chline}.} 
\label{fig:f20}
\end{figure*}

A key question concerns the location of the CH column around Orion BN/KL, and its relative abundance compared to H$_2$.  Previous observations show that CH traces H$_2$ well in the diffuse ISM. Steady-state models by Levrier et al. (2012) reproduce the observed trend of CH {\em{vs.}} H$_2$, without the need to invoke turbulence dissipation. This might be understood from the fact that the observations sample CH in regions with a significant fraction of hydrogen in H$_2$, $f({\rm{H}}_2) > 0.1$ up to $\sim$1, while CH$^+$ prefers regions of the ISM with low hydrogen fractions as H$_2$ $f$(H$_2$) $< 0.5$.  In the region around Orion BN/KL, however, we are in a regime of strong FUV irradiation.

The measured intensities of the 1656/1661 lines and 1471/1477 lines, $-$51 K {\kms} and $-$12.6 K {\kms}, respectively, give the total column of the $\Pi_{3/2}$  $J= 3/2$ level of 1.25 $\times \; 10^{14}$ cm$^{-2}$.  Using the same arguments for the rotation (excitation) temperature of CH$^+$ in Sections~\ref{sec:chpabs} and \ref{sec:chpem}, $T_{\rm{rot}} = 35$~K is a reasonable upper limit for CH, and agrees well with a one-component model in LTE carried out with the XCLASS package,\footnote{P. Schilke, XCLASS public home page http://www.astro.uni-koeln.de/projects/schilke/XCLASS/.} and checked with RADEX, yielding a total column from emission of 1.7 $\times \; 10^{14}$ cm$^{-2}$ towards the BN/KL region.   The total column decreases with higher values of $T_{\rm{rot}}$.

\begin{figure}
\centering
\includegraphics[width=\columnwidth, angle=0]{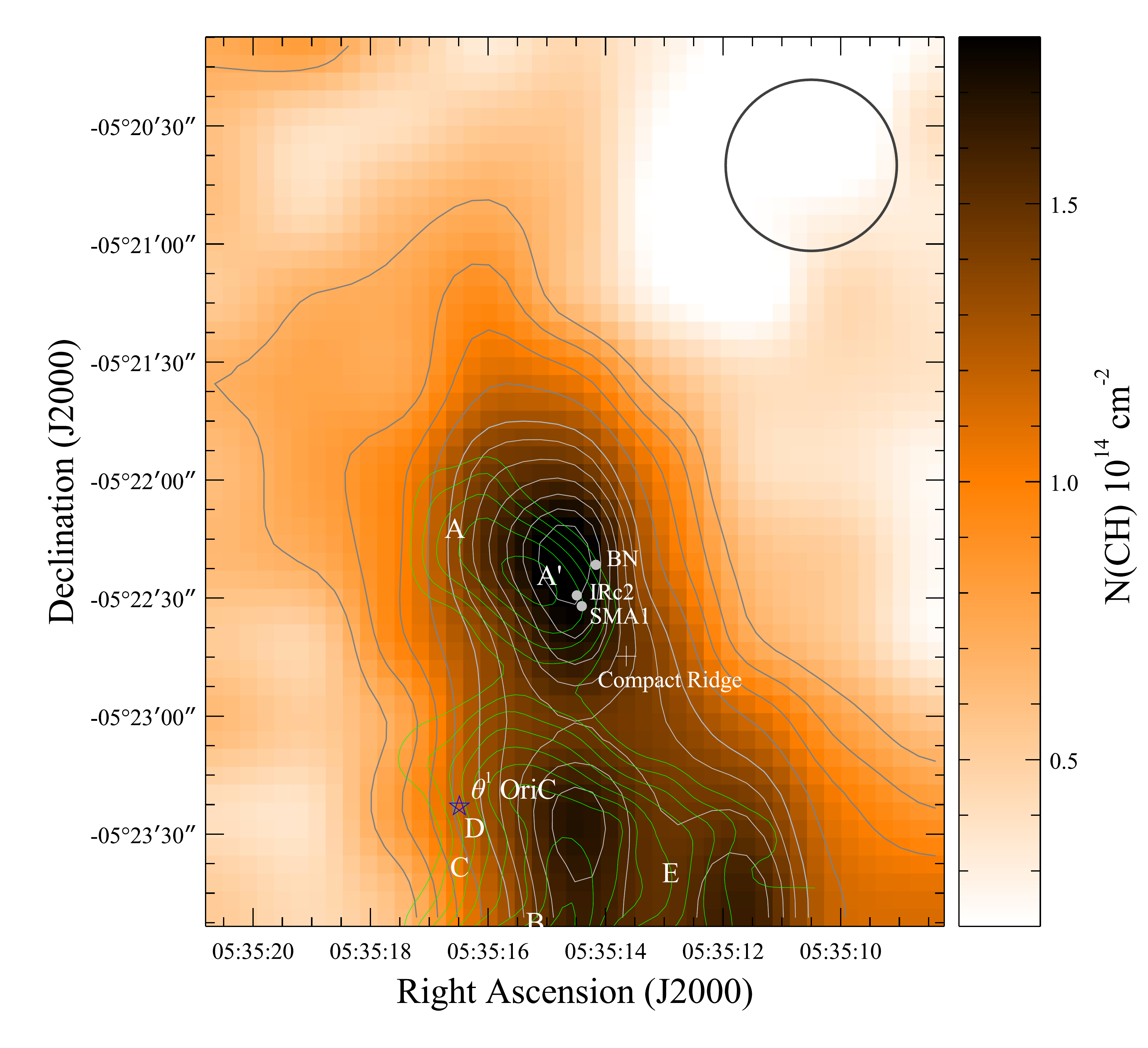}
\caption{CH column densities at $T_{\rm{rot}} = 35$~K in LTE.  The calculation is based on measurements of the unblended $F = 2^- - 1^+$ triplet line, and scaled according to partition with the other two $\Pi_{3/2}$ $J= 3/2-1/2$ hyperfine transitions.   Gray contours follow the $N$(CH) column density distribution on intervals of $0.1\times10^{14}$ cm$^{-2}$ where light gray contours indicate the 50\% densest columns.  Green contours are of the 50\% highest columns of $N$(CH$^+$) on intervals of 0.15 $\times$ 10$^{13}$ cm$^{-2}$.  Labeled positions are the same as in Fig.~\ref{chline}.} 
\label{fig:f21}
\end{figure}

Figure~\ref{fig:f20} 
shows the integrated intensity distribution of CH 537 GHz in the unblended and strongest $F = 2^- - 1^+$ transition.   The distribution is quite similar to that of CH$^+$, although the CH peak intensity is just slightly closer (in projection) to the BN/KL complex.   In relation to \thet1 with its strong ionizing UV field, the highest CH intensities are radially exterior to those of CH$^+$, as we might expect.  

CH column densities for the $^2\Pi_{3/2}$ state have been estimated from measurements of the 537 GHz $F = 2^- - 1^+$ line, and scaled using the partition function to include the $F = 1^- - 1^+$ and $1^- - 0^+$ contributions.  Figure~\ref{fig:f21} shows $N$(CH) calculated in LTE at 35~K.  We find that $N$(CH) towards the BN/KL complex is exactly as predicted using the 1471/1477 and 1656/1661 GHz absorption pairs at 35~K, which provides confirmation of the consistency of the measurements, not necessarily the LTE assumption and adopted $T_{\rm{rot}}$.  Nonetheless, at positions where all three lines are unaffected by emission of other species (namely SO$_2$ 10$_{4,6} - 9_{3,7}$ 549.3 GHz observed towards the Hot Core in the opposite sideband and especially blending in the wings of the $F = 1^- - 0^+$ line; see Fig.~\ref{fig:f19}), 
the observed $1-0$:$1-1$:$2-1$ ratio is in excellent agreement with the theoretical partitioning 0.25:0.125:0.625.  The CH column densities are generally lower than $N$(CH$^+$) by factors of $\approx$1.2 to 3 at specific locations, and $\approx$5 weaker on average over the area mapped in common.   The usual relationship [CH]/[H$_2$] = 3.5 $\times$ 10$^{-8}$ (Sheffer et al.  2008) gives a total column of H$_2$ of $\sim$5 - 12 $\times$ 10$^{22}$ cm$^{-2}$,  which is consistent with that of the Compact Ridge (Plume et al. 2012)---higher than would be observed in a boundary layer, and lower than would be expected from the Hot Core.  The emission in the 537 GHz lines suggest that the critical density of $\sim10^6$ cm$^3$ has been reached, but not by enough to cause emission in the higher lying transitions at sub-thermal excitation temperatures of $T_{\rm{ex}}$(CH) $\lesssim$ 35~K. 

\subsection{Inferred C$^+$/CH$^+$/CH cloud interfaces}\label{clouds}

Our HIFI spectral maps reveal two separate features visible in CH$^+$ and C$^+$: one which is absorbing thermal radiation toward the BN/KL complex; and a structured cloud distribution of emission over the entire region enveloping BN/KL.    Only the first excited level line profile of CH$^+$ reveals the absorption component.  The ground state profile shows no indication of absorption, and with the constraint $I_{\rm{abs}}({\rm{J}}=1-0)/I_{\rm{abs}}({\rm{J}}=2-1) < 1$, the excitation temperature of the absorbing gas is in the range of 10 to 25 K, similar to the absorbing component of C$^+$, but at lower optical depths.   The excitation temperatures of the emitting CH and CH$^+$ gas are similarly sub-thermal, while C$^+$ excitation temperatures are much higher, $\approx$175 K on average.   Our derived C$^+$ excitation temperatures for the inner $3' \times 3'$ around BN/KL are higher by almost a factor 2 compared to  that  towards the BN/KL as well as the overall OMC~1 average, as estimated by Goicoechea et al. (2015b), resulting in nearly the same factor increase in column densities.  Nonetheless, the description by Goicoechea et al. based on similar PDR model predictions are confirmed here, namely that the C$^+$ emission across OMC~1 is  tracing only the narrow cloud surface layers to $A_V \; <$ 4 mag and not the full line of sight.  Quantitative differences in the maximum penetration depth of C$^+$ in our modeling are small (peaking in abundance at $A_V \; \lesssim$ 3 mag), arising partly from a more detailed treatment of the dust extinction properties by Goicoechea et al.   Our PDR models predict abundances of CH to peak at comparable penetration depths, while CH$^+$ peaks at much lower depths, 0.2 mag $\lesssim \; A_V \; \lesssim$ 0.6 mag.  This is consistent with the predicted increase of electron densities at depths similar to the C$^+$ peak abundance, where CH$^+$ will easily react to form neutral CH.  

\begin{figure}
\centering
\includegraphics[width=\columnwidth, angle=0]{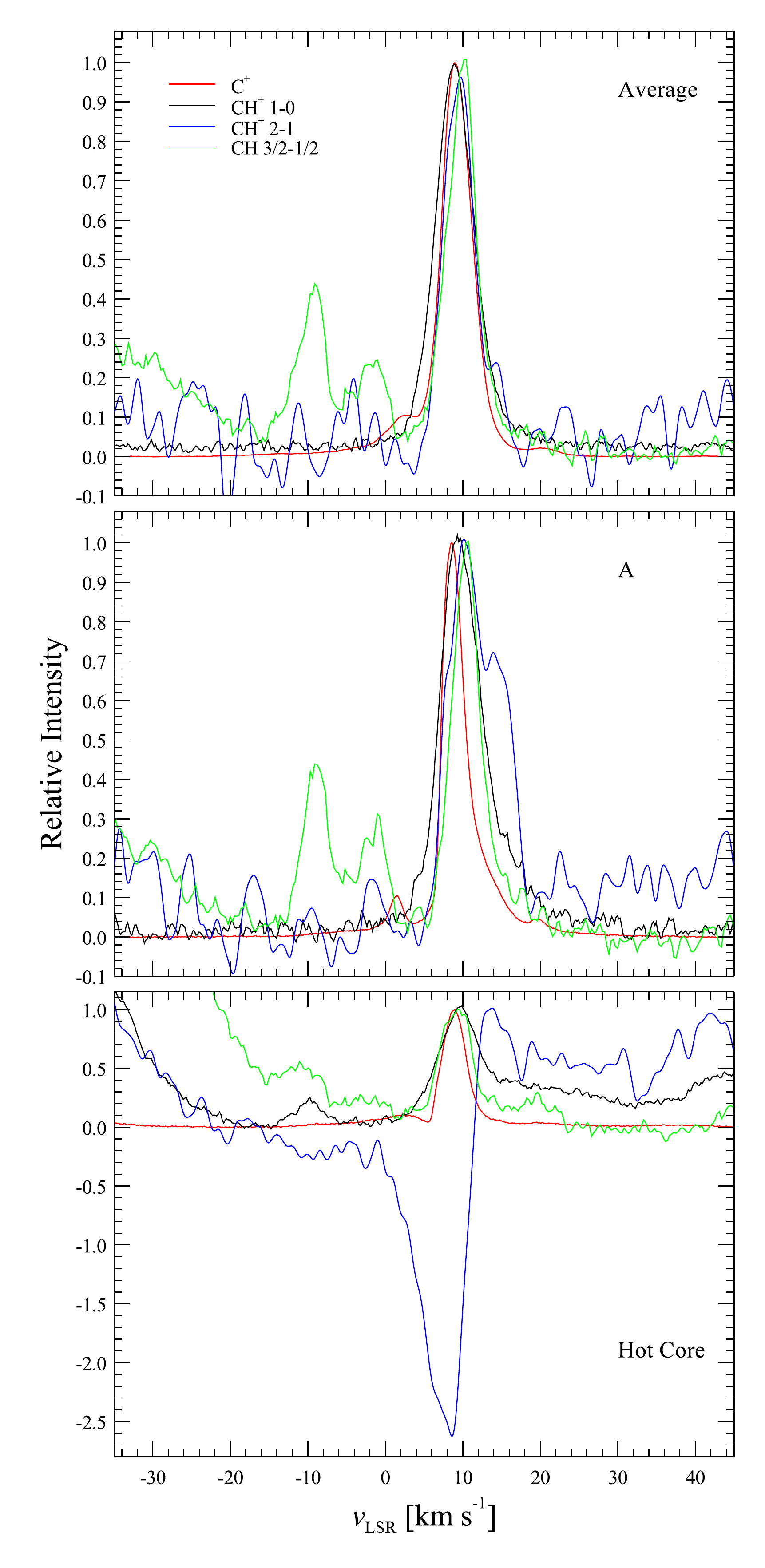}
\caption{Comparison of the C$^+$, CH$^+$ $J=1-0$ and $2-1$, and CH $^2\Pi_{3/2} - ^2\Pi_{1/2}$ profiles extracted from the Hot Core, Position A, and averages as labeled, on a normalized intensity scale.  The LSR velocity for CH is based on the $F = 1^+$ - 2$^-$ line.}
\label{fig:f22}
\end{figure}

Since the brightness distributions of C$^+$, CH$^+$, and CH are very similar, and indeed all profiles outside of the Hot Core are very similar (see Fig.~\ref{fig:f22}), it is reasonable to think of these molecular components as situated in a halo around the Trapezium, with the kinematically more quiescent CH gas exterior to the CH$^+$ and C$^+$.  From the PDR model we expect an anti-correlation in an edge-on geometry at modest to low values of $A_V$. The good spatial correspondence of each species in emission may have three explanations: a) face-on-geometry of a veil illuminated by {\thet1}; b) very clumpy picture with surfaces everywhere; or c) efficient dynamical mixing. Case (a) would be most consistent with conclusions for the larger OMC~1 cloud by Goicoechea et al. (2015b).  The picture is more or less consistent with the ionized/PDR/molecular gas interfaces illustrated by Goicoechea et al., however the high reactivity of CH$^+$ allows CH to coincide with C$^+$ at higher $A_V$ where electron densities are also higher.  The picture illustrated by Goicoechea et al. is suited at a schematic level to the larger OMC~1 region, while the smaller region we have mapped in C$^+$, CH$^+$, and CH more deeply around BN/KL is inhomogeneous, with numerous projection effects from different kinematic components of C$^+$ and evidence for a mix of gas temperatures towards the Hot Core.


\subsection{CH$_3$OH K = 5$\leftrightarrow$4 (E)}
\label{methenol}  

At the HIFI LO frequency setting used to map CH$^+$ $J=1-0$ we have also detected a dozen of the strongest $\Delta$J=0, K = 5 $\leftrightarrow$ 4, v$_t$=0 E-symmetry Q-branch methanol transitions centered at 834.8 GHz, with upper state energies ranging between 171~K to 452~K  This is the only band in Orion which has been mapped with HIFI, and is ideal for studying the kinematics and physical conditions in the dense BN/KL core region, particularly since the observed transitions ($J_{u,l}$ 5 through 15) are isolated from CO, H$_2$O, OH, or other broad features.   Only the $J_{u,l}$ 14 and 16 transitions in our detected set of lines are blended with SO$_2$ and HDO (respectively) from the opposite sideband.   

The CH$_3$OH emission spectra are shown in Figure~\ref{fig:f23}.   We have taken $\sim$1/3 beam-sized spectral extractions at positions centered on SMA~1 near the 835~GHz thermal continuum peak, and at the Compact Ridge.  These locations each have their own distinctive kinematic features within a single HIFI beam (e.g. Wang et al.  2011; see also Fig.~\ref{fig:f2}, and Persson et al.  2007 for a summary of the velocity features within 1$'$ of IRc~2).  Two velocity components are easily detected in each spectrum shown in the top panel, and in profile decompositions at the two positions shown in the middle and lower panels of Figure~\ref{fig:f23}.    The fits are derived using Gaussian profiles which are initially constrained only by the velocity at the observed emission peak;  the profile center frequency, width, and intensity are treated as free parameters using a Levenberg-Marquardt fitting engine to minimize $\chi^2$ between the observed and modeled spectrum.  A best-fit model derived from the average spectrum is then supplied as an initial guess for subsequent fitting at each map point.     The fit results yield two ranges of LSR velocities and widths, \vlsr(broad) = 6.5-8.5 \kms, $\Delta v$(broad) = 7.5-14.5 \kms, and \vlsr(narrow) = 7.3-8.2 \kms, $\Delta v$(narrow) = 1.5-2.7 \kms.

The velocities and widths of the two distinctive kinematic components of the methanol line map are generally consistent with their first detection in the earliest methanol observations by Menten et al.  (1986) and Wilson et al.  (1989), of a single transition of the 23 GHz 10$-$9 A$^-$ band.  The present results are also consistent with HIFI observations of the $\Delta$J=0, $K =-4 \rightarrow -3$ and $K = 7 \rightarrow 6$ v$_t$=0 Q branch transitions at 524 GHz and 1061 GHz by Wang et al.  (2011), at a fixed position centered on the Hot Core.   These distinct components were demonstrated by Wang et al. to probe gas with different temperatures.

Observations of several complex organic molecules with the Submillimeter Array (SMA) and the IRAM 30-m telescope over 220 to 230 GHz by Feng et al. (2015) also show a similar LSR velocity range across numerous detected methanol lines.  At the higher angular resolution of the interferometric data, $\simeq$3$''$.0, emission peaks are resolved around five narrow LSR velocity ranges centered at 3, 5, 8, 11, and $>$ 12 {\kms}.   Feng et al. quote that these lines are all broad, $\sim$ 5 $-$ 7 {\kms}, apparently lacking the additional narrow $<$ 3 {\kms} component observed by Menten et al. (1986), Wilson et al. (1989), Wang et al. (2011), and in our data.  

\begin{figure}
  \begin{center}
  \includegraphics[width=\columnwidth]{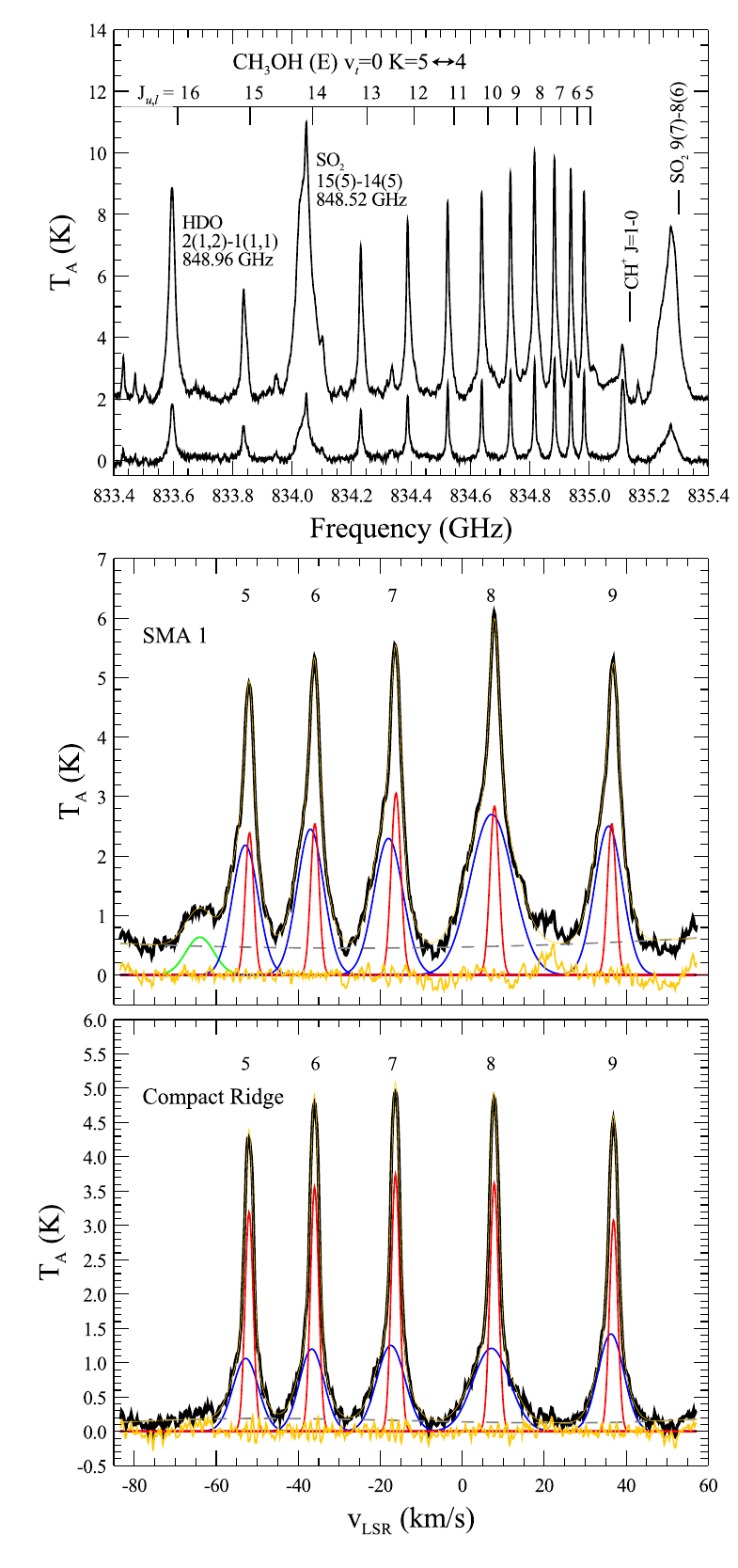}
  \end{center}
\vspace*{-6.0mm}
  \caption{Continuum-subtracted spectra of CH$^+$ J = 1-0 and methanol observed towards SMA~1 and the Compact Ridge.   In the top panel on a frequency scale in GHz, the SMA~1 spectrum is offset by $+$2.0 K for clarity.  The SO$_2$ and HDO transitions labeled with frequencies outside the plotted range are from the opposite sideband.  The middle and lower panels are on a \vlsr \ scale in \kms, referenced to the strongest $J_{u,l}$=8 transition.   Gaussian profile fits are shown for the first five E-branch methanol transitions, indicated by narrow (red) and broad (blue) components; the total model fits and residual fluxes are shown in orange.  The component in green in the fit to the SMA~1 spectrum is included to improve the overall fit.   
\label{fig:f23} }
\end{figure}

Concerning spatial distribution, Wang et al.  (2011) associated the narrow component with signatures of the Compact Ridge, which is usually approximated as a spherical clump of molecular gas 5$''$-10$''$ in diameter  (an idea put forth by Blake et al.  1987), and externally heated to produce the relatively low velocity widths.  Menten et al.  (1986, 1988) had previously identified the the narrow component with a larger region $\approx$25$''$ in size, centered 5$''$-10$''$ south of the Hot Core where the broad emission component peaks. The distributions resulting from our profile decomposition analysis are shown in Figure~\ref{fig:f24}, where it is seen that the narrow component peaks $\sim$8$''$ to the south of the broad emission center, and 12$''$ east and north of the Compact Ridge.  The broad and narrow emission regions are roughly the same size, 30$''$ to 35$''$ ($\approx$ 0.08 pc) after correcting for the HIFI beam, somewhat more extended than observed in the radio frequency transitions.  This difference might be attributed to higher sensitivities in our combined mapping observations, or a difference in gas temperatures being probed by the different sets of transitions. 

\begin{figure*}
  \begin{center}
  \includegraphics[width=\textwidth]{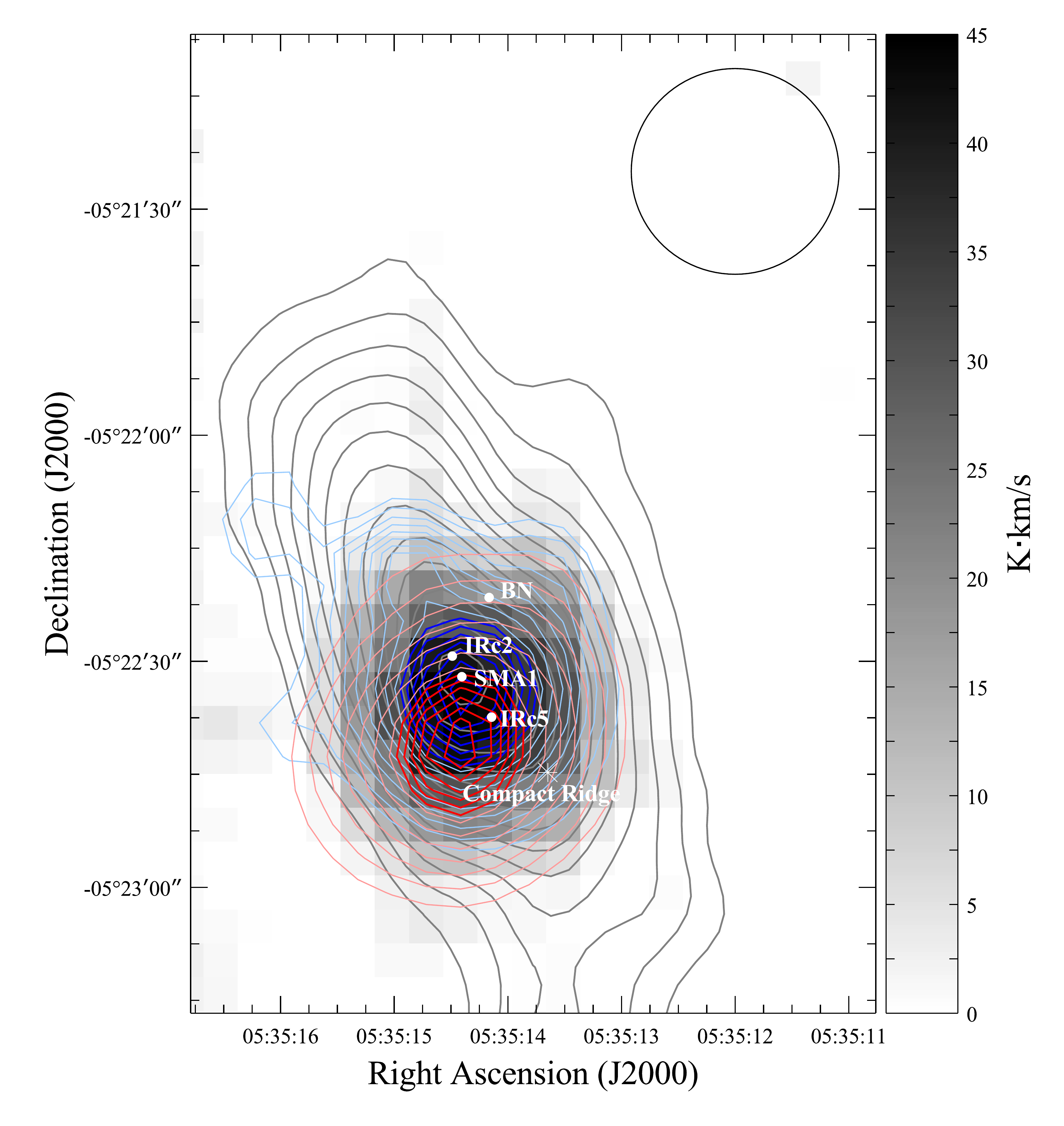}
  \end{center}
  \caption{Average intensity in reverse gray scale of the CH$_3$OH K = 5 $\leftrightarrow$ 4, v$_t$=0 E branch transitions between $J_{u,l}$ 5 and 15 (excluding 14 due to blending with SO$_2$), integrated over \vlsr \ between $-$7 {\kms} to $+$25 {\kms}  for each line.  Overlaid 835 GHz continuum contours in gray are the same as in Fig.~\ref{fig:f3}.  Blue and red contours are integrated intensities of the fitted broad and narrow components, respectively, at each position as shown for the explosive outflow and Compact Ridge spectra in Fig.~\ref{fig:f23}.  Broad (narrow) component contours in light blue (red) are on a linear scale between 3 to 11 {\kms} (0.5 to 1.9 {\kms}) on 1 {\kms} (0.5 {\kms}) intervals, and in dark blue between 11.5 to 15.5 {\kms} (2 to 2.5 {\kms}) on 1 {\kms} (0.5 {\kms}) intervals.   The angular separation between peak broad and narrow component emission positions is 8$''$.5.  The 27$''$.5 HPBW of HIFI at 835 GHz is indicated.
\label{fig:f24} }
\end{figure*}

\begin{figure}
  \begin{center}
  \includegraphics[width=\columnwidth]{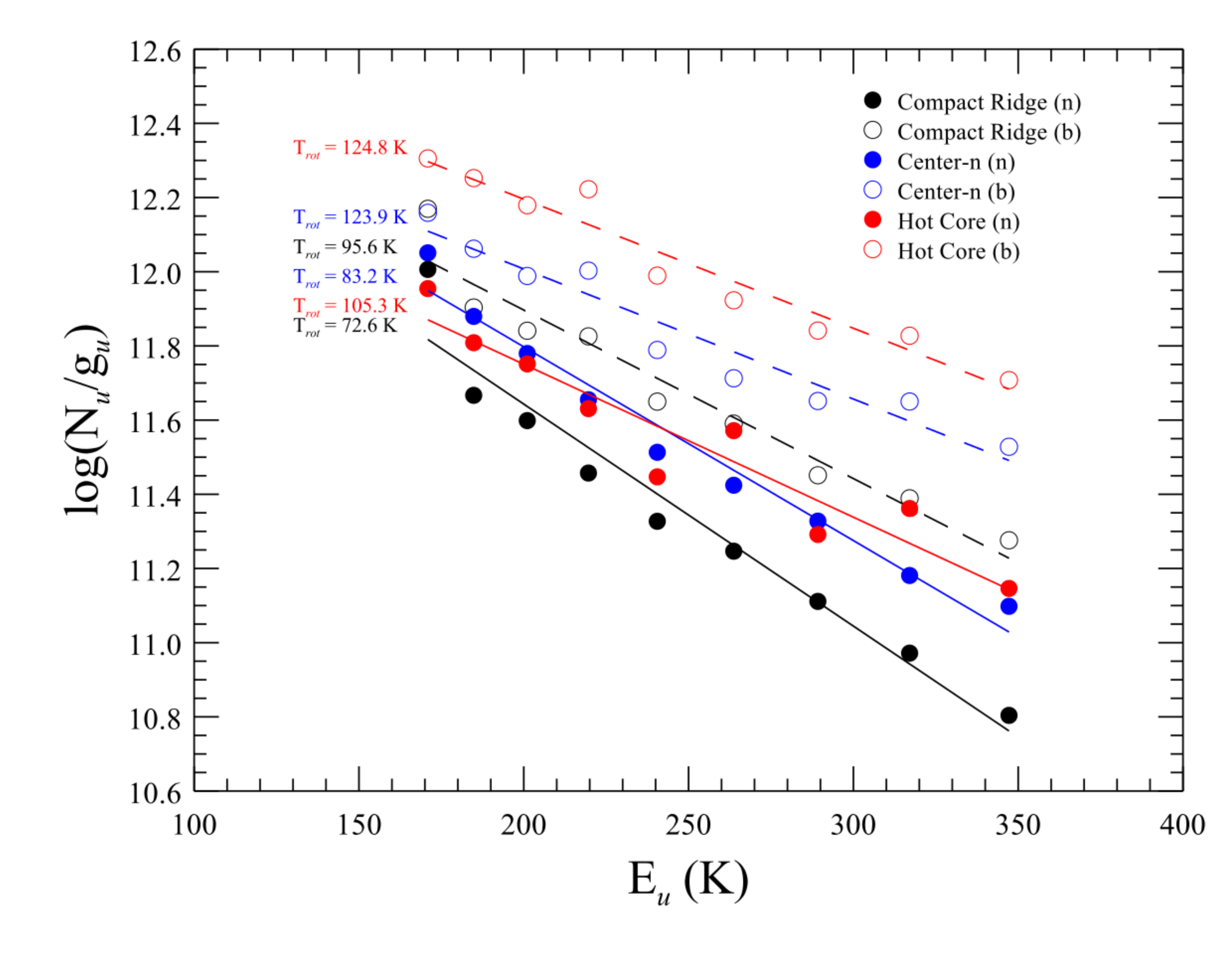}
  \end{center}
  \caption{Population diagram of the CH$_3$OH 835 GHz band for 9 of 12 detected transitions, $J_{u,l}$ 5 through 13, at labeled positions.  LTE is assumed for the linearly fitted excitation temperatures and column densities (see Table~\ref{methanoltable}) of the narrow line (n) and broad line (b) emission components.
\label{fig:f25} }
\end{figure}

In the SMA observations carried out by Feng et al. (2015), which indicate only a broad emission component, dual velocity-dependent emission peaks at the Hot Core and near the Compact Ridge (their ``mm3a'' and ``mm3b'' positions) are resolved.   Feng et al. also point out emission to the south of the Compact Ridge, and newly-discovered tails of emission extending towards the SE that they suggest are a part of the high velocity outflow from the Hot Core, in part because one of their observed lines 8$_{-1,8} \rightarrow 7_{0,7}$ is known to be a potential Class I maser that traces shocks.  Neither feature is detected in our map of K = 5 $\leftrightarrow$ 4 E-symmetry lines.   

An excitation diagram analysis of the maps resulting from our Gaussian profile decompositions yield rotational temperatures and column densities of the isolated E-type transitions of the 834 GHz methanol band to probe conditions in the broad and narrow line components.   Following a similar LTE starting point analysis as carried out by Wang et al.  (2011),  the population diagram for the isolated 835 GHz lines observed towards the broad and narrow centers of emission and towards the Compact Ridge is shown in Figure~\ref{fig:f25}; the total column density and excitation temperature at each position are given in Table~\ref{methanoltable}.    

Our derived excitation (rotation) temperatures are lower overall compared to those estimated by Wang et al. (2011) for the Hot Core, and by Feng et al. (2015) for the Hot Core and their position mm3a ($\sim$5$''$ to the east of the Compact Ridge).  $T_{\rm{rot}}$ increases if some of the methanol lines involved in the fitting are optically thick, as pointed out by Feng et al. from their spectral synthesis results.  Uncertainties on our results may also be higher due to the smaller set of available lines, at least compared to Wang et al. who measured 33 total E-symmetry lines in the 524 GHz and 1061 GHz bands in their HIFI observations of the Hot Core.

It can be noted that the lowest lying methanol transition is systemically offset to higher column densities (higher line intensities) at all positions where emission is detected --- an effect which can be seen in data presented by Wang et al. and Feng et al. as well, but influencing our slopes more strongly due to our smaller set of lines involved in the fitting.   This level population behavior could be explained by optical pumping by thermal emission from the Hot Core of either the torsional state or the rotational R-branch of much colder gas.   These Q-branch transitions mostly trace warmer gas but for the lowest $J=K$ line there is also an optically-pumped cold component, which is a non-LTE condition leading to deviation from linearity of the population diagrams.  Non-linearity is difficult to discern in Figure~\ref{fig:f25} (first and second order fits have equal RMS of the residuals).  If we exclude the $J_{5,5}$ line from fitting, $T_{rot}$ would increase by 10 K to 15 K, i.e., closer but still somewhat low compared to Wang et al. 

With a larger set of observed transitions, curvature in the population diagram for narrow velocity component has been discerned by Wang et al. (2011).  They calculated a model spectrum that gives a reasonably good match to the observed spectrum by approximating the emitting material to be confined to a clump of gas 7$''$.5 in diameter and externally heated, using a temperature profile with constant 30 K for $r$ < 5$''$.0, $T\sim$r$^{3.8}$ outside $r$ > 5$''$.0, and a constant methanol abundance of 3 $\times$ 10$^{−6}$ and 3 $\times$ 10$^{−7}$, corresponding to column densities of of 9.5 $\times$ 10$^{18}$ cm$^{-2}$ and 9.5 $\times$ 10$^{17}$ cm$^{-2}$ in the inner and outer part of the clump, respectively. This gives a temperature of 120 K on the clump surface.   The column densities we estimate for the Hot Core are comparable to those of Wang et al. in the outer part of their modeled clump but a factor $\sim$ 20 low at the inner radius.  Feng et al. (2015)  show that the emission is quite a bit more extended in some transitions in the 220 $-$ 230 GHz range, leading (geometrically) to lower column densities.  

\begin{deluxetable}{l c c c c}

  \tablewidth{0pc}
  \tablecaption{CH$_3$OH 835 GHz excitation temperature and column density \label{methanoltable}}

\tablehead{\colhead{Position} & \multicolumn{2}{c}{$T_{ex}$} &  \multicolumn{2}{c}{N} \\
 & \multicolumn{2}{c}{ (K) } & \multicolumn{2}{c}{ (10$^{17}$ cm$^{-2}$) }  \\
 &  b & n & b & n 
} 
 \startdata

SMA~1 / Hot Core & 125 & 105 & 7.8 & 3.8 \\
Center-n &  124 & 83 & 5.1 & 6.9 \\
Compact Ridge & 96 & 73 &  6.4 &  \ 6.9 
 \enddata
\tablecomments{Positions correspond in Fig.~\ref{fig:f24} to labeled sources, where ``Center-n'' refers to the center of narrow-line emission.  Columns ``b'' and ``n'' are for the broad and narrow line components.}
\end{deluxetable}

In summary, the 835 GHz methanol band in our mapping data indicate spatially extended and overlapping broad and narrow line components in each transition that trace different gas kinematics and temperatures, coinciding with the Hot Core (near SMA~1) and a center of emission 9$''$ to the south, respectively.  The southern emission peak is offset to the east of the Compact Ridge to a position which is consistent with the mm3a/b peak positions identified in SMA  observations of methanol and other complex organic molecules by Feng et al. (2015),  resolving the Hot Core and Compact Ridge centers of emission into dual {\vlsr}-dependent peaks.  The kinematics are clearly distinct from those of the CH$^+$ and CH emission, which trace more quiescent gas and probe different physical and chemical conditions outside the BN/KL outflow.  Our calculated methanol column densities and rotation temperatures are somewhat low compared to those derived by Wang et al. (2015), but (a) optical depth corrections may be needed for certain sub-mm lines that could invalidate the assumption of a single temperature for the emission, and (b) their results depend on an approximation of the emitting region as a $\sim$7$''$.5-diameter clump of gas, which appears too confined compared to our HIFI mapping and recent SMA observations by Feng et al.

\section{Discussion}\label{discussion}

The principal focus of this paper is on the endothermic formation of the CH$^+$ ion and its excitation in Orion~KL.  In the diffuse ISM, CH$^+$ has column densities in the range of 10$^{12}$ to a few 10$^{13}$ cm$^{-2}$ (e.g. Black, Dalgarno \& Oppenheimer 1975).  The columns around Orion~KL traced by line emission are about an order of magnitude higher, and $I_{\rm{em}}({\rm{J}}=1-0)/I_{\rm{em}}({\rm{J}}=2-1)$ where $I_{\rm{em}}$ is the integrated line intensity, is everywhere $> 1$, pointing to lines which are not optically thick and not formed in a region where turbulent dissipation is the main gas heating mechanism driving the CH$^+$ formation.  Characterization of the emission in the Orion Bar by Nagy et al.  (2013) and Orion BN/KL (this study, Sec.~\ref{PDRmodels} and \ref{radex}) point entirely to PDR characteristics.  In fact, the kinematic properties and $I_{\rm{em}}(n_{\rm{upper}})/I_{\rm{em}}(n_{\rm{upper}}+1)$ line ratios are very similar between the two regions, and we can conclude that the same material extending around these OMC~1 features is responsible for the CH$^+$  emission.  We can similarly conclude based our models that electron collisions affect the excitation of CH$^+$ and that reactive collisions also need to be taken into account in the calculations. 

The models show that CH$^+$ is a tracer of the warm surface region ($A_{\rm{V}} < 0.6$ mag) of the PDR with kinetic temperatures between 500 and 1000 K, and excitation temperatures of 23~K but possibly as low as 10~K in the absorbing cloud of gas in front of the Hot Core.  At velocity widths of $\sim$5 \kms, the CH$^+$ emission is broader than the average line width of other molecules and C$^+$ observed in the Orion Bar (Nagy et al.  2013).  It is clear from the models that the extended CH$^+$ gas is dense and under comparably high pressures (in the range of $1-2 \times 10^8$ cm$^{-3}$ K) in both regions.

The kinematics of the C$^+$ emission in the $\sim$11 arcmin$^2$ area around Orion~KL are considerably more complex than in the Orion Bar.  In our mapped region, the two strongest C$^+$ components probe a range of $A_V$ up to around 3 mag.  The two components have velocity widths of 2 - 3 {\kms} and 3.7 - 6.6 \kms,  are highly variable in their relative intensities (see Fig.~\ref{fig:f4}) indicating large optical depth variations of these regions, somewhat wider than calculated for the Orion Bar by Ossenkopf et al.  (2013).  Measurements of $^{13}$C$^+$ at selected peaks of C$^+$ intensity around Orion BN/KL indicate a range on $\tau({^{12}C^+})$ of 1.4 to 2.6 outside of the Hot Core, and from 4 to 6 towards the Hot Core. 

\begin{figure*}
  \begin{center}
  \includegraphics[width=8.5cm]{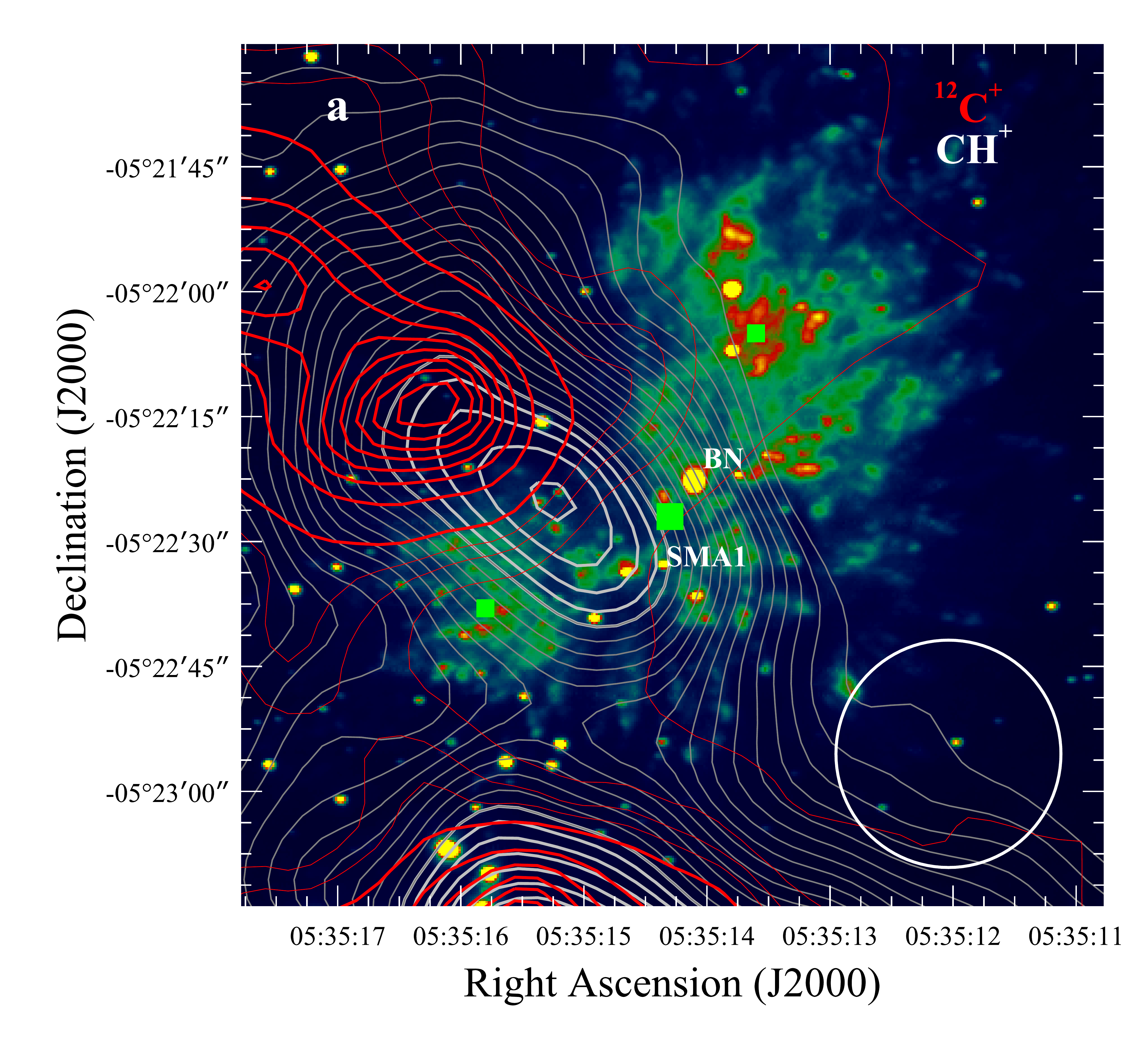}  \includegraphics[width=8.5cm]{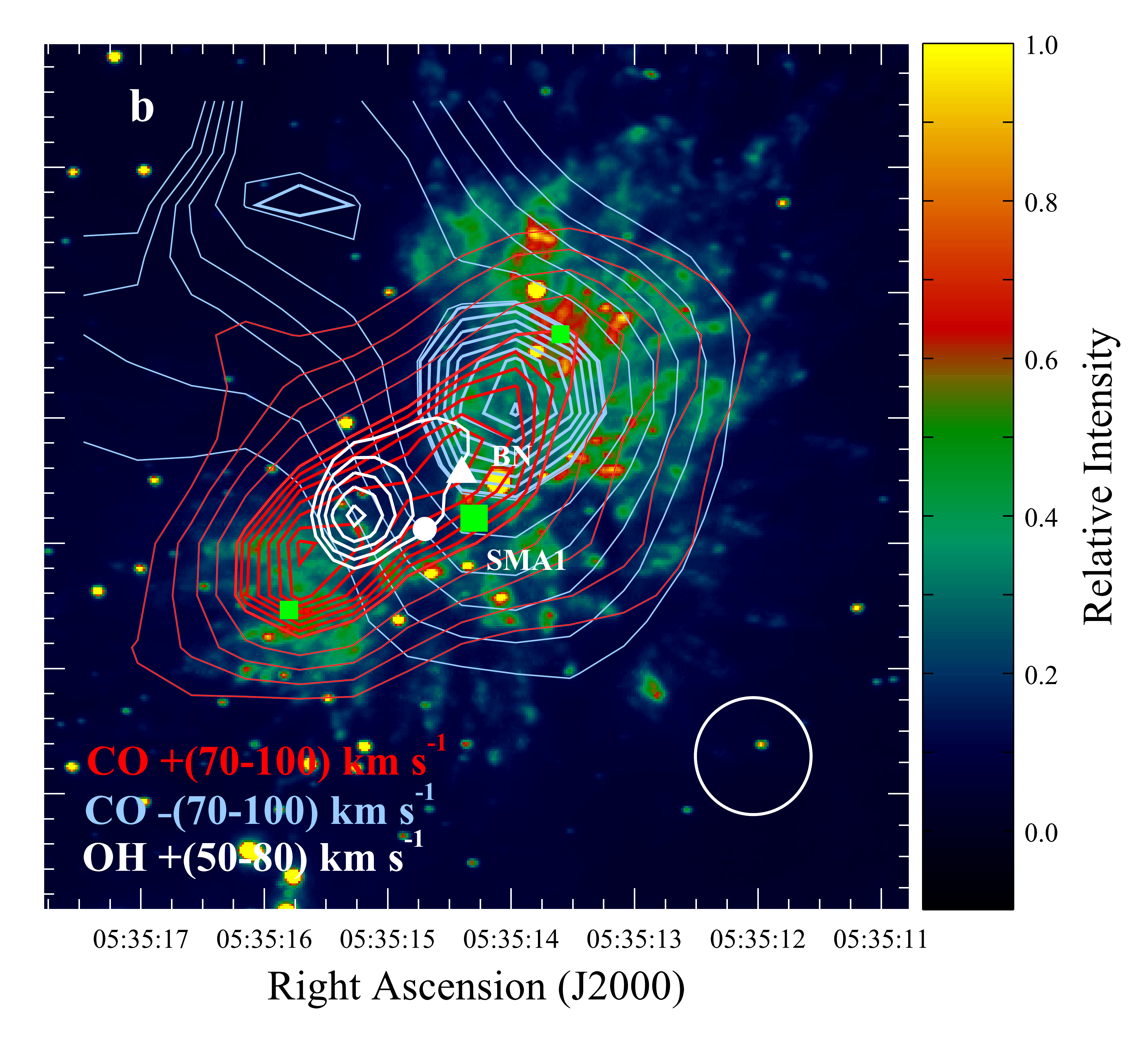}
  \end{center}
  \caption{The energetic BN/KL outflow imaged in the H$_2$ S(1) 2.12 $\mu$m filter with the Near-Infrared Camera and Fabry-Perot
Spectrometer on the ARC 3.5 m telescope at the Apache Point Observatory in 2004 December, published by Cunningham (2006) and Bally et al. (2011), compared with: ({\it{a}}) contours of CH$^+$ (white and gray) and C$^+$ (red); ({\it{b}}) CO $J=10-9$ (blue and red) and OH 1837 GHz triplet transitions between $^2\Pi_{1/2} \; J = 1/2$ states (white).  Thick contours represent the upper 60\% of radiated power, thin lines correspond to the 30\% to 60\% power range, and all are on intensity scales normalized to peak integrated emission on intervals of 5\% relative flux.  The C$^+$ contours are for emission at the system velocity of 9.7 {\kms}.   The red- and blue-shifted emission from CO integrated between $\pm$(70 $-$ 100) {\kms} (offset by $-$9.7 {\kms} for the mean system {\vlsr}) are indicated by corresponding colored contours.  The white filled triangle indicates the peak location of total integrated CO $J=10-9$ emission.   White contours indicate OH red-shifted emission, integrated over $+$(50 $-$ 80) {\kms}.  Blue-shifted OH could not be measured due to confusion with CO $J=16-15$ from the image sideband of the HIFI observations. The filled white circle indicates the peak of OH emission integrated over $-$10 to $+$80 {\kms}.    In both panels, the filled large and small green squares indicate the explosion center and H$_2$ emission peaks, respectively (e.g., Peng et al. 2012b). The HIFI beam sizes at 835 GHz and 1840 GHz are shown in panels ({\it{a}}) and ({\it{b}}), respectively.
}
\label{fig:f26} 
\end{figure*}

R{\"o}llig \& Ossenkopf (2013) have shown that isotopic fractionation can be significant for C$^+$, C, and CH$^+$ in PDRs with high densities at moderate radiation fields. It requires a significant amount of ionized carbon at temperatures on the order of the barrier of the isotope-selective reaction of 35~K. Most gas in Orion BN/KL is, however, much hotter (see Fig.~\ref{fig:f9}) so that no measurable C$^+$ fractionation is expected, in agreement with our results from Sec.~\ref{CplusTau} and those of Ossenkopf et al. (2013) for the Orion Bar.  Moreover, the chemical models show that CH$^+$ only partially inherits the isotopic fractionation from C$^+$, so that a noticeable isotopic fractionation in CH$^+$ only occurs at column densities well below the detectable level (Fig. 16 from R{\"o}llig \& Ossenkopf 2013). Hence, both model computations and our $^{13}$C$^+$ measurements indicate that the isotopic ratio in CH$^+$ follows the elemental abundance ratio of $^{12}$C/$^{13}$C$\approx 67$.

While we could successfully characterize the physical conditions leading to the formation and excitation of CH$^+$  around Orion BN/KL in the framework of a PDR, with parameters describing physical and chemical properties similar to those of the Orion Bar,  the reservoir of vibrationally excited H$_2$ needed to react with C$^+$ is not the most observationally evident source---namely in H$_2$ tracing shocked gas from the energetic BN/KL outflow.   Figure~\ref{fig:f26}a shows H$_2$ 2.122 $\mu$m  1 - 0 S(1) tracing the BN/KL outflow in observations published by Cunningham (2006) and Bally et al.  (2011), with contour overlays of our CH$^+$ J $= 1-0$ and C$^+$ observations on a relative intensity scale.   The origin of the outflow is attributed to a recent ($\sim$500 - 1000 years) stellar merger associated with the IRc~2 source or the dynamical decay of non-hierarchical multiple system, with velocities in the 10 - 30 \kms range.   Neither the distribution nor kinematics of these features traced in H$_2$ are consistent with our CH$^+$ mapping data, nor can we identify a comparable morphology of C$^+$ in any observed velocity channel.  The C$^+$ contours in Figure~\ref{fig:f26}a are from the $+$9.7 {\kms} velocity channel, and it might be tempting to suggest some boundary interactions particularly on the eastern side outflows where the strongest C$^+$ emission is indicated. However, this may be purely coincidental structure of the C$^+$ gas, and we detect no CH$^+$ emission in this velocity range, particularly in the region not affected by the confusion from the strong SO$_2$ and CH$_3$OH emission lines from the dense Hot Core. 

Moreover, spectral observations do not indicate formation being traced in the shocked gas, because the profiles of CH$^+$ and CH show no significant broadening or Doppler-shifted emission associated with shocks and outflows.  This is consistent with the lack of a correlation between CH$^+$ and the shock-excited H$_{2}(v=1)$ (Fig.~\ref{fig:f4}).  These observations do no rule out interactions or shock chemistry involving C$^+$ in the outflow, simply that we cannot detect such interactions through the gas in which the outflow is embedded, indicating that the chemistry which dominates this region is from UV irradiation.    These points lead us to conclude that the relatively quiescent CH$^+$ gas cloud envelops OMC~1 and traces the warm surface regions of molecular clumps of gas over an extended region, from Orion North to the Orion Bar, at the 8 $-$ 10 {\kms} system velocity of the nebula, but is not physically associated with the outflow source of higher velocity vibrationally excited H$_2$ to a level we can detect with HIFI.

UV-irradiation and chemical formation pumping can supply the vibrationally excited H$_{2}$ required for efficient CH$^+$ formation, and evidence favors UV-irradiation as a source of vibrationally excited H$_{2}$ in the regions of the nebula where CH$^+$ is observed. A significant fraction of the CH$^+$ emission in the Orion nebula probably arises in photon-dominated gas of modest density ($n_{\rm{H}} \sim 10^4$ cm$^{-3}$) that is not prominent in the vibrational lines, but still contains sufficient vibrationally excited H$_2$ to efficiently form CH$^+$. Where the vibrational lines are measured, they can be used to assess the gas properties. Emission from the $v=1-0$ and $v=2-1~S(1)$ ro-vibrational lines of H$_2$ near $2~\mu$m has been detected in the Orion Bar PDR and OMC~1 (Beckwith et al. 1978; van der Werf et al. 1996).  As discussed by \citet{SD89}, the intensity ratio $r_{21}$ ($\equiv I_{10}/I_{21}$) of the two lines depends on the UV flux ($\chi$), as well as the gas density ($n_\mathrm{H}$).  In the Orion Bar, $r_{21}$ has been found to range from a low value ($3.4\pm1.9$)---attributed to purely radiative fluorescence of H$_2$ from relatively low-density ($n_\mathrm{H} \leq 10^{4}$~cm$^{-3}$) gas of the so-called ``inter-clump medium'' (far from the ionization front illuminated by the intense UV flux of the Trapezium stars), to a high value ($8.1\pm0.7$)---attributed to thermal H$_2$ line emission in warm dense ($T\sim2000$~K, $n_\mathrm{H} \geq 10^{5}$~cm$^{-3}$) molecular gas heated by intense UV radiation \citep{VSS96}.  Similarly large values of $r_{21}$ have been reported for Orion BN/KL ($\sim10$; Beckwith et al. 1978) over a region which overlaps with much of the region mapped in the present study.  As demonstrated by \citet{SD89}, while a large $r_{21}$ ($\geq 10$) is usually attributed to shock-heating, it does not prove the presence of shock waves, particularly in the vicinity of strong ultraviolet sources, because H$_2$ line emission produced by shock-heating of the gas may be quite difficult to distinguish from similar thermal H$_2$ line emission from quiescent dense gas heated by UV photons.  The overall lack of correspondence in the observed CH$^+$ and shock-excited H$_2$ distribution (Fig.~\ref{fig:f26}a), and the similarity of PDR model parameters which reproduce the CH$^+$ line intensities in Orion BN/KL with those of the Orion Bar (Fig.~\ref{fig:f16}), support this view, and are strongly suggestive of the principle role of UV-irradiation in the production of CH$^+$ in Orion BN/KL.

Further support for UV-driven rather than shock-driven production of CH$^+$ is provided by a straightforward estimate of the production timescales of CH$^+$, which takes into account all formation and destruction processes of CH$^+$.  For example, obtaining $n$(CH$^+$) = $1.86\times10^{-2}$ cm$^{-3}$ at $0.3~A_\mathrm{V}$ and at a net production rate of $1.2\times10^{-11}$ cm$^{-3}$s$^{-1}$ (at $T = 800$~K; cf. Fig.~\ref{fig:f17}) requires approximately 49 yr---more than an order of magnitude shorter than the 500 $-$ 1000 yr dynamical age estimated for BN/KL outflow.  Thus, it seems far more likely that the CH$^+$ abundance is maintained by a steady-state UV-driven PDR chemistry than by shocks.  Undoubtedly, the effect of shocks as well as UV must be considered to explain the molecular emission and abundances in Orion BN/KL and similar regions, but a quantitative evaluation of the relative contributions of shocks and UV to the production and excitation of H$_2$ and other molecules requires sophisticated models combining UV and shock excitation, which are still quite uncertain (e.g., Kristensen et al. 2008; Visser et al. 2012), and are beyond the scope of the present work. 

The timescale argument that supports our observations and modeling of the CH$^+$ as arising in the diffuse gas with a UV-driven chemistry does not explain why, on the contrary, no CH$^+$ is observed in the outflow, where excited H$_2$ is abundant. This is possibly due in part to differences in the degree of excitation in the two environments.  While C$^+$ coexists with H$_2$ in multiple vibrationally excited levels in PDRs where fluorescence populates the higher levels,  propensity rules suggest that the H$_2$ goes mostly to the lowest vibration level in shocks, i.e., H$_2$ is thermalized at $\nu$ = 1.  This provides the required energy for C$^+$ to react with the H$_2$, but leaves little additional energy for the CH$^+$ to be hot enough to emit, in contrast with the average amount of vibrational heating of $\approx$9920 K available for all 15 vibrational levels of H$_2$ (R\"ollig et al. 2006; Nagy et al. 2013),  some 5300 K above the activation barrier. 

We should nonetheless expect some CH$^+$ emission where both H$_2$($\nu$=1) and C$^+$ are abundant in a shocked environment, such as observed in the DR 21 outflow (Falgarone et al. 2010).  Hence the apparent lack of a correlation of C$^+$ at any velocity with the Orion~KL outflow provides another clue for the absence of CH$^+$ there.  A plausible mechanism for C$^+$ depletion in the outflow is in one of the CO production pathways.  In a standard gas phase chemistry (e.g., Liszt, Lucas, \& Pety 2005), CO is formed by first producing CO$^+$, via C$^+$ + OH $\rightarrow$ CO$^+$ + H, followed by either CO$^+$ + H $\rightarrow$ CO + H, or CO$^+$ + H$_2$ $\rightarrow$ HCO$^+$ + H and HCO$^+$ + {\em{e}} $\rightarrow$ CO + H.  Orion BN/KL is known to be abundant in the products of these reactions, which all proceed relatively fast.  Furthermore, high velocity ($\sim$ 70 $-$ 100 {\kms}) CO from low to very high $J$ levels is observed to correlate well with each of the H$_2$($\nu$=1) outflow peaks (Zapata et al. 2009; Peng et al. 2012b; Goicoechea et al. 2015a).  For contrast with the distributions of CH$^+$ and C$^+$ in relation to the outflow in Figure~\ref{fig:f26}a, we also show contours of CO ($J=10-9$) and OH (1834 GHz triplet) observed with HIFI in Figure~{\ref{fig:f26}}b, overlaid onto the H$_2$ 2.12 $\mu$m imaging.  The CO was integrated between $\pm$(70 $-$ 100) {\kms}, where blue- and red-shifted emission contours are correspondingly color coded.  The OH red-shifted emission was integrated between $+$(50 $-$ 80) {\kms}; blue-shifted emission could not be measured due to substantial blending with CO $J=13-12$ emission occurring in the opposite sideband.   The distributions of this material agrees will with the previous published results of the fast-moving CO in relation to the H$_2$ fingers, indicating that they are dynamically coupled (e.g., Peng et al. 2012b).   Hence  C$^+$ in the outflow may be chiefly consumed in the production of CO, with higher efficiency compared to the reaction with H$_2$($\nu$=1).   This hypothesis needs to be tested with  further modeling that incorporates shock chemistry to account for absent or suppressed CH$^+$ emission from the outflow (see Sec.~\ref{sec:future}).

\section{Summary and Conclusions}

We have presented high SNR velocity-resolved spectral mapping observations of $^{12}$C$^+$, $^{13}$C$^+$, CH$^+$ rotational transitions, CH $\Lambda$-doubling transitions, and CH$_3$OH torsional lines for the purposes of characterizing their distributions and kinematic properties around the dynamic Orion BN/KL SFR, and have focused on the formation and excitation of the CH$^+$ ion.

\begin{enumerate}

\item  The C$^+$ observations reveal a complex ionized cloud structure of a PDR surrounding BN/KL, with multiple LSR velocity components between $-$15 to $+$20 {\kms} following distinct spatial distributions.  The two principal emission components show narrow ($2 - 3$ \kms) and broad ($4 - 7$ \kms) velocity widths that we interpret to trace (respectively) higher density material and the lower density inter-clump medium.  Significant variation of the relative strengths of each velocity components indicates a corresponding spread in optical depth conditions in the nebula.  The average column density and optical depth of $^{12}$C$^+$ outside the Hot Core region is comparable to that of the Orion Bar peak reported by Ossenkopf et al. (2013), but at the emission peaks around Orion BN/KL the values of $N$(C$^+$) and $\tau_{12}$ are factors of 2 to 3 higher compared to the Orion Bar.  Excitation temperatures  of $200-300$ K also tend to be higher. The average $\langle T_{\rm{ex}} \rangle$ of the C$^+$ emitting gas outside of the Hot Core is almost a factor 2 higher than the estimated average for OMC~1 by Goicoechea et al. (2015b).

\item Absorption of background continuum radiation from heated dust peaking at the Hot Core source SMA~1 is observed in the CH$^+$ $J=2-1$ transition, and in C$^+$ by decomposition of the composite profile.  No absorption by the CH$^+$ $J=1-0$ line in the 850 GHz continuum is detected (but weak absorption is not ruled out), setting limits of $T_{\rm{ex}}$(CH$^+$) = $10-25$ K in the absorbing gas.  The CH$^+$ emission material has a similar (low) excitation temperature.  The gas with C$^+$ in absorption similarly has $T_{\rm{ex}}$(C$^+$) $<$ 28~K, but has higher optical depths compared to CH$^+$, while $\langle T_{\rm{ex}}({\rm{C}}^+) \rangle \simeq$ 175~K for the emitting gas.       

\item On average the kinematics of the C$^+$ gas around Orion BN/KL are consistent with earlier large-beam observations of Orion A (Balick et al. 1974), which is in turn similar to CH$^+$ absorption line observations of OMC-2 FIR4 and leading L{\'o}pez-Sepulcre et al. (2013) to conclude that this northern cloud region is an extension of the C$^+$ interface between Orion BN/KL and the Trapezium.  On the finer spatial scale of our HIFI maps of this region, the C$^+$, CH$^+$, and CH emission profiles share similar LSR velocities, but with distinct differences in kinematic complexity---the CH and CH$^+$ gas is more quiescent, exhibiting much lower variations in LSR velocities and line broadening.  All CH and CH$^+$ lines can be fit with a single Gaussian profile at most positions, with average widths $\langle \Delta v \rangle \simeq 5.3 \pm 0.6$ {\kms} 

\item The distribution of CH$^+$ is similar to that of the main emission component of C$^+$, but there is a tendency for CH$^+$ emission to be strongest where intensity gradients in C$^+$ are highest.  At all locations, the measured {\vlsr} of CH$^+$ at peak line center is red-shifted by $\approx$ $+$2 {\kms} relative to the main C$^+$ emission feature.  The distribution of C$^+$ over the range of CH$^+$ velocity channels is the same.   Secondary kinematic structures of C$^+$ (outside the range of $+6.0$ to $+10.0$ {\kms}) do not have any spatial CH$^+$ correspondences.   A weak red-shifted and broader ($\Delta v \approx$12 \kms) secondary component to the CH$^+$ profile is detected at the brightest peaks nearest to the Hot Core and towards the Hot Core itself, at \vlsr \ $\approx$13 \kms.  

\item Measurements of $^{13}$C$^+$ hyperfine lines and the total $^{12}$C$^+$ emission at several positions of peak intensity yield $\tau({^{12}C^+})$ in the range of 1.4 to 2.6 along lines of sight of strong integrated emission, and up to 6.2 towards the Hot Core.  The estimates are based on assuming a constant $^{12}$C$/^{13}$C abundance ratio of $67\pm10$, and our estimates $T_{\rm{ex}}$(C$^+$) = $200-300$ K outside the Hot Core and $\approx$120 K in the Hot Core region.  Our approach and results support conclusions for the Orion Bar PDR by Ossenkopf et al.  (2013) that chemical fractionation of carbon does not play a detectable role in influencing the isotopic balance. 

\item CH$^+$ line intensities can be reproduced by PDR and non-LTE radiative transfer models, without the need to introduce shock models for the high activation barrier of CH$^+$ to be overcome in the reaction between C$^+$ and vibrationally excited H$_2$. Our conclusion is that PDR conditions describing the Orion Bar in which electron collisions affect the excitation of CH$^+$, requiring also the inclusion of reactive collisions, are similar in Orion BN/KL.  Unlike the Orion Bar, however, the strong continuum emission from KL causes the CH$^+$ excitation temperatures to increase by factors of 1.2-2.9, while integrated intensities decrease for the three lowest transitions by factors of 1.5-1.0, and decrease by 8-21\% for the $J=4-3$ and $J=5-4$ transitions.   

\item The line shapes and excitation of CH $\Lambda$-doubling lines in the $^2{\Pi}_{3/2}$ and $^2{\Pi}_{1/2}$ states can be modeled well using a one component model with an excitation temperature of $\sim35$~K.  The CH abundance relative to H$_2$ appears to agree with the Sheffer et al. (2008) relationship, and does not seem to decline as suggested by Polehampton et al. (2010).  We conclude that the CH originates in gas where the density is a few 10$^6$ cm$^{-3}$, is not in thermal equilibrium, and is concentrated in the $^2{\Pi}_{1/2} {\rm{J}}=1/2$ and $^2{\Pi}_{3/2}$ $J = 3/2$ levels. The population of the $^2{\Pi}_{1/2} {\rm{J}}=3/2$ and $^2{\Pi}_{3/2} {\rm{J}}=5/2$ levels is dominated by absorption of the continuum. However the 1661 GHz transition, observed only toward the BN/KL complex, is in the emission wing of the $2_{2,1}-2_{1,2}$ water line ($E_u=194$~K) making the analysis more uncertain.  Assuming that the continuum level is set by the water wing, the absorption is less deep than would be expected, implying that the absorption and emission originate at least in part from the same gas, similar with our conclusions for the absorbing and emitting CH$^+$ gas towards the BN/KL complex.

\item Similar distributions and line ratios of CH and CH$^+$ explained by special geometry (face-on or clumpy) suggest that whatever the amount of CH$^+$ synthesized via non-equilibrium processes in these lines of sight, an amount of CH that is a constant fraction of the amount of CH$^+$ is also formed. If this result were true in general, then we would expect some minimum amount of CH to be present whenever CH$^+$ is detected.  Analyses like those presented here, but for other SFR lines of sight, are needed to assess how narrow a range in $N$(CH)/ $N$(CH$^+$) (which is less than unity in Orion BN/KL) is generally present in the region of CH$^+$ synthesis.

\item We showed that the spatial morphology of the prominent BN/KL outflow traced by shock-excited H$_2$ does not correlate well with the observed distribution of CH$^+$.   Our excitation and PDR modeling demonstrates that UV irradiation plays the critical role of providing a distributed reservoir of H$_2$ to react with C$^+$ in the extended PDR to produce CH$^+$.   This is supported by earlier observations of $\nu = 1-0$ and $2-1$ $ S$(1) ro-vibrational lines of H$_2$ in thermal emission close to the Trapezium stars and in fluorescence further away.  Timescales to reach equilibrium production of CH$^+$ are at least an order of magnitude shorter than the dynamical age of the outflow.  Hence, as indicated by homogeneous line profiles across the region, production of CH$^+$ is more likely to be maintained by a steady-state UV-driven PDR chemistry rather than by a shock chemistry due to the large disparities between the production rate needed to reach observed abundances compared to the approximate age of the Orion BN/KL outflow.  Furthermore, C$^+$ does not correlate well at any observed velocity with the shock-excited H$_2$ features, indicating that a shock chemistry involving C$^+$ may be competing with, or inhibiting, production of CH$^+$.  We suggest that C$^+$ is most likely being consumed in the production of CO, which is observed to have high-velocity wings with a spatial distribution that is consistent with the outflow, in reactions involving OH, CO$^+$, HCO$^+$, and H$_2$.   Radiative association between C$^+$ and H$_2$ to form CH is not favored, due to the same lack fo spatial correspondance of observed CH as CH$^+$ with the outflow.

\item From our PDR models for this region C$^+$ is found to trace outer cloud boundary layers to depths of  $A_V < 3$ mag, consistent with analysis of lower sensitivity mapping of C$^+$ over the larger OMC~1 region by Goicoechea et al. (2015b).  In models by Goicoechea et al. the C$^+$ abundances peak about 1 mag deeper in $A_V$, a difference which is expected from the larger dust grain sizes built into their adopted extinction parameters.  We also estimate higher C$^+$ excitation temperatures and optical depths  for the inner 10 arcmin$^2$ around the BN source.  Electron densities peak at similar depths as C$^+$, giving the interesting result that CH abundances also peak as deeply into the clouds as C$^+$, while the highly reactive CH$^+$ molecules trace the warmest layers only to $A_V \; <$ 0.6 mag.  Nonetheless it is clear that CH$^+$ is formed in the transition zone between the atomic and molecular gas.  In this layer, H$_2$ is not optically thick to the UV field, and as a result, it is strongly excited through its electronic states.  Due to the lack of an electronic dipole moment, H$_2$ is inefficient in cooling itself.  The net result is highly excited H$_2$, both rotationally and vibrationally.  

\item CH$_3$OH transitions extracted towards SMA~1 and the Compact Ridge, tracing the densest molecular clouds to higher $A_V$ in the dusty core region, exhibit two distinct kinematic components at both locations.  A broad $\Delta v$ = 7.5 - 14.5 \kms set of lines dominates narrow $\Delta v$ = 1.5 - 2.7 \kms at SMA~1, and vice versa at the Compact Ridge.  These kinematic results are generally consistent with previous investigations by Menten et al. (1986, 1988), Wilson et al. (1989), and Wang et al. (2011).  The separate velocity components follow a distribution consistent with results by Wilson et al. (1989), but are indicated to be more extended in the sub-mm transitions than at 23 GHz (possibly due to higher sensitivities in the HIFI observations),  generally more consistent with recent 220$-$230 GHz SMA observations by Feng et al. (2015).    The excitation temperatures we have estimated are lower than the results by Wang et al. (2011) for the Hot Core, possibly indicating that level populations are not in statistical equilibrium or that a single excitation temperature is invalid in the population diagram analysis.  Similarly, column densities are somewhat higher according to Wang et al., who approximated the emitting region to be confined to a 7$''$.5-diameter gas clump (compared to the $\sim$30$''$ diameter regions of narrow and broad emission in our maps).

\end{enumerate}

The basic modeling we presented in Sections~\ref{radex} and \ref{PDRmodels} captures the essential elements of the chemical and physical conditions in the regions of CH$^+$ and CH formation, but other processes such as shocks and intermittent turbulent dissipation may need to be included in a complete model of Orion~KL.   Observations do not indicate detectable levels of CH$^+$ that we can associate with shock chemistry, and the levels of intermittency required to reproduce the observed abundances through the dissipation of turbulence await elucidation by three-dimensional simulations (Godard, Falgarone, \& Pineau des For{\^e}ts 2014).   Even where shocks are known to be present, the importance of a UV driven chemistry for CH$^+$ production should not be underestimated, especially in the case of actively star forming galaxies where CH$^+$ has been detected , for example in Arp~220, NGC~253, and M~83 where the $J=1-0$ line is observed in absorption (Rangwala et al. 2011, 2014) and in Mrk~231 where it is in emission (van der Werf et al. 2010). 

\section{Future Work}\label{sec:future}

In order to improve our understanding of the relative roles of UV and shock chemistries in Orion BN/KL with application to other SFRs, we can study other endothermic processes to test the consistency of our models.  In particular the SH$^+$ ion shares a production pathway similar to that of CH$^+$ but with a factor $\simeq$2 higher energy barrier, S$^+$ + H$_2$ $\rightarrow$ SH$^+$ + H with $\Delta E$ = 9860~K.  This could be tested with detections of the rotational transitions of SH$^+$ $N_J = 1_2 - 0_1$ towards the bright CO peak in the Orion Bar by Nagy et al. (2013).  SH$^+$  was not mapped across Orion BN/KL, however we can exploit HIFI spectral scans of the Hot Core and Compact Ridge (see Crockett et al. 2014) and observations of transitions accessible from the ground (e.g. M{\"u}ller et al. 2014 observed the Orion Bar)  which we are following up at the Atacama Pathfinder Experiment (APEX) telescope.  CO+ transitions provide another means to explore the production
mechanisms of molecules in highly UV irradiated environments.  In addition, a test for the role of the radiative association between C$^+$ and H$_2$  involving the molecular ions CH$_2^+$ and CH$_3^+$  as intermediate products would be their detection in any environment where CH$^+$ is abundant (cf. a first attempt by Indriolo et al. 2010). The CH$_3^+$ ion is particularly intesresting and has ro-vibrational transitions that can be observed in the near-IR from the ground.  CF$^+$ is another interesting ion involving C$^+$ reactions with HF, and has low-$J$ rotational transactions which have been observed from ground towards the Orion Bar (Neufeld et al. 2006) and the Horsehead Nebula (Guzman et al. 2012). 














\smallskip
\smallskip

\centerline{{\em{Acknowledgments}}}

This work is based on observations made with the HIFI instrument on the {\em{Herschel}} Space Observatory, which was designed and built by a consortium of institutes and university departments from across Europe, Canada and the United States (the National Aeronautics and Space Administration, NASA) under the leadership of the Netherlands Institute for Space Research (SRON), Groningen, The Netherlands, and with major contributions from Germany, France and the US.  We express our gratitude to Nathan Crockett for helpful discussions and providing the Orion BN/KL continuum data, and to John Bally and Nathan Cunningham for providing the H$_2$ observations of the BN/KL outflow.  We also thank Octavio Roncero and Alexandre Zanchet for providing their CH$^+$ state-to-state formation rates, and we especially appreciate support with the Meudon PDR code from Frank Le Petit.  We are grateful to the HIFI Instrument Control Center team for its many years of dedicated work and support.  We also thank an anonymous referee for thorough review of the manuscript and thoughtful comments to improve its quality.   Support for this work was provided by NASA ({\it{Herschel}} GT funding) through an award issued by JPL/Caltech.  VO acknowledges support by the Deutsche Forschungsgemeinschaft (DFG) via the collaborative research grant SFB 956, project C1.    EF and MG thank the CNES and the INSU program PCMI for funding.  A part of this research was performed at the Jet Propulsion Laboratory, California Institute of Technology, under contract with NASA.  HG acknowledges support from the National Science Foundation (NSF).  Any opinions, findings, and conclusions in this article are those of the authors, and do not necessarily reflect the views of the NSF.














\end{document}